\documentclass[12pt]{article}

\usepackage{tipa}
\usepackage{amsmath}
\usepackage{amssymb}
\usepackage{authblk}

\usepackage{natbib}
\usepackage{graphicx}
\usepackage{xr}
\usepackage{bbm}
\usepackage{algorithm, algorithmic}

\usepackage{blindtext}
\usepackage{hyperref}

\usepackage{amsfonts}
\usepackage{multirow}
\usepackage[deletedmarkup=xout, final]{changes}

\usepackage{calrsfs}
\usepackage{cancel}
\DeclareMathAlphabet{\pazocal}{OMS}{zplm}{m}{n}
\providecommand{\keywords}[1]{\textbf{\textit{Keywords:}} #1}

\newtheorem{defn}{Definition}

\newcommand{\vect}[1]{\boldsymbol{#1}}
\newcommand{\matr}[1]{\boldsymbol{#1}}
\newcommand{\sample}[1]{\pazocal{#1}}
\newcommand{\E}{\operatorname{\mathbb{E}}}

\newcommand{\cov}{\operatorname{\mathbb{C}ov}}
\newcommand{\trace}{\operatorname{tr}}

\newcommand{\EM}{\operatorname{\mathsf{EM}}}

\graphicspath{
{./figure/}
{./figure/ncomp/}
{./figure/multivariate/}
}

\begin{document}

\title{Torus Probabilistic Principal Component Analysis}

\author{Anahita Nodehi$^{1,2,\dagger}$, Mousa Golalizadeh$^{1,2,\dagger}$, Mehdi Maadooliat$^3$, Claudio Agostinelli$^4$ \\
\small{
$^1${Department of Statistics, Tarbiat Modares University, Tehran, Iran}\\
$^2${School of Biological Science, Institute for Research in Fundamental Sciences (IPM), Tehran, Iran}\\
$^3${Department of Mathematical and Statistical Sciences, Marquette University, Milwaukee, USA}\\
$^4${Department of Mathematics, University of Trento, Trento, Italy}
}
}

 \markboth
 {A. Nodehi and others}
{Torus Probabilistic Principal Component Analysis}

\footnotetext[1]{{\bf Corresponding author:} 
Anahita Nodehi \& Mousa Golalizadeh, \\ Department of Statistics, Tarbiat Modares University, Tehran, Iran. \\ 
Email: ana\_nodehi@yahoo.com - golalizadeh@modares.ac.ir.
}
\date{~}
\maketitle

\begin{abstract}

Analyzing data in non-Euclidean spaces, such as bioinformatics, biology, and geology, where variables represent directions or angles, poses unique challenges. This type of data is known as circular data in univariate cases and can be termed spherical or toroidal in multivariate contexts. In this paper, we introduce a novel extension of Probabilistic Principal Component Analysis (PPCA) designed for toroidal (or torus) data, termed Torus Probabilistic PCA (TPPCA). We provide detailed algorithms for implementing TPPCA and demonstrate its applicability to torus data. To assess the efficacy of TPPCA, we perform comparative analyses using a simulation study and three real datasets. Our findings highlight the advantages and limitations of TPPCA in handling torus data. Furthermore, we propose statistical tests based on likelihood ratio statistics to determine the optimal number of components, enhancing the practical utility of TPPCA for real-world applications.
\end{abstract}


\section{Introduction}
\label{sec:introduction}

Angular data, characterized by a large number of variables, presents significant
challenges when applying statistical methods. These challenges include difficulties in data visualization, increased computational demands, and a higher risk of over- or under-fitting during modeling. To address these issues, dimension reduction methods are commonly employed. Principal Component Analysis (PCA) is a well-known technique in classical Euclidean space, with origins dating back to \citet{Pearson1901} and \citet{Hotelling1933}. The main idea of PCA is to reduce the dimensionality of a data set by uncorrelated components, retaining as much of the variation present in the data set as possible \citep{Jolliffe2002}. The PCA is also one of the main tools extracting information from large datasets in bioinformatics; to name a few, we can mention problems in genetics \citep{PriceETAL:2006, NovembreStephens:2008, PriveETAL:2020}, in meta-analysis \citep{kimETAL:2018} and shape analysis \citep{Wiseman2021,moghimbeygi2022}. Classic PCA lacks a probabilistic framework and associated likelihood measures, which can limit specific applications. A probabilistic formulation is often preferred because of its capacity to enable comparisons with other probabilistic techniques, facilitate statistical testing, and support the application of Bayesian methods. It also improves the handling of missing data \citep{TippingBishop1997}. Probabilistic PCA (PPCA) is an extension of PCA, offering a maximum likelihood solution within a probabilistic latent variable model. PPCA was independently introduced by \citet{TippingBishop1997, TippingBishop1999} and \citet{Roweis1998} and is closely related to factor analysis \citep{Basilevsky1994}.

There are numerous situations in applied sciences, such as biology, bioinformatics, astronomy, and geology, where the data are supported in non-Euclidean spaces. For example, wind directions (in meteorology), animal navigation (in biology), and the epicenter of an earthquake (in earth sciences) are three classical instances of direction measurements in applied sciences.
It is common to specify a measure as an angle on a unit circle once the circle's initial direction and orientation have been chosen.
In bioinformatics, angles are the suitable components to represent the protein structure. In detail, there are two torsional angles ($\phi, \psi$) in the protein backbones for each amino acid residue \citep{Maadooliat2016}. In RNA, the dimensionality is much more complex. For each nucleotide residue, there are seven independent torsion angles and one angle that describes the rotation of the base relative to the sugar \citep{Eltzner2018}. The statistical analysis of such angular data is known as directional statistics \citep{Mardia1972}. Some authors named directional data as ``circular data'' and ``angular data'' in univariate cases and ``spherical data" or ``torus data" in multivariate cases. Accordingly, these data types are also part of the so-called ``non-Euclidean data'' or ``manifold-valued data''. There have been many attempts to extend PCA on manifold data. To name a few: Principal Curve \citep{HastieStuetzle1989}, Principal Geodesic Analysis (PGA) \citep{Fletcher2004}, Geodesic Principal Component Analysis (GPCA) \citep{HuckemannZiezold2006}, dihedral angles Principal Component Analysis (dPCA) (\citet{Mu2005} and \citet{Altis2007}), and angular PCA (aPCA) \citep{Riccardi2009}, Principal Arc Analysis (PAA) \citep{Jung2010}, Principal Nested Analysis (PNS) \citep{Jung2012}, Principal Geodesic Analysis on the shape space \citep{Fotouhi2012,Fotouhi2015}, Principal Flows (PF) \citep{Panaretos2014}, dihedral angles Principal Geodesic Analysis \citep{Nodehi2015}, PCA on torus (dPCA$+$) \citep{Sittel2017}, Torus Principal Component Analysis (T-PCA) \citep{Eltzner2018} and Scaled Torus PCA (ST-PCA) \citep{zoubouloglou2021}.

In the literature exploring torus-specific PCA techniques for analyzing toroidal data, we explore modern methods designed for practical application. 
The dPCA, introduced by \citet{Mu2005} and \citet{Altis2007}, transforms dihedral angles into sinusoidal and cosinusoidal representations. Furthermore, angular PCA (aPCA) and Cartesian PCA (cPCA) apply PCA to torus data, centered on their circular means \citep{Riccardi2009}. \citet{Altis2007} proposes Complex dPCA as an alternative extension to the sin/cos transformation in dPCA, transforming angles into complex numbers using Euler’s formula. Further advancements include dihedral angles Principal Geodesic Analysis \citep{Nodehi2015}, which is an extension of PGA to dihedral angles in protein structures. The PCA on the torus (dPCA$+$) \citep{Sittel2017} is a customized solution that minimizes projection errors in the torus data due to periodicity. Torus Principal Component Analysis (T-PCA) by \citet{Eltzner2018} deforms tori into spheres using PNS and introduces data-adaptive pre-clustering. In addition, \citet{Eltzner2018} proposed two data-driven orderings of variables, \textit{SI ordering} and \textit{SO ordering}, corresponding to the order of the variables in terms of the decreasing and increasing amount of circular variability, respectively. Lastly, Scaled Torus Principal Component Analysis (ST-PCA) by \citet{zoubouloglou2021} is a recent method based on spherical embeddings, using Spherical MultiDimensional Scaling (SMDS) to map from a torus to a sphere. In ST-PCA, the analyses are done on spheres, and after finding the best fit of data, it can be inverted back to the torus. For detailed technical information, comprehensive descriptions of these methods can be found in the Supplementary Material.


Despite advances in extending PCA methods, a significant gap remains in the application of probabilistic modeling to manifold-valued datasets, particularly in the context of torus data. While recent developments have extended PCA to non-Euclidean spaces using probabilistic approaches, such as the Probabilistic PGA for spherical data introduced by \citet{ZhangFletcher2013} and the method for genotype data proposed by \citet{AgrawalETAL:2020}, there is still no established probabilistic PCA framework specifically designed for toroidal data. This gap is critical, given the unique properties of toroidal manifolds that standard PCA techniques cannot adequately address. To address this need, we introduce Torus Probabilistic PCA (TPPCA), a novel extension of Probabilistic PCA designed explicitly for torus data. Our approach includes an efficient iterative algorithm that fully utilizes the unique structure of toroidal manifolds, providing a robust solution to the challenges of feature extraction in this space. TPPCA not only allows for comparisons with other probabilistic methods but also enables the application of statistical testing to determine the optimal number of components. This novel technique provides a valuable tool for analyzing toroidal data.

The remainder of this paper is organised as follows: In Section \ref{sec:tppca}, we introduce the methodology of TPPCA and a discussion on the selection of the number of components. The empirical contributions of the paper are highlighted in Section \ref{Sec3}, where we present the results of our simulation study and explore three real-world applications, demonstrating the practical implications of our proposal. Lastly, in Section \ref{sec:conclusion}, we present our concluding thoughts and remarks, wrapping up the key insights and contributions of the paper.

\section{Torus Probabilistic Principal Component Analysis}
\label{sec:tppca}
We introduce an extension of PPCA for toroidal data. Let $X \in \mathbb{R}$ be a real random variable, a circular random variable $Y$ might be written as $Y = X \bmod 2\pi \in [0, 2\pi)$ or $X = Y + 2\pi K$ for some $K \in \mathbb{Z}$. Let $\mathbb{T}^D = [0, 2\pi)^D$ be the $D$-torus. The TPPCA is a latent variable model that seeks to relate a $D$-dimensional observation vector $\vect{Y} \in \mathbb{T}^D$ to a corresponding $d$-dimensional vector of latent (or unobserved) variables $\matr{Z} \in \mathbb{R}^d$ ($d<D$), that is,
\begin{align} \label{WNYX}
\vect{Y} & = \matr{X} \bmod 2 \pi \nonumber \\
\matr{X} & = \vect{\mu}+ \matr{W} \matr{Z} + \vect{\epsilon},
\end{align}
where $\matr{W}$ is a $(D\times d)$ matrix that relates the two sets of variables,  
$\matr{X} = (X_1, \ldots, X_D)^\top, \matr{Z} = (Z_1, \ldots, Z_d)^\top$, while the parameter vector $\vect{\mu}$ allows the model to have a non-zero mean. The vector $\vect{Z} \sim N_d(\vect{0}, \vect{I}_d)$ represents a $d$-dimensional Gaussian latent variable, and $\vect{\epsilon} \sim N_D(\vect{0}, \sigma^2 \vect{I}_D)$ is a $D$-dimensional zero-mean Gaussian noise variable with covariance $\sigma^2 \vect{I}_D$. We assume $\cov(\vect{Z}, \vect{\epsilon}) = \vect{0}$ and there exists a random vector $\vect{K}=\vect{K}(\vect{X})$ such that $\vect{Y} = \vect{X} - 2\pi\vect{K} \in \mathbb{T}^D$. We recall the following basic results \citep{Mardia1979}
\begin{align} \label{XZ}
\matr{X} & \sim N_D(\vect{\mu}, \matr{W} \matr{W}^\top + \sigma^2 \vect{I}_D), \nonumber \\
\matr{X}|\matr{Z} & \sim N_D (\vect{\mu} + \matr{W} \matr{Z}, \sigma^2 \vect{I}_D), \nonumber \\
\matr{Z}|\matr{X} & \sim N_d (\matr{M}^{-1} \matr{W}^\top (\matr{X} - \vect{\mu}), \sigma^2 \matr{M}^{-1}) \ , \nonumber \\
\matr{Z}|\vect{Y}  & \sim N_d (\matr{M}^{-1} \matr{W}^\top (\vect{Y} + 2\pi\vect{K} - \vect{\mu}), \sigma^2 \matr{M}^{-1}),
\end{align}
where $\matr{M} = \matr{W}^\top \matr{W} + \sigma^2 \vect{I}_d$. For a given point $\vect{x} \in \mathbb{R}^D$ we define a parameter vector $\vect{k} \in \mathbb{Z}^D$ so that $\vect{y} = \vect{x} - 2 \pi \vect{k} \in \mathbb{T}^D$, then
\begin{equation*}
f(\vect{y},\vect{x},\vect{z},\vect{k}) = f(\vect{y},\vect{k}|\vect{x},\vect{z}) f(\vect{x}|\vect{z}) f(\vect{z}) = f(\vect{y}, \vect{k}|\vect{x}) f(\vect{x}|\vect{z}) f(\vect{z}),
\end{equation*}
where
\begin{equation*}
f(\vect{y},\vect{k}|\vect{x}) = \mathbbm{1}(\vect{y}, \vect{k},\vect{x}) = \left\{
\begin{array}{ll}
1 & \text{ if } \vect{y} \in \mathbb{T}^D ~\mbox{and}~\vect{k}=\dfrac{\vect{x}-\vect{y}}{2\pi}  \\
0 & \text{ otherwise},
\end{array}
\right.
\end{equation*}
which leads to the joint distribution
\begin{align*}
& f(\vect{y}, \vect{x}, \vect{z}, \vect{k})  \propto \mathbbm{1}(\vect{y}, \vect{x}, \vect{k}) (\sigma^2)^{-\frac{D}{2}} \times \nonumber \\
&  \exp{\{ -\frac{(\vect{y}+2\pi \vect{k}-\matr{W}\vect{z}-\vect{\mu})^\top (\vect{y}+2\pi \vect{k}-\matr{W}\vect{z}-\vect{\mu})}{2\sigma^2}
  -\frac{\vect{z}^\top \vect{z}}{2} \}} .
\end{align*}
Let $\sample{Y} = (\vect{y}_1, \ldots, \vect{y}_N)$ be a sample of size $N$ and let $\sample{K} = (\vect{k}_1, \ldots, \vect{k}_N)$ a set of missing values from the random vector $\vect{K}$, then 
the logarithm of the likelihood function, say $\ell(\vect{\mu}, \matr{W}, \sigma^2, \sample{K}),$  is given by
\begin{align*}
& \ell(\vect{\mu}, \matr{W}, \sigma^2, \sample{K})  = \sum_{j=1}^{N} \ln f(\vect{y}_j ,\vect{x}_j ,\vect{z}_j , \vect{k}_j ) \\
& \propto \sum_{j=1}^{N} \left[ -\frac{D}{2}\ln{\sigma^2} - \frac{(\vect{x}_j - \matr{W} \vect{z}_j - \vect{\mu})^\top (\vect{x}_j - \matr{W} \vect{z}_j - \vect{\mu})}{2 \sigma^2} - \frac{\vect{z}_j ^\top\vect{z}_j }{2} \right] \nonumber\\
& \times \mathbbm{1}(\vect{y}_j ,\vect{k}_j,\vect{x}_j) \nonumber\\
& \propto \sum_{j=1}^{N} \left[ -D \ln \sigma^2 - \frac{\trace[(\vect{y}_j +2\pi \vect{k}_j -\vect{\mu})(\vect{y}_j +2\pi \vect{k}_j -\vect{\mu})^\top]}{\sigma^2} \right. \nonumber \\
& + 2 \left. \frac{(\vect{y}_j +2\pi \vect{k}_j -\vect{\mu})\vect{z}_j ^\top \matr{W}^\top}{\sigma^2} - \frac{\trace[(\matr{W} \vect{z}_j )(\matr{W}\vect{z}_j )^\top]}{\sigma^2} - \trace(\vect{z}_j \vect{z}_j ^\top) \right] \nonumber\\
& \times \mathbbm{1}(\vect{y}_j ,\vect{k}_j ,\vect{x}_j ),
\end{align*}
where in the last line we use the fact $\trace(\vect{AB})=\trace(\vect{BA})$. We are going to take expectation of the log-likelihood over the latent variables $\matr{Z}_1$, $\ldots$, $\matr{Z}_N$ given the data sample $\sample{Y}$, that is,
\begin{align*}
&\Gamma(\vect{\mu}, \matr{W}, \sigma^2, \sample{K} | \sample{Y})  = \E(\ell|\sample{Y}) = \sum_{j=1}^{N} \E(\ell(\vect{\mu}, \matr{W}, \sigma^2, \sample{K})|\vect{Y}_j = \vect{y}_j) \nonumber \\ 
& \propto \sum_{j=1}^{N} \Big\{ - D \ln{\sigma^2} - \trace[\E(\matr{Z}_j \matr{Z}_j^\top|\vect{Y}_j = \vect{y}_j )] \nonumber \\
& -\frac{(\vect{y}_j +2\pi \vect{k}_j - \vect{\mu})(\vect{y}_j +2\pi \vect{k}_j -\vect{\mu})^\top}{\sigma^2} \nonumber \\
& + 2 \frac{(\vect{y}_j -\vect{\mu}+2\pi \vect{k}_j ) \E(\matr{Z}_j ^\top|\vect{Y}_j = \vect{y}_j ) \matr{W}^\top}{\sigma^2} \nonumber \\
& -\frac{\trace[\matr{W} \E(\matr{Z}_j \matr{Z}_j ^\top|\vect{Y}_j = \vect{y}_j ) \matr{W}^\top]}{\sigma^2} \Big\} \nonumber \\
& \times \mathbbm{1}(\vect{y}_j ,\vect{k}_j,\vect{x}_j) \nonumber \\
& = \sum_{j=1}^{N} \Big\{ - D \ln{\sigma^2} - \trace[\E(\matr{Z}_j \matr{Z}_j ^\top|\matr{X}_j =\vect{y}_j + 2\pi \vect{k}_j )] \nonumber \\
& - \frac{(\vect{y}_j +2\pi \vect{k}_j -\vect{\mu})^\top (\vect{y}_j +2\pi \vect{k}_j -\vect{\mu})}{\sigma^2} \nonumber \\
& - \frac{\trace[\matr{W} \E(\matr{Z}_j \matr{Z}_j^\top|\matr{X}_j = \vect{y}_j +2\pi \vect{k}_j) \matr{W}^\top]}{\sigma^2} \nonumber \\
& + 2 \frac{(\vect{y}_j +2\pi \vect{k}_j -\vect{\mu}) \E(\matr{Z}_j^\top|\matr{X}_j =\vect{y}_j +2\pi \vect{k}_j ) \matr{W}^\top}{\sigma^2} \Big\} \nonumber \\
& \times \mathbbm{1}(\vect{y}_j ,\vect{k}_j,\vect{x}_j),
\label{Estep}
\end{align*}
where in the last line, given a function $g(\cdot)$ and the definition of $\vect{k}$ we see that
\begin{equation*}
\E(g(\matr{Z})|\vect{Y}=\vect{y}) = \E(g(\matr{Z})|\matr{X}=\vect{y}+2\pi\vect{k}) .
\end{equation*}

To provide an overview of TPPCA, the proposed algorithm can be summarized in one initial step (Step 0) followed by an iterative procedure consisting of two steps (Steps 1 and 2), which are repeated until convergence:

\subsection*{Step 0: Initial Estimates}
\label{sec:initial}

The initial estimates for the parameters $\vect{\mu}$, $\matr{W}$, and $\sigma^2$, as well as a reasonable imputation of $\sample{K}$, are crucial for the convergence and performance of the proposed algorithm. Initially, an unstructured variance-covariance matrix $\matr{\Sigma}$ is assumed, such that $\vect{Y} \sim WN_{D}(\vect{\mu}, \matr{\Sigma})$. The initial estimates $\hat{\sample{K}}_0 = (\hat{\vect{k}}_1, \ldots, \hat{\vect{k}}_N)$ are obtained using the Classification Expectation-Maximization (CEM) algorithm proposed by \citet{Nodehi2018}. Subsequently, a strategy similar to that used by \citet{TippingBishop1999} is employed to maximize the log-likelihood $\ell(\vect{\mu}, \matr{W}, \sigma^2, \hat{\sample{K}}_0)$ with respect to the parameters, yielding the initial estimates $(\hat{\vect{\mu}}_0, \hat{\matr{W}}_0, \hat{\sigma}^2_0)$. These estimates are then used to initiate the iterative algorithm (Steps 1 and 2).

\subsection*{Step 1: Updating $\vect{\mu}$ and $\sample{K}$}
\label{Sub:step1}

In this step, $\vect{\mu}$ is updated and the missing values of $\sample{K}$ are imputed using the CEM algorithm, as described by \citet{Nodehi2018}. The CEM algorithm is iteratively applied to maximize the conditional expectation of the log-likelihood, consisting of the following steps:
\begin{itemize}
    \item \textbf{E-step (Expectation):} Compute the expectation $\Gamma(\hat{\vect{\mu}}_{i-1}, \hat{\matr{W}}_{i-1}, \hat{\sigma}^2_{i-1}, \sample{K} | \sample{Y})$.
    \item \textbf{C-step (Classification):} Update $\hat{\sample{K}}_{i}$ by maximizing the expected complete-data log-likelihood with respect to $\sample{K}$.
    \item \textbf{M-step (Maximization):} Update $\hat{\vect{\mu}}_{i}$ by solving:
    \[
    \hat{\vect{\mu}}_{i} = \frac{1}{N} \sum_{j=1}^{N} (\vect{y}_j + 2\pi \hat{\vect{k}}_{i,j}).
    \]
\end{itemize}
This process is repeated until convergence of $\hat{\vect{\mu}}_i$ and $\hat{\sample{K}}_i$.

\subsection*{Step 2: Updating $\matr{W}$ and $\sigma^2$}
\label{Sub:step2}

After obtaining updated values for $\vect{\mu}$ and $\sample{K}$, the next step is to update the estimates of $\matr{W}$ and $\sigma^2$. This is accomplished by maximizing the log-likelihood function with respect to these parameters. The updated estimates are given by:
\[
\hat{\matr{W}}_i = \hat{\matr{S}}_i \hat{\matr{W}}_{i-1} \hat{\matr{M}}_{i-1}^{-1}(\sigma^2 \matr{I}_d + \hat{\matr{M}}_{i-1}^{-1} \hat{\matr{W}}_{i-1}^\top \hat{\matr{S}}_i \hat{\matr{W}}_{i-1})^{-1},
\]
where $\hat{\matr{S}}_i = \frac{1}{N}\sum_{j=1}^{N} (\hat{\vect{x}}_j - \hat{\vect{\mu}}_i)(\hat{\vect{x}}_j - \hat{\vect{\mu}}_i)^\top$ and $\hat{\matr{M}}_{i-1} = \hat{\matr{W}}_{i-1}^\top \hat{\matr{W}}_{i-1} + \hat{\sigma}^2_{i-1} \matr{I}_d$. The estimate for $\sigma^2$ is then updated using:
\[
\hat{\sigma}^2_i = \frac{1}{D} \trace[\hat{\matr{S}}_i(\matr{I} - \hat{\matr{W}}_i \hat{\matr{M}}_i^{-1} \hat{\matr{W}}_i^\top)],
\]
where $\hat{\matr{M}}_i = \hat{\matr{W}}_i^\top \hat{\matr{W}}_i + \hat{\sigma}_{i-1}^2 \matr{I}_d$.

This iterative algorithm, which integrates the CEM algorithm in the first step, refines the estimates for $\vect{\mu}$, $\sample{K}$, $\matr{W}$, and $\sigma^2$ until convergence. Unlike the approach in \citet{Nodehi2018}, which focused solely on estimating $\vect{\mu}$ and $\sample{K}$ with a general covariance structure, the present work extends this by also estimating the loading matrix $\matr{W}$ associated with the principal components of the wrapped normal model and $\sigma^2$. This extension enables a more comprehensive analysis of toroidal data using TPPCA. The details of this algorithm can be found in the Supplementary Material.


\subsection*{Selection of the number of components}
\label{sec:ncomp}

In PCA and Factor Analysis (FA), one of the key challenges is determining the optimal number of components to retain. Over the past several decades, various methods have been developed to tackle this problem. These include: \textbf{(a) proportion of variance} \citep{Jackson1991} which evaluates the cumulative variance explained by each component, \textbf{(b) Kaiser-Guttman:} a criterion that involves counting the number of eigenvalues of the sample correlation matrix that exceed one, as initially proposed by \citet{Guttman1954} and \citet{Kaiser1960}, \textbf{(c) eigenvalue-based methods:} the number of positive eigenvalues of the sample correlation matrix whose $j^{th}$ diagonal element is replaced by the squared multiple correlations between the $j^{th}$ variable and the rest of the $D-1$ variables, $j=1,\ldots,D$ \citep{Hayashi2007}, 
\textbf{(d) Likelihood Ratio Test (LRT):} a method that employs maximum likelihood analysis to evaluate the significance of additional components \citep{Bartlett1950,Joreskog1967}, \textbf{(e) Akaike information criterion (AIC)} \citep{Akaike1987} and its variations, including CAIC \citep{Bozdogan1994} and BIC \citep{Schwarz1978}, \textbf{(f) Cross-Validation (CV)} which is proposed by \citet{Krzanowski1987}.

Given the probabilistic nature of our model, this paper developed the use of the LRT, specifically for torus data. We explore two different LRT-based tests to assess the number of factors, both of which are conventional in covariance structure analysis \citep{LawleyMaxwel1971, Bartlett1950}.
The first approach is the standard model goodness-of-fit test (Type 1), while the second is a Chi-square difference test (Type 2). In Factor Analysis, the Bartlett correction is generally not applied to the difference test. Therefore, we employ the Chi-square difference test (Type 2) in our analysis, taking into account the differences between PPCA and Factor Analysis. This section explores the LRT and its extension to TPPCA.

If the distribution of the random samples $\sample{X}=(\matr{X}_{1},\ldots,\matr{X}_{n})^\top$ depends on a parameter vector ${\vect{\theta}} \in \Theta$ and if $H_{0}: {\vect{\theta}} \in \Theta_{0}$ and $H_{1}: {\vect{\theta}} \in \Theta_{1} = \Theta - \Theta_0$ are related hypotheses, then the likelihood ratio statistics for testing $H_{0}$ against $H_{1}$ is defined as
\begin{equation*}
\lambda = \frac{L_{0}^{*}}{L_{1}^{*}} \ ,
\end{equation*}
where $L^{*}_{i}$ is the largest value which the likelihood function takes in region $\Theta_{i}$, $i=0,1$. Equivalently, we might use the log-likelihood statistics
\begin{equation*}
-2\log{\lambda} = -2(l_{0}^{*}-l_{1}^{*})
\end{equation*}
where $l_{i}^{*}=\log{L_{i}^{*}}$, $i=0,1$ \citep{Mardia1979}. Thus, to perform the LRT, we have
\begin{align*}
l_{0}^{*} & \propto -\frac{n}{2}\log{|\hat{\matr{C}}|}-\frac{n}{2}\trace(\hat{\matr{C}}^{-1}\matr{S}) \\
l_{1}^{*} & \propto -\frac{n}{2}\log{|\matr{S}|}-\frac{n}{2}\trace(\matr{S}^{-1}\matr{S}) = -\frac{n}{2}\log{|\matr{S}|}-\frac{nD}{2}\,
\end{align*}
where $\hat{\matr{C}}=\hat{\matr{W}}\hat{\matr{W}}^\top +\hat{\sigma}^2 \matr{I}_{D}$  is an estimate of $\matr{C}$. Then the statistical test is given by
\begin{align}
U_{d}= -2\log{\lambda} & = -2\log(L_{0}^{*}-L_{1}^{*}) \nonumber\\
& = -n\log{|\hat{\matr{C}}^{-1}\matr{S}|}+n\trace(\hat{\matr{C}}^{-1}\matr{S})-nD \nonumber\\
& = -n\log{(g^{D})}+n(Da)-nD \nonumber\\
& = n D(a-\log{g}-1), \ 
\label{hypo1}
\end{align}
where $U_d$ is calculated based on the eigenvalues of $\hat{\matr{C}}^{-1}\matr{S}$. To clarify, let $a$ represent the arithmetic mean and $g$ the geometric mean of these eigenvalues. The relationship can be expressed as $\trace(\hat{\matr{C}}^{-1}\matr{S}) = Da$ and $\det(\hat{\matr{C}}^{-1}\matr{S}) = g^{D}$. For toroidal data, we utilized the two-step iterative algorithm in TPPCA to estimate the parameters $\matr{W}$ and $\sigma^2$. With these estimates, $U_d$ was subsequently computed as per equation \eqref{hypo1}.

Following the Chi-square difference test's approach for testing the $d$-component model against the $(d+1)$-component model, we consider the following hypothesis
\begin{align*}
H_{0} & : \matr{\Sigma}=\matr{C} \text{ with $\matr{W}$ is $D \times d$ matrix} \\
H_{1} & : \matr{\Sigma}=\matr{C} \text{ with  $\matr{W}$ is $D \times (d+1)$ matrix}. 
\end{align*}
It is easy to see that the LRT statistic is given by
\begin{equation*}
V_{d} = U_{d}- U_{d+1} \ ,
\end{equation*}
where $U_{d}$ and $U_{d+1}$ are computed using \eqref{hypo1}. In this regard, $V_{d}$ is approximately distributed as a Chi-square variate with $(D-d)$ degrees of freedom.

It is important to note that in either test, both \textit{backward} and \textit{forward stepwise} methods can be used. However, in the absence of prior knowledge, \citet{Hayashi2007} recommend using the forward stepwise method as the standard approach for implementing these tests.
This approach begins with a one-component model, estimating it and subsequently testing the null hypothesis $H_{0}$ with $d=1$ component. If $H_{0}$ is rejected, the process advances by incrementing $d$ by 1, leading to the estimation of a two-component model and a corresponding test for $H_{0}$ with $d=2$ components. Should the rejection of $H_{0}$ persist, the process continues with $d=3$ components, and so on. At each step in this sequence, either or both versions of the LRT can be employed. This iterative process of estimation and testing persists until the null hypothesis $H_{0}$ cannot be rejected, at which point the current value of $d$ is taken as the estimated number of components.

\section{Numerical Studies}
\label{Sec3}

This section aims to strengthen the credibility of our proposal by presenting its reliability across a wide range of relevant scenarios. To accomplish this, we utilize essential numerical studies, including ``simulation studies" and ``real data", to assess various properties, including bias. Given the scope of this paper, our simulation study primarily focuses on assessing the performance of TPPCA and PPCA, while performing sensitivity analyses through methods like Cross-Validation (CV), Kaiser-Guttman (KG), and the LRT for estimating the number of components. In contrast, the real-world applications are aimed at evaluating the performance of our proposed method relative to other PCA-specific extensions for torus data. The insights derived from these numerical analyzes enable researchers to make well-informed decisions when selecting the most appropriate methods to address their specific research questions within their datasets.

\subsection{Monte Carlo experiment}
\label{sec:simulation}

Simulation studies are included for various reasons, such as validating theoretical models, evaluating TPPCA algorithm performance, and conducting sensitivity analysis, particularly by utilizing the CV, KG, and LRT for estimating the optimal number of components. 

\added{Our goal was to conduct simulation studies with dimensions typically seen in real-world applications. According to the literature (see \citet{Sargsyan2012, Eltzner2018, Mardia2021,zoubouloglou2021}), RNA datasets are the most relevant, with dimensions up to seven in a hyper-torus. Beyond RNA datasets, we are not aware of any angular real data with dimensions greater than seven. Additionally, due to the high computational cost (See Figures SM-7--SM-8 in Supplementary Material)  and limited resources when the dimension exceeds five, we restricted our simulation studies to dimensions of five.}

\subsubsection*{Simulation Design}

We evaluate the performance of TPPCA in comparison to PPCA through an extensive Monte Carlo experiment. Our simulation design covers several scenarios, including sample sizes of $N=(50, 100, 500)$, upper dimensions $D=5$, and lower dimensions $d=(2, 3)$. \added{For simplicity, we set $\vect{\mu} = 0$ and use an orthonormal matrix $\matr{W}$ (generated using the \textit{orth} command in the $\mathbb{R}$ package \texttt{pracma}) to relate the variables $\matr{X}$ and $\matr{Z}$. Using equations \eqref{WNYX} and \eqref{XZ}, we generate $\matr{X}$ across a range of noise levels, denoted by $\sigma^2 = (\pi/8, \pi/4, \pi/2, \pi, 3\pi/2, 2\pi)$.} For each of these scenarios, we conduct a total of $1000$ Monte Carlo replications.

The simulation study consists of two parts: In the first part, $d$ is constant at its \textit{true value}. In the second part, we estimate the number of components using various methods such as CV, KG, and LRT.

To assess the performance, we employ several statistical criteria, including \textbf{(i)} Mean Square Error of the reconstructed $\matr{X}$ (denoted as $\matr{X}^{recons}$), i.e., $$\operatorname{MSE}(\matr{X}^{recons})= \dfrac{1}{N} \sum_{i=1}^{N}(\vect{X}_{i}^{recons}-\vect{X}_{i}^{orig})^2,$$ \textbf{(ii)} Mean Absolute Error of $\matr{X}^{recons}$, i.e., $$\operatorname{MAE}(\matr{X}^{recons})= \dfrac{1}{N}\sum_{i=1}^{N}|\vect{X}_{i}^{recons}-\vect{X}_{i}^{orig}|,$$ \textbf{(iii)} Mean Square Error of the reconstructed latent variable $\matr{Z}$ (noted as $\matr{Z}^{recons}$) as 
$$\operatorname{MSE}(\matr{Z}^{recons})= \dfrac{1}{N} \sum_{i=1}^{N}(\vect{Z}_{i}^{recons}-\vect{Z}_{i}^{orig})^2,$$
\textbf{(iv)} Mean Absolute Error of $\matr{Z}^{recons}$ as  
$$\operatorname{MAE}(\matr{Z}^{recons})= \dfrac{1}{N} \sum_{i=1}^{N}|\vect{Z}_{i}^{recons}-\vect{Z}_{i}^{orig}|.$$
\added{It is important to note that the term ``orig" in $\matr{X}^{orig}$ (or $\matr{Z}^{orig}$) denotes the original dataset (or latent variable) used in the simulation study, while ``recons" in $\matr{X}^{recons}$ (or $\matr{Z}^{recons}$) refers to the reconstructed data (or latent variable) obtained through the TPPCA algorithm, as applied using \eqref{WNYX}.}

\subsubsection*{Simulation Results}

We present a summary of the results in Tables \ref{tab:X}--\ref{tab:Z}. Notably, TPPCA consistently outperforms PPCA. As the standard deviation increases, the Mean Square Error (MSE) of $\matr{X}^{recons}$ in TPPCA also exhibits an upward trend. Furthermore, the sample size plays a crucial role in enhancing the accuracy of this method. In essence, by accounting for the periodic nature of angles and extending PPCA to TPPCA, significant improvements in the results are achieved. Additionally, as the lower dimension ($d$) increases, both methods exhibit an increase in these two criteria (i.e., MSE and MAE).

\begin{table}
\begin{minipage}[t]{0.75\textwidth}
\caption{\scriptsize{Performance of the TPPCA and PPCA on the reconstruction of $X$ measured by the MSE and MAE measures using true value of $d$ and $D=5$.}}
\resizebox{\columnwidth}{!}{
 \begin{tabular}{cccccccccc}
\hline
&& \multicolumn{4}{c}{$\operatorname{MSE}(\matr{X}^{recons})$} & \multicolumn{4}{c}{$\operatorname{MAE}(\matr{X}^{recons})$ } \\
\cline{4-6}
\cline{8-10}
Lower dim. & Methods& $\sigma^2$ & $n=50$ & $n=100$ & $n=500$ && $n=50$ & $n=100$ & $n=500$ \\
\hline
\multirow{12}{*}{$d=2$} &&$\frac{\pi}{8}$&      1.589 & 1.654 & 0.879 && 0.657 & 0.637 & 0.460 \\
&\multirow{4}{*}{TPPCA}&$\frac{\pi}{4}$&            2.385 & 2.152 & 1.500 && 0.930 & 0.845 & 0.671 \\
&&$\frac{\pi}{2}$&       3.285 & 3.068 & 3.068 && 1.247 & 1.190 & 1.145 \\
&&$\pi$&                      4.230 & 4.305 & 4.159 && 1.581 & 1.582 & 1.529 \\
&&$\frac{3\pi}{2}$&           4.498 & 4.634 & 4.606 && 1.663 & 1.684 & 1.676 \\
&&$2\pi$&                     4.950 & 4.960 & 4.901 && 1.747 & 1.756 & 1.752 \\
\cline{2-10}
&\multirow{6}{*}{PPCA}&$\frac{\pi}{8}$&            16.430 & 16.324 & 16.519 && 3.294 & 3.289 & 3.297 \\
&&$\frac{\pi}{4}$&            15.688 & 15.465 & 15.575 && 3.290 & 3.253 & 3.269 \\
&&$\frac{\pi}{2}$&        14.906 & 14.395 & 14.475 && 3.262 & 3.224 & 3.234 \\
&&$\pi$&                      14.289 & 13.915 & 13.808 && 3.264 & 3.240 & 3.224 \\
&&$\frac{3\pi}{2}$&           13.917 & 13.842 & 13.487 && 3.250 & 3.245 & 3.216 \\
&&$2\pi$&                     13.755 & 13.620 & 13.156 && 3.243 & 3.236 & 3.211 \\
\hline
\multirow{12}{*}{$d=3$} &&$\frac{\pi}{8}$&        3.738 & 2.868 & 2.397 && 1.152 & 0.953 & 0.788 \\
&\multirow{4}{*}{TPPCA}&$\frac{\pi}{4}$&            4.309 & 4.060 & 3.092 && 1.400 & 1.289 & 1.039 \\
&&$\frac{\pi}{2}$&       5.392 & 5.120 & 4.512 && 1.708 & 1.625 & 1.448 \\
&&$\pi$&                      5.884 & 6.073 & 5.919 && 1.890 & 1.910 & 1.865 \\
&&$\frac{3\pi}{2}$&           6.117 & 6.415 & 6.326 && 1.950 & 1.990 & 1.975 \\
&&$2\pi$&                     6.376 & 6.338 & 6.927 && 1.997 & 1.996 & 2.081 \\
\cline{2-10}
&\multirow{6}{*}{PPCA}&$\frac{\pi}{8}$&            18.018 & 17.852 & 17.988 && 3.368 & 3.344 & 3.365 \\
&&$\frac{\pi}{4}$&            17.257 & 17.459 & 17.260 && 3.345 & 3.368 & 3.339 \\
&&$\frac{\pi}{2}$&        16.519 & 16.448 & 16.262 && 3.340 & 3.341 & 3.314 \\
&&$\pi$&                      16.005 & 15.562 & 15.406 && 3.361 & 3.317 & 3.306 \\
&&$\frac{3\pi}{2}$&           15.370 & 15.405 & 15.184 && 3.328 & 3.325 & 3.318 \\
&&$2\pi$&                     15.413 & 15.232 & 15.037 && 3.335 & 3.329 & 3.310 \\
\hline
\hline
\end{tabular}} \\
\\
\label{tab:X}
\end{minipage}
\begin{minipage}[t]{.75\textwidth}
\caption{\scriptsize{Performance of the TPPCA and PPCA on the reconstruction of $Z$ measured by the MSE and MAE measures using true value of $d$ and $D=5$.}}
\resizebox{\columnwidth}{!}{
\begin{tabular}{cccccccccc}
\hline
&&  \multicolumn{4}{c}{$\operatorname{MSE}(\matr{Z}^{recons})$} & \multicolumn{4}{c}{$\operatorname{MAE}(\matr{Z}^{recons})$ } \\
\cline{4-6}
\cline{8-10}
Lower dim. & Methods& $\sigma^2$&    $n=50$ &  $n=100$ & $n=500$ && $n=50$ & $n=100$ & $n=500$ \\
\hline
\multirow{12}{*}{$d=2$} &&$\frac{\pi}{8}$&           1.943 & 2.049 & 1.946 && 0.991 & 1.010 & 0.927 \\
&\multirow{4}{*}{TPPCA}&$\frac{\pi}{4}$&               2.460 & 2.355 & 1.952 && 1.197 & 1.134 & 0.980 \\
&&$\frac{\pi}{2}$&          2.664 & 2.658 & 2.533 && 1.285 & 1.275 & 1.224 \\
&&$\pi$&                         2.560 & 2.592 & 2.538 && 1.290 & 1.294 & 1.269 \\
&&$\frac{3\pi}{2}$&              2.559 & 2.467 & 2.398 && 1.290 & 1.267 & 1.243 \\
&&$2\pi$&                        2.620 & 2.410 & 2.427 && 1.307 & 1.253 & 1.252 \\
\cline{2-10}
&\multirow{6}{*}{PPCA}&$\frac{\pi}{8}$&               9.327 & 9.222 & 9.388 && 2.574 & 2.560 & 2.570 \\
&&$\frac{\pi}{4}$&               7.849 & 7.936 & 7.722 && 2.336 & 2.358 & 2.329 \\
&&$\frac{\pi}{2}$&           6.753 & 6.659 & 6.425 && 2.159 & 2.136 & 2.108 \\
&&$\pi$&                         5.956 & 5.573 & 5.202 && 2.001 & 1.935 & 1.872 \\
&&$\frac{3\pi}{2}$&              5.539 & 5.218 & 4.863 && 1.928 & 1.865 & 1.798 \\
&&$2\pi$&                        5.395 & 5.073 & 4.687 && 1.901 & 1.838 & 1.758 \\
\hline
\multirow{12}{*}{$d=3$} &&$\frac{\pi}{8}$&           2.343 & 2.354 & 2.118 && 1.168 & 1.146 & 1.031 \\
&\multirow{4}{*}{TPPCA}&$\frac{\pi}{4}$&               2.642 & 2.438 & 2.553 && 1.285 & 1.200 & 1.185 \\
&&$\frac{\pi}{2}$&          2.579 & 2.721 & 2.696 && 1.280 & 1.308 & 1.278 \\
&&$\pi$&                         2.483 & 2.520 & 2.491 && 1.264 & 1.276 & 1.259 \\
&&$\frac{3\pi}{2}$&              2.509 & 2.456 & 2.499 && 1.275 & 1.256 & 1.267 \\
&&$2\pi$&                        2.522 & 2.439 & 2.419 && 1.279 & 1.257 & 1.250 \\
\cline{2-10}
&\multirow{6}{*}{PPCA}&$\frac{\pi}{8}$&               6.228 & 6.208 & 5.731 && 2.059 & 2.063 & 1.970 \\
&&$\frac{\pi}{4}$&               5.956 & 5.641 & 5.559 && 2.006 & 1.953 & 1.944 \\
&&$\frac{\pi}{2}$&           5.475 & 5.350 & 5.161 && 1.922 & 1.898 & 1.870 \\
&&$\pi$&                         5.190 & 5.008 & 4.792 && 1.857 & 1.827 & 1.787 \\
&&$\frac{3\pi}{2}$&              5.021 & 4.877 & 4.637 && 1.829 & 1.796 & 1.748 \\
&&$2\pi$&                        4.949 & 4.754 & 4.553 && 1.817 & 1.776 & 1.731 \\
\hline
\hline
\end{tabular}}
\label{tab:Z}
\end{minipage}
\end{table}

In practical applications, the selection of the lower dimension is challenging. To address this, we conducted a secondary simulation study where we employed CV, KG, and  LRT to estimate the number of components (here $d$), as elaborated in Section \ref{sec:ncomp}. The performance of TPPCA based on these three different procedures is detailed in Table \ref{tab:2XestimatedD}.
Our analysis shows minor discrepancies in the performance measures across the three methods, with a slight overall preference for LRT. Detailed results are provided in Section SM-6 of the Supplementary Material. These findings indicate that LRT is the most effective method among those considered for estimating the lower dimension.

\begin{table}
\begin{minipage}[t]{0.95\textwidth}
\caption{\small{Performance of the TPPCA on the reconstruction of $\matr{X}$ measured by the MSE and MAE measures using an estimated value of $d$ by several methods, $D=5$, true value of $d$.}}
\resizebox{\columnwidth}{!}{
 \begin{tabular}{cccccccccc}
\hline
&&  \multicolumn{4}{c}{$\operatorname{MSE}(\matr{X}^{recons})$} & \multicolumn{4}{c}{$\operatorname{MAE}(\matr{X}^{recons})$ } \\
\cline{4-6}
\cline{8-10}
Lower dim. & Method & $\sigma^2$&    $n=50$ &  $n=100$ & $n=500$ && $n=50$ & $n=100$ & $n=500$ \\
\hline
\multirow{18}{*}{$d=2$} &\multirow{6}{*}{CV}&$\frac{\pi}{8}$&       1.76 & 1.68 & 1.09 && 0.73 & 0.69 & 0.53 \\
&&$\frac{\pi}{4}$&                      2.56 & 2.29 & 1.67 && 0.98 & 0.89 & 0.71 \\
&&$\frac{\pi}{2}$&                     3.48 & 3.26 & 2.58 && 1.31 & 1.23 & 1.03 \\
&&$\pi$&                                    3.88 & 3.87 & 3.92 && 1.50 & 1.49 & 1.46 \\
&&$\frac{3\pi}{2}$&                  4.54 & 4.25 & 4.32 && 1.65 & 1.60 & 1.61  \\
&&$2\pi$&                                 4.56 & 4.51 & 4.68 && 1.67 & 1.67 & 1.70\\
\cline{2-10}
 &\multirow{6}{*}{KG}&$\frac{\pi}{8}$&          2.18 & 1.84 & 1.10 && 0.80 & 0.71 & 0.51 \\
&&$\frac{\pi}{4}$&              2.74 & 2.16 & 1.48 && 1.00 & 0.84 & 0.68 \\
&&$\frac{\pi}{2}$&             3.39 & 3.16 & 2.74 && 1.29 & 1.22 & 1.06 \\
&&$\pi$&                            4.48 & 4.30 & 4.12 && 1.61 & 1.57 & 1.53 \\
&&$\frac{3\pi}{2}$&            4.62 & 4.81 & 4.72 && 1.69 & 1.72 & 1.70 \\
&&$2\pi$&                           4.82 & 4.73 & 5.09 && 1.74 & 1.72 & 1.79 \\
\cline{2-10}
 &\multirow{6}{*}{LRT}&$\frac{\pi}{8}$&         1.75 & 1.27 & 1.09 && 0.72 & 0.57 & 0.50 \\
&&$\frac{\pi}{4}$&             2.72 & 2.41 & 1.82 && 1.02 & 0.92 & 0.75 \\
&&$\frac{\pi}{2}$&             3.44 & 3.41 & 3.10 && 1.32 & 1.29 & 1.20 \\
&&$\pi$&                            4.56 & 4.56 & 4.79 && 1.66 & 1.66 & 1.67  \\
&&$\frac{3\pi}{2}$&            5.12 & 5.24 & 5.41 && 1.79 & 1.81 & 1.84 \\
&&$2\pi$&                           5.48 & 5.38 & 5.74 && 1.86 & 1.86 & 1.92  \\
\hline
\multirow{18}{*}{$d=3$} &\multirow{6}{*}{CV}&$\frac{\pi}{8}$&  4.02 & 3.18 & 2.70 && 1.29 & 1.09 & 0.93 \\
&&$\frac{\pi}{4}$&                 4.14 & 3.96 & 3.31 && 1.38 & 1.30 & 1.12\\
&&$\frac{\pi}{2}$&                 4.67 & 4.64 & 4.07 && 1.57 & 1.54 & 1.38 \\
&&$\pi$&                                 5.23 & 4.92 & 5.18 && 1.75 & 1.69 & 1.72 \\
&&$\frac{3\pi}{2}$&                5.45 & 5.52 & 5.56 && 1.83 & 1.83 & 1.83 \\
&&$2\pi$&                               5.57 & 5.65 & 5.88 && 1.85 & 1.85 & 1.90\\
\cline{2-10}
&\multirow{6}{*}{KG}&$\frac{\pi}{8}$&             3.85 & 3.19 & 2.82 && 1.21 & 1.06 & 0.94 \\
&&$\frac{\pi}{4}$&                 4.52 & 3.73 & 3.27 && 1.46 & 1.25 & 1.10 \\
&&$\frac{\pi}{2}$&                 4.74 & 4.82 & 4.13 && 1.59 & 1.56 & 1.38 \\
&&$\pi$&                                5.32 & 5.46 & 5.33 && 1.77 & 1.79 & 1.75 \\
&&$\frac{3\pi}{2}$&               5.77 & 5.91 & 5.78 && 1.88 & 1.91 & 1.88\\
&&$2\pi$&                              5.75 & 6.02 & 6.37 && 1.88 & 1.93 & 1.99 \\

\cline{2-10}
&\multirow{6}{*}{LRT}&$\frac{\pi}{8}$&      3.40 & 3.19 & 2.35 && 1.12 & 1.01 & 0.78\\
&&$\frac{\pi}{4}$&            4.41 & 3.73 & 3.39 && 1.38 & 1.23 & 1.11\\
&&$\frac{\pi}{2}$&            5.02 & 5.07 & 4.63 && 1.63 & 1.61 & 1.49\\
&&$\pi$&                            5.75 & 6.03 & 6.33 && 1.86 & 1.90 & 1.92  \\
&&$\frac{3\pi}{2}$&           6.12 & 6.19 & 6.67 && 1.96 & 1.97 & 2.04\\
&&$2\pi$&                          6.25 & 6.58 & 7.02 && 1.98 & 2.04 & 2.11 \\
\hline
\hline
\end{tabular}}
\label{tab:2XestimatedD}
\end{minipage}
\end{table}

\subsection{Real Application}
\label{sec:rna}

To conduct a thorough comparative analysis between our proposed method and various existing approaches in the literature, we selected three distinct datasets: \textit{Sunspots}, as studied by \citet{zoubouloglou2021}; \textit{small RNA}, and \textit{large RNA (or bigRNA)}, both of which have been previously analyzed by researchers including \citet{Eltzner2018}, \citet{Nodehi2018}, and \citet{zoubouloglou2021}. Notably, for the first two real datasets (\textit{Sunspots}, \textit{small RNA}), we conducted a performance comparison of TPPCA against existing methods. However, in the case of \textit{bigRNA}, our aim is to validate simulation results in real-world scenarios. Each of these datasets exhibits unique characteristics and complexities, making them particularly well-suited for the thorough evaluation of our proposed approach in contrast to the methodologies investigated in the studies conducted by the aforementioned researchers. 

\subsubsection{Sunspots}

Sunspots are cooler areas on the Sun's surface and are closely linked to concentrated magnetic fields on the Sun. These magnetic fields that give rise to sunspots follow a nearly regular 11-year pattern known as the solar cycle. In a study by \citet{Garcia2020}, they provided a well-organized collection of data on the central points of sunspot groups. Our study specifically examines the most recent solar cycle with comprehensive curated data, which is the 23$^{rd}$ solar cycle. This dataset spans from August 1996 to November 2001 and contains 5373 records. \added{The comprehensive description of sunspots can be found in the paper published by \citet{zoubouloglou2021}. We investigate the dependence structure of sunspot birth longitudes, denoted by $\theta \in [-\pi,\pi]$. Since the Earth's rotation around the Sun affects sunspot visibility, we expect a deviation from a linear dependence structure in the series of sunspot longitudes if there are significantly preferred longitudes. To explore this serial structure, we consider the series of longitudes along with its lagged versions of order one and two:
\begin{eqnarray*}
\Theta_{i}=(\theta_{i}, \theta_{i+1}, \theta_{i+2})^{\top} \in \mathbb{T}^3, ~~~\theta_{i} \in [-\pi,\pi]~~\mbox{for}~~ i=1,\cdots, 5373.
\end{eqnarray*}
Since the overall dimension of the Sunspots dataset is three, we have three principal component (PC) scores in this dataset. Using a randomly selected sample of 1000 sunspots on the torus, we computed the first PC score based on various methods and determined the proportion of variability explained by each of the three components. This enabled us to provide a detailed comparison between TPPCA and nine other dimension reduction techniques.}

\begin{figure}
\begin{minipage}{\textwidth}
\centering 
\includegraphics[width=0.75\textwidth]{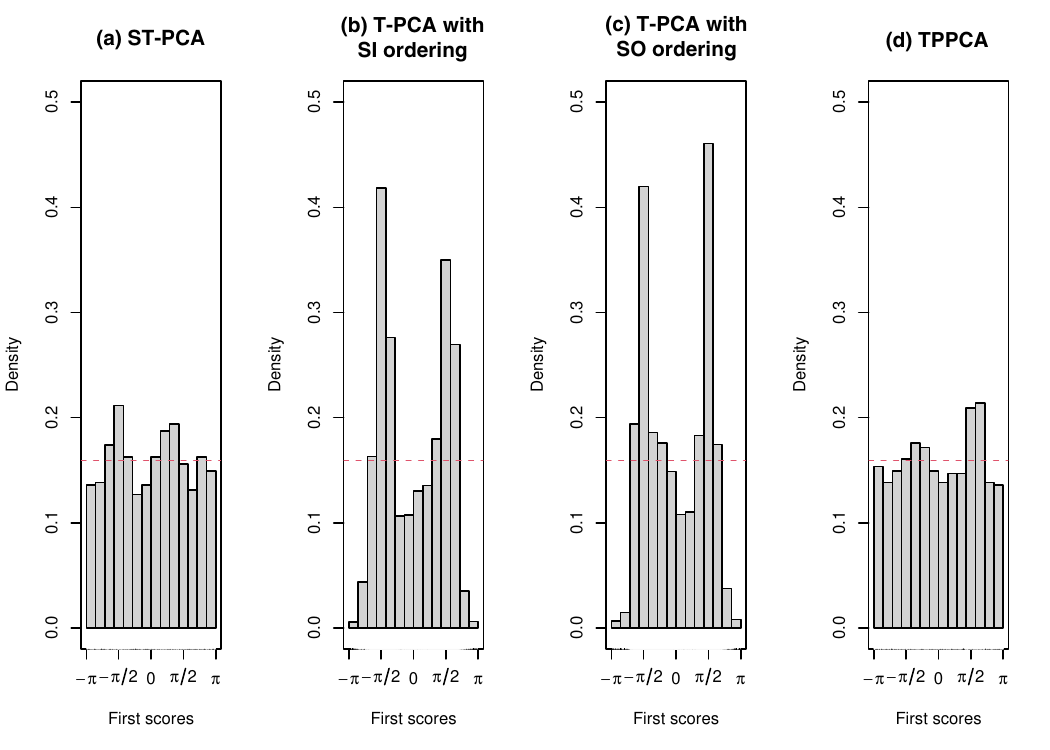}
\caption{Histograms representing the first score of different methods are shown as follows:  (a) ST-PCA; (b) T-PCA with SI ordering; (c) T-PCA with SO ordering; (d) TPPCA. Both versions of T-PCA reveal the presence of unexpected clusters, despite the marginal distributions of the data not significantly deviating from uniformity. In contrast, ST-PCA and TPPCA do not display these spurious clusters.}
\label{fig:hist}
\end{minipage}
\end{figure}

One limitation of T-PCA \citep{Eltzner2018} is that it introduces distortion when it transforms the hyper-torus ($\mathbb{T}^3$) into the hypersphere ($\mathbb{S}^3$). This distortion leads to an artificial clustering pattern in the first scores, as illustrated in Figures \ref{fig:hist}(b) and \ref{fig:hist}(c), even though the marginal values of $\Theta_{i}$ are uniformly distributed. In contrast, our proposed method (Figure \ref{fig:hist}(d)) and ST-PCA (Figure \ref{fig:hist}(a)) demonstrate a uniform distribution.  \added{As recommended by \citet{zoubouloglou2021}, we conducted a formal comparison of the distributions of the first scores from TPPCA, T-PCA (SI), T-PCA (SO), and ST-PCA. This was done through three omnibus circular uniformity tests, using their asymptotic distributions as implemented in the \texttt{sphunif} package in $\mathbbm{R}$ \citep{Garcia2021}. The results, summarized in Table \ref{tab:Sunspots_uniformity}, show that the uniformity hypothesis cannot be rejected for the marginal distribution of the original longitudes ($\Theta_{i}$), as well as the corresponding scores for TPPCA and ST-PCA. However, the hypothesis is rejected for the scores obtained from the two possible orderings of T-PCA.}

We also conducted a comprehensive comparison between TPPCA and nine other dimension reduction techniques as suggested by \citet{zoubouloglou2021}. For each method, we calculated the percentage of variability explained by their three components. Our findings are summarized in Table \ref{tab:3Sunspots}. It is important to note that existing software for GeoPCA primarily employs PGA and provides information on the variance explained by the first two components rather than the total variance explained. Consequently, we omitted GeoPCA from Table \ref{tab:3Sunspots}. Notably, TPPCA and ST-PCA stand out for explaining the highest proportion of variance through their first principal components compared to their competitors.

\begin{table}
\begin{minipage}{\textwidth}
\caption{p-values for various circular uniformity tests.}
\label{tab:Sunspots_uniformity}
\resizebox{\columnwidth}{!}{
 \begin{tabular}{lccccccc}
\hline
Test &&  Longitudes &&   ST-PCA & TPPCA & T-PCA (SO) & T-PCA (SI) \\
\cline{1-2}
\cline{3-8}
Giné's Fn                          && 0.32 &&   0.08  & 0.30 &  $<$0.001  & $<$0.001\\
Watson                             && 0.35 &&   0.18  & 0.40 &  $<$0.001 &  $<$0.001\\
Projected Anderson–Darling         && 0.75 &&   0.13  & 0.40 &  $<$0.001 &  $<$0.001\\
\hline
\hline
\end{tabular}}
\footnotesize{\textbf{Note}:  From left to right column: name of the test; p-values for testing uniformity on the original longitudes; p-values for testing uniformity on the first scores of
the corresponding method. The original longitudes, ST-PCA and TPPCA’s first scores are
not significantly nonuniform, while uniformity is rejected in T-PCA’s first scores.}
\end{minipage}
\end{table}

\begin{table}
\begin{minipage}{\textwidth}
\caption{Percentage of variance explained by components of dimension reduction methods on the torus,
sorted decreasingly according to the percentage of variance explained by their first components.}
\label{tab:3Sunspots}
\resizebox{\columnwidth}{!}{
 \begin{tabular}{lcccccccc}
\hline
Methods &&  Component 1 &&   Component 2 && Component 3 && Total percentage explained\\
\cline{1-2}
\cline{3-9}
                   &&      &&       &&    && \\
ST-PCA             && 90.51 &&    3.07  &&  6.41  && 100\\
TPPCA              && 90.28 &&    5.19  &&   4.53 && 100\\
T-PCA (SO)        && 89.32 &&   6.43    &&  4.25 && 100\\
T-PCA (SI)         && 88.27 &&   6.97    &&  4.76 && 100\\
Complex dPCA    && 72.51 &&   14.59   &&   12.90 && 100\\
PCA                  && 55.43  &&   24.48   &&  20.09 && 100\\
aPCA                && 43.27  &&   31.28   && 25.45 && 100\\
dPCA+              &&  42.69 &&  30.53   &&  26.78 && 100\\
dPCA               &&  38.48 &&  34.06   &&  7.67 && 80.21\\
PPCA                &&  37.44 &&  37.44   &&  25.12 && 100\\
\hline
\hline
\end{tabular}}
\footnotesize{\textbf{Note}: TPPCA and ST-PCA explain the most percentage of variance in their first component. The dPCA gives a six-dimensional embedding and, as a result, the sum of the variance explained by the first three dimensions does not add up to 100$\%$.}
\end{minipage}
\end{table}


\subsubsection{Small RNA data sets}
\label{sm:rna:small}
\begin{figure}
\begin{center}
\includegraphics[width=0.48\textwidth]{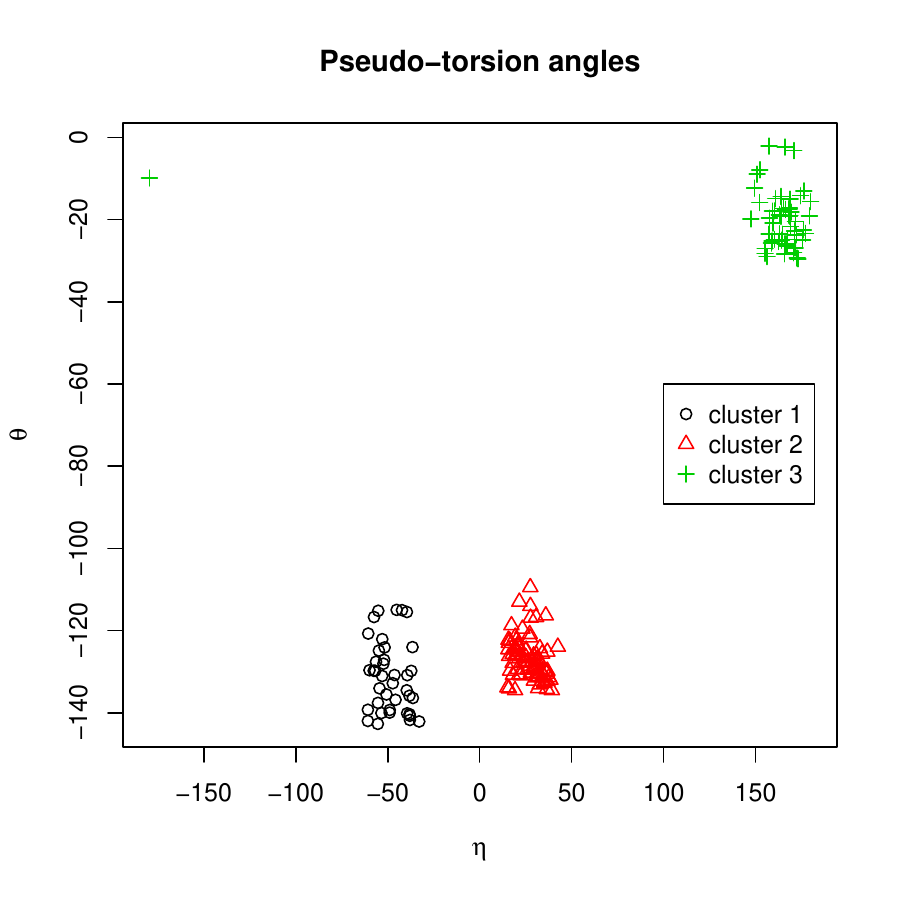}
\includegraphics[width=0.48\textwidth]{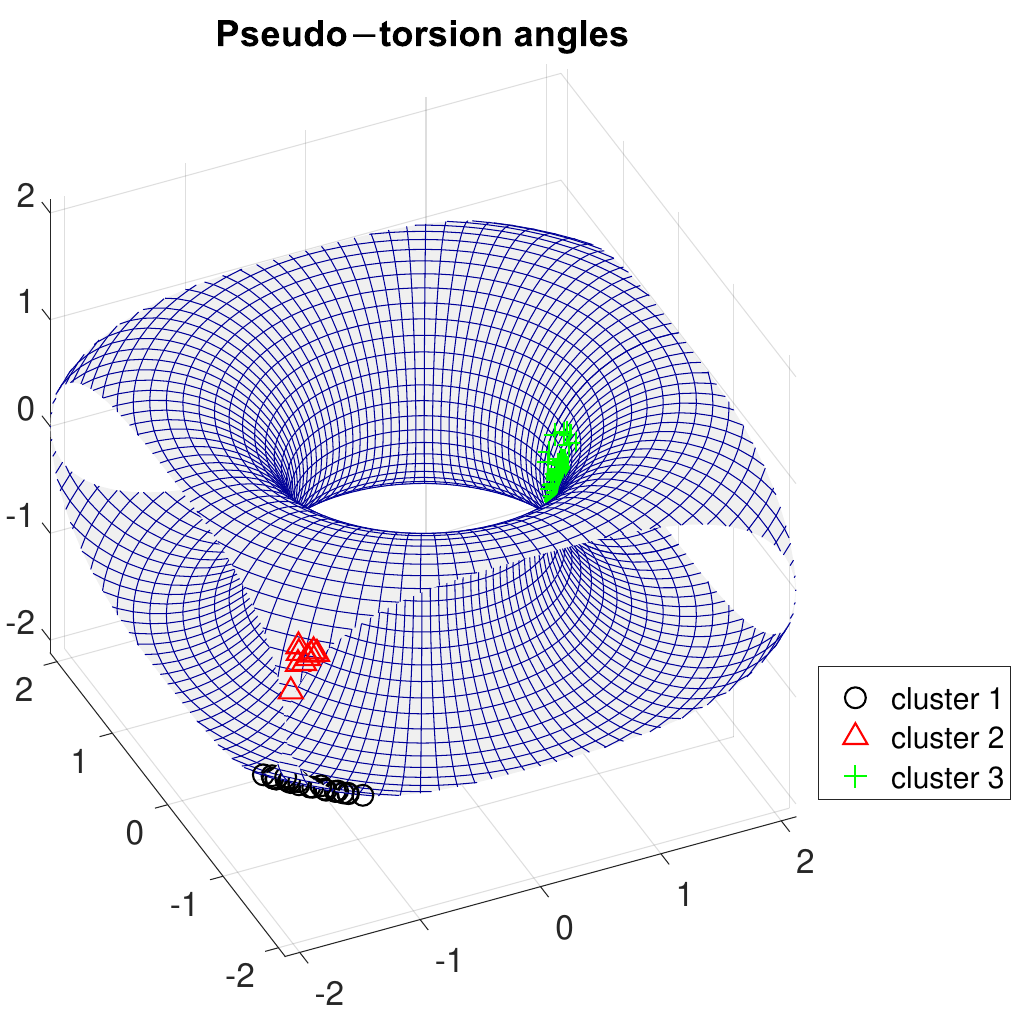}
\end{center}
\caption{RNA data set. Their three preselected clusters in the $\eta-\theta$ plot in $2D$ and on the torus.}
\label{fig:small:rna}
\end{figure}

The RNA datasets have been frequently employed in the literature \citep{Sargsyan2012,Eltzner2018,Mardia2021,zoubouloglou2021} to assess and validate PCA methods designed for toroidal data. In RNA data, each nucleic base corresponds to a backbone segment described by $6$ dihedral angles and one angle for the base, giving a total of $7$ angles.  The small RNA data set contains $181$ observations, which form three clusters in the $\eta-\theta$ plot as shown in Figure \ref{fig:small:rna}. \added{The $\eta$-$\theta$ plot is a graphical tool used to capture the conformational landscape of RNA molecules by plotting two pseudo-torsion angles, $\eta$ and $\theta$. Each point on the plot represents a specific nucleotide position, with clusters of points indicating regions of similar conformations. This plot is particularly useful for identifying structural patterns and differences within the RNA molecule \citep{Duarte1998, Sargsyan2012}.}

\begin{figure}
\centering
\includegraphics[width=0.85\textwidth]{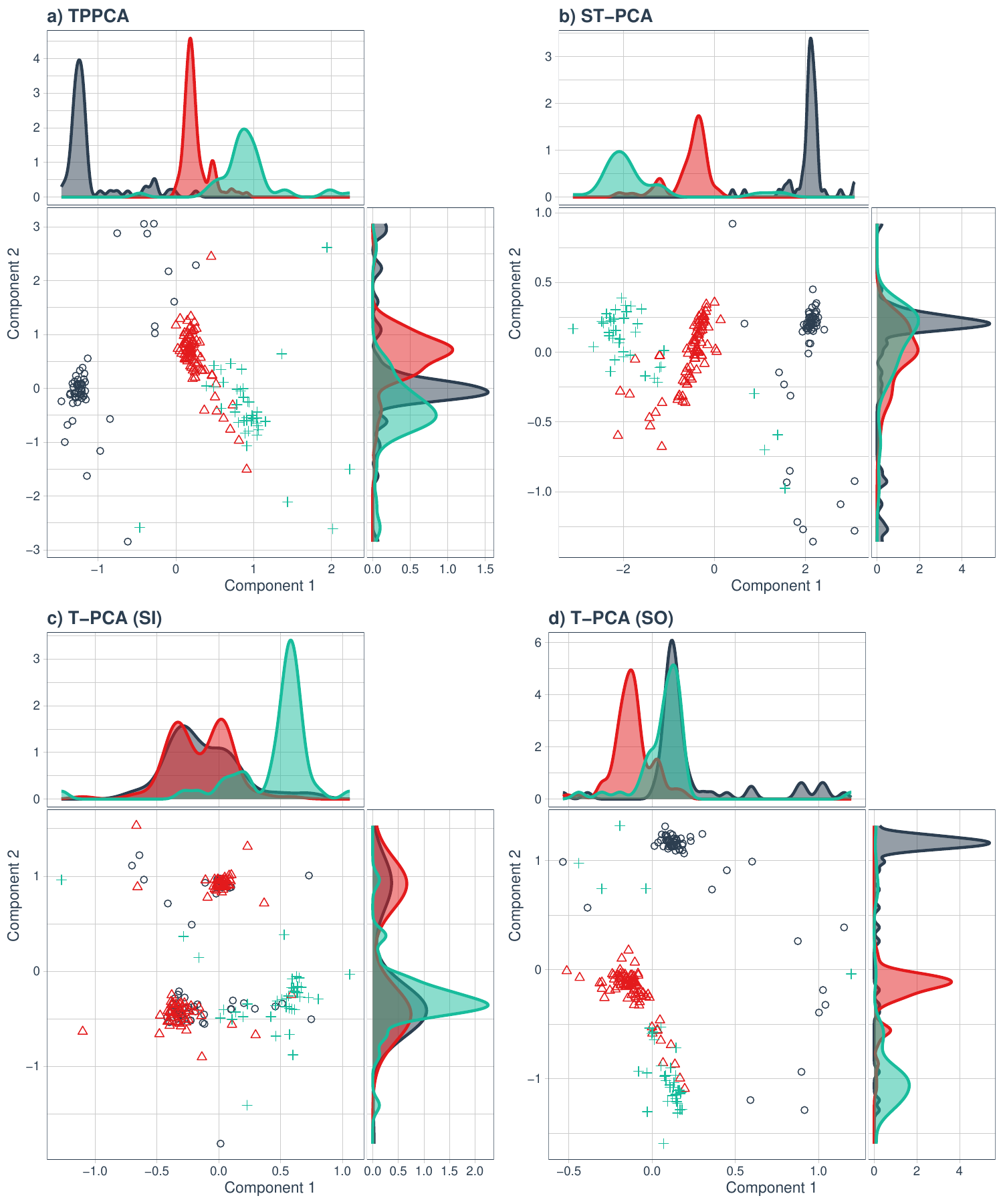}
\caption{Visualizations of the two-dimensional representation of the first two components are provided for a) TPPCA, b) ST-PCA, c) T-PCA (SI), and d) T-PCA (SO) applied to the RNA dataset. Kernel density estimation is utilized to examine modality, with the analysis focusing on the distribution of the first and second scores presented on the horizontal and vertical axes, respectively.}
\label{fig:small:rna:PPCA:TPPCA}
\end{figure}

In this section, our primary focus centers on examining four distinct methods that exhibit the highest proportion of variance through their first component in \textit{Sunspots}. These methods of interest include (a) TPPCA, (b) ST-PCA, (c) T-PCA with SI ordering, and (d) T-PCA with SO ordering. These methods were selected based on their ability to capture the most variance within their initial components.
A thorough analysis was performed, considering all four algorithms, and the findings are presented in Figure \ref{fig:small:rna:PPCA:TPPCA}. Remarkably, following the identification of three clusters in the $\eta-\theta$ plot, it is worth highlighting that both TPPCA and ST-PCA consistently group the data into three distinctive clusters when evaluating the first two principal components. \added{To quantify the overlap between the clusters identified in the $\eta$-$\theta$ plot and those found by the first two principal components, we calculated the Adjusted Rand Index (ARI) for each method \citep{HubertArabie1985} The ARI values are as follows: ST-PCA (ARI=0.71), TPPCA (ARI=0.64), T-PCA (SI) (ARI=0.22), and T-PCA (SO) (ARI=0.53). These results indicate that ST-PCA and TPPCA show good agreement with the clusters from the $\eta$-$\theta$ plot, suggesting a high degree of consistency. In contrast, T-PCA (SI) and T-PCA (SO) show lower overlap, with T-PCA (SI) demonstrating the least agreement. The high ARI values for TPPCA and ST-PCA underscore their robustness in accurately capturing the underlying cluster structure in RNA data.} 

To assess the quality of our clustering results, follow the recommendations provided by  \citet{Brock2008}, \citet{Koutroumbas2008} and \citet{charrad2014}. These authors suggest employing clustering validation statistics, including direct comparisons and statistical testing methods. The following three criteria are used for this purpose:

\begin{itemize}
\item[1.] {\bf Elbow plot} \citep{kaufman2009} seeks to find clusters in a manner that minimizes the cumulative sum of squares within each cluster, often referred to as the within-cluster sum of squares (WSS). The point at which an ``elbow" is observed in the plot is conventionally considered an indicator of the most suitable number of clusters. However, there are some limitations. The exact elbow point can sometimes be ambiguous or subjective. In some datasets, especially those with clusters of varying density or size, the elbow method might not yield a clear point of inflection. This is pretty common in real-world datasets.
\item[2.] {\bf Average silhouette} \citep{kaufman2009} computes the mean silhouette value for data points across various cluster numbers, denoted as $k$. The optimal number of clusters, represented by $k$, is the one that maximizes the average silhouette across a range of potential cluster values.  
\item[3.] {\bf The gap statistic} \citep{Tibshirani2001} provides a robust method of determining the optimal number of clusters. It involves comparing the total within-cluster variation for different values of $k >1$ with their expected values under a null reference distribution, \added{which is generated by randomly sampling data points from a uniform distribution over the range of the observed data. By measuring how much better the observed clustering is compared to random clustering, the gap statistic identifies the optimal number of clusters as the value that maximizes the gap statistic. This method is considered robust for determining the number of clusters, as it distinguishes meaningful clustering structures from those expected by chance.} 
\end{itemize}

\begin{figure}
\begin{center}
\includegraphics[width=0.85\textwidth]{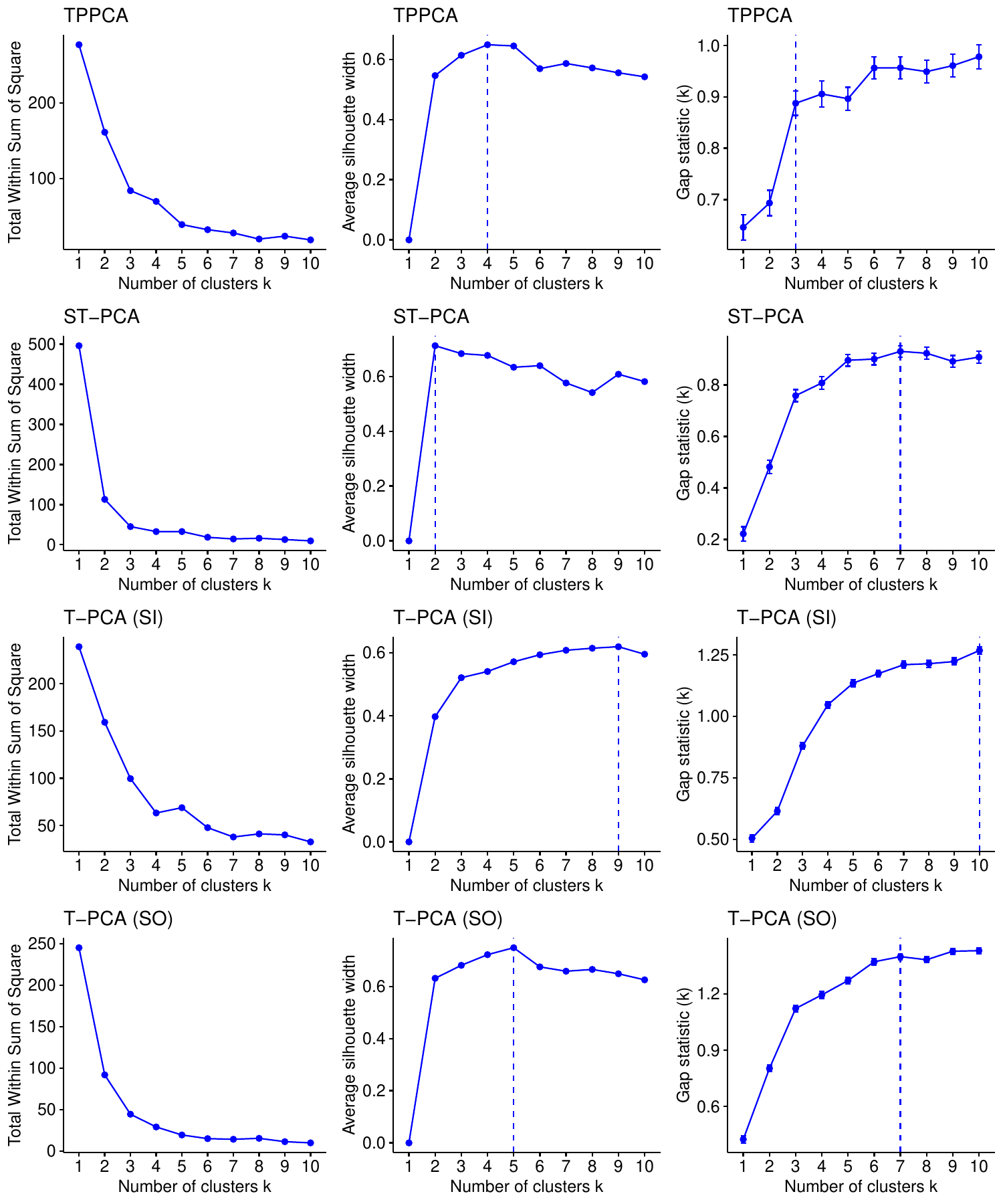}
\end{center}
\caption{Visualization of the optimal number of clusters based on the first two components in TPPCA, ST-PCA, T-PCA with SI ordering (denoted as T-PCA (SI)), and T-PCA with SO ordering (denoted as T-PCA (SO)) for the RNA dataset.}
\label{fig:small:rna:criteria}
\end{figure}

Figure \ref{fig:small:rna:criteria} provides a visualization of the optimal number of clusters for the RNA dataset using the first two principal components of TPPCA, ST-PCA, T-PCA (SI), and T-PCA (SO). We used three criteria to determine the optimal number of clusters: the elbow method, the average silhouette, and the gap statistic.

The elbow method, presented in the first column of Figure \ref{fig:small:rna:criteria}, involves identifying the point where the within-cluster sum of squares starts to level off, although this method is somewhat subjective. The second column of Figure \ref{fig:small:rna:criteria} shows the results for the average silhouette method, which identifies the optimal number of clusters by maximizing the average silhouette score across various potential cluster counts, with the optimal value indicated by the vertical line. Notably, TPPCA and ST-PCA were the only methods that suggested a number of clusters closely matching the true value. Finally, the third column of Figure \ref{fig:small:rna:criteria} illustrates the results of the gap statistic method, which identifies the number of clusters that maximize the gap statistic value, as indicated by the vertical line. According to \citet{Tibshirani2001}, the gap statistic is considered the most robust criterion for determining the optimal number of clusters.

\subsubsection{Large RNA data set (\textit{bigRNA})}
\label{sec:rna:big}

In our previous practical applications, we compared the performance of TPPCA against other established methods. This section now aims to validate the simulation results in real-world scenarios. \added{Our objective is to apply TPPCA and compare the performance of the LRT, KG, and CV criteria in terms of explaining a higher proportion of variability}. To achieve this, we used a large RNA dataset, also known as \textit{bigRNA}, which has been previously analyzed in the studies by \citet{Eltzner2018} and \citet{Nodehi2018}. The original data set initially consisted of $8301$ observations in seven dimensions. \added{For the bigRNA dataset, a deviation of $50^\circ$ in torus distance from the nearest neighbor, based on visual inspection, was used to identify outliers. Observations deviating by more than $50^\circ$ in angular distance were labeled as outliers and excluded from the analysis. The dataset was divided into 23 clusters using single linkage hierarchical clustering with a branch-cutting heuristic.} 
Consequently, the final dataset comprises $7390$ observations distributed across the $23$ clusters.

We applied the proposed TPPCA algorithm to this data set and the results to select the number of components across the $23$ clusters are presented in Table SM-1 (in the Supplementary Material). Consistent with our simulation studies, the LRT criterion outperformed the others in this real-world scenario. Table SM-1 highlights the effectiveness of LRT, showing a higher cumulative percentage of explained variance compared to the KG and CV criteria. These findings further support the robustness and practical use of the LRT, as demonstrated in both our simulations and real-world applications.

\section{Conclusions}
\label{sec:conclusion}

We have developed a novel algorithm for extracting information from torus data, thoroughly evaluating its performance through extensive simulation studies and practical real-world applications.

\added{When applied to Euclidean data, both probabilistic and classical PCA are valid and yield the same results. However there are a few distinct advantages of employing probabilistic PCA, such as its potential to work with  probabilistic models, perform statistical testing, and apply Bayesian approaches. Matrix decomposition and the $\EM$ algorithm are the two parameter estimation methods that \citet{TippingBishop1997} suggested. In TPPCA, the proposed iterative algorithm, which integrates the CEM algorithm in the first step, refines the estimates for $\vect{\mu}$, $\sample{K}$, $\matr{W}$, and $\sigma^2$ until convergence. Unlike the approach in \citet{Nodehi2018}, which focused solely on estimating $\vect{\mu}$ and $\sample{K}$ with a general covariance structure, the present work extends this by also estimating the loading matrix $\matr{W}$ associated with the principal components of the wrapped normal model and $\sigma^2$. This extension enables a more comprehensive analysis of toroidal data using TPPCA. }

\added{The implications for the toroidal variables are significant. Specifically, while the eigenvalue-based solution still holds for unwrapped variables in the Euclidean setting, for toroidal data, the estimation process must account for the periodicity and the manifold structure. This difference necessitates a modified approach in TPPCA, where the wrapping of variables is taken into consideration during inference. As a result, the interpretation of the loading matrix $\matr{W}$ in the context of toroidal variables differs from that in the classical Euclidean setting.}

To determine the optimal number of components, we introduced a novel approach based on LRT. This statistical framework allows us to make informed decisions about the most suitable number of components to retain. To validate the effectiveness of TPPCA, we used it in real-world data analysis. Specifically, we applied TPPCA to three distinct datasets: one dealing with the occurrence of sunspots and two RNA datasets. The results from these applications indicated that both TPPCA and ST-PCA outperformed existing methods across various evaluation criteria. This demonstrates the advantages and potential of our approach for real-world datasets, emphasizing its capability to enhance data analysis tasks.

Like any framework, our proposal is not without limitations. Techniques like PCA, while beneficial, do involve a trade-off, namely, information loss. Striking a balance between feature extraction and preserving critical information represents a crucial compromise inherent to utilizing PCA on a torus. Moreover, it is essential to know that the $\EM$ algorithm can be sensitive to local extrema. To mitigate this sensitivity, it is recommended to run the algorithm with a range of initial values, improving its robustness and reliability. Overall, despite its limitations, our proposal provides a valuable tool for analyzing toroidal data and is applicable across a wide range of fields.

\section*{Acknowledgements}
The authors extend their gratitude to \citet{Eltzner2018} and \citet{zoubouloglou2021} for their contribution in preparing the real dataset.

\section*{Data Availability}
The datasets employed in this study are publicly accessible through $\texttt{R}$ and have been sourced from previously published works \citep{Eltzner2018, zoubouloglou2021}. Specifically, the Sunspots dataset is available within the \texttt{rotasym} package on $\texttt{R}$, while the Small RNA dataset can be found in the supplementary material of \citep{zoubouloglou2021} (please refer to \href{https://www.tandfonline.com/doi/full/10.1080/10618600.2022.2119985?scroll=top&needAccess=true}{here}). The large RNA (\textit{bigRNA)} dataset is accessible via the following link: \href{https://projecteuclid.org/journals/annals-of-applied-statistics/volume-12/issue-2/Torus-principal-component-analysis-with-applications-to-RNA-structure/10.1214/17-AOAS1115.full?tab=ArticleLinkSupplemental}{here} \citep{Eltzner2018}. Additionally, the $\texttt{R}$ codes used in this study are available upon request from the first author.

\newpage
\section*{Supplementary Material}
\setcounter{section}{0}
\setcounter{table}{0}
\setcounter{figure}{0}
\setcounter{equation}{0}
\renewcommand\thesection{SM--\arabic{section}}
\renewcommand\thetable{SM--\arabic{table}}
\renewcommand\thefigure{SM--\arabic{figure}}
\renewcommand\theequation{SM--\arabic{equation}}

The Supplementary Material provides extensive additional content, including a review of the Probabilistic Principal Component Analysis (PPCA) in Section \ref{Sec2} and a thorough examination of manifold structures in Section \ref{sec:manifold}. Section \ref{sm:reviewPCA} provides a detailed discussion of PCA techniques specifically adapted for toroidal data. Furthermore, Section \ref{SecSM-5} expands on the methodology behind Torus Probabilistic PCA (TPPCA). The theoretical foundations for selecting the optimal number of components using Cross-Validation (CV) are explored in Section \ref{sm:sec:cv}. Section \ref{sm:sec:simulation} presents comprehensive results from simulation studies, while Section \ref{sm:sec:rna} highlights a real-world application involving an RNA dataset with 23 groups across 7 variables and varying sample sizes. This supplementary material is a valuable resource for a deeper understanding of the concepts and methodologies discussed in the main text.

\section{A review on Probabilistic Principal Component Analysis (PPCA)}
\label{Sec2}
There are two formulations of PCA found in the literature. The first defines PCA as an orthogonal projection of the data onto a lower-dimensional linear space, termed the principal subspace, with the objective of maximizing the variance of the projected data, a concept introduced by \citet{Hotelling1933}. The second formulation characterizes PCA as a linear projection that minimizes the average projection cost. This cost is determined by the mean squared distance between the data points and their respective projections. Furthermore, PCA is a linear transformation of variables to create new uncorrelated combined variables. The discussed definition of PCA is rooted in linear projection of the data onto a subspace with lower dimensionality than the original data space, a concept initially articulated by \citet{Pearson1901}.

The Probabilistic Principal Component Analysis (PPCA) is an extension of PCA that provides several advantages compared to classical PCA. Incorporating a probabilistic model with a measure of likelihood enables the capacity to conduct comparisons with other probabilistic methodologies and enables the application of statistical testing and Bayesian approaches. In addition, the use of a probabilistic model and Expectation-Maximization ($\EM$) algorithm improves the ability of the PPCA to effectively address the issue of missing values in datasets. Further, the PPCA can also be used to model class-conditional densities, and hence can be applied to classification problems.

A latent variable model seeks to relate a $D$-dimensional observation vector $\vect{X}$ to a corresponding $d$-dimensional vector of latent (or unobserved) variables $\vect{Z}$ ($d<D$), using the model
\begin{equation*}
\vect{X} = \vect{\mu} + \matr{W} \vect{Z} + \vect{\epsilon}
\end{equation*}
where $\matr{W}$ is a $(D\times d)$ matrix that relates the two sets of variables,  
$\matr{X} = (X_1, \ldots, X_D)^\top, \matr{Z} = (Z_1, \ldots, Z_d)^\top$, while the parameter vector $\vect{\mu}$ allows the model to have a non-zero mean. The vector $\matr{Z} \sim N_d(\vect{0}, \vect{I}_d)$ is an $d$-dimensional Gaussian latent variable, and $\vect{\epsilon} \sim N_D(\vect{0}, \sigma^2 \vect{I}_D)$ is a $D$-dimensional zero-mean Gaussian-distributed noise variable with covariance $\sigma^2 \vect{I}_D$. We assume $\cov(\matr{Z}, \vect{\epsilon}) = \vect{0}$. Then, we obtain,
\begin{equation*}
\E(\vect{X}) = \E(\vect{\mu}+ \matr{W} \vect{Z} + \vect{\epsilon}) = \vect{\mu}
\end{equation*}
and
\begin{equation*}
\cov(\vect{X}) = \E[(\vect{\mu} + \matr{W} \vect{Z} + \vect{\epsilon}) (\vect{\mu} + \matr{W}
\vect{Z} + \vect{\epsilon})^\top] = \matr{W} \matr{W}^\top + \sigma^2 \matr{I}_D = \matr{C}.
\end{equation*}
\citet{TippingBishop1997} introduced two methods to estimate the parameters: matrix decomposition and the Expectation-Maximization ($\EM$) algorithm. Notably, the $\EM$ algorithm offers a notable advantage, particularly in large-scale applications, due to its computational efficiency, as highlighted by \citet{Roweis1998}. Furthermore, the $\EM$ algorithm considers an interesting limit where $\sigma^2 \longrightarrow 0$ corresponds to the classical PCA, as noted by \citet{Roweis1998}.

\section{A review on manifold}
\label{sec:manifold}

Manifolds are a fundamental concept in mathematics and geometry, playing a crucial role in various scientific disciplines, including differential geometry, topology, and data analysis. In this section, we briefly review some relevant concepts of manifolds based on \citet{Karcher1977}, \citet{Boothby1986}, \citet{Fletcher2004} and \citet{Pennec2006}. 

\begin{defn}\label{def:homeomorphic}
Let $\mathcal{X}$ and $\mathcal{Y}$ be topological spaces. A mapping $f : \mathcal{X} \rightarrow \mathcal{Y}$ is a
homeomorphism if it is bijective and both $f$ and $f^{-1}$ are continuous. In this case $\mathcal{X} $ and $\mathcal{Y}$ are said to be \textit{homeomorphic}.
\end{defn}

\begin{defn}\label{def:Hausdorff}
A topological space $\mathcal{X}$ is said to be \textit{Hausdorff} if for any two distinct points $x, y \in \mathcal{X}$, there exist disjoint open sets $U$ and $V$ with $x \in U$ and $y \in V$.
\end{defn}

A \textit{manifold} is a topological space that is locally equivalent to Euclidean space. More precisely, 
\begin{defn}[Manifold]\label{def:mfd}
A manifold $M$ is a Hausdorff space with a countable basis such that for each point $p \in M$ there is a neighborhood $U$ of $p$ that is homeomorphic to Euclidean space $\mathbb{R}^n$ for some integer $n$.
\end{defn}

Lines and circles are classic examples of one-dimensional manifolds, while two-dimensional manifolds are commonly referred to as surfaces. Notable instances include the plane, the sphere, and the torus. The concept of a manifold is foundational in geometry and non-Euclidean spaces, as it enables the representation of intricate structures in terms of well-understood topological properties from simpler spaces. Manifolds serve as a fundamental framework, facilitating the understanding and characterizing complex geometric spaces.

According to \citet{Karcher1977}, for each manifold $M \subset \mathbb{R}^d$, it is possible to associate a linear subspace of $\mathbb{R}^d$ to each point $x \in M$. 
This space is called the \textit{tangent space} at $x$, and is written as $T_{x} M$.
\begin{defn}[Tangent space]
The tangent space $T_x M$ is the set of tangent vectors of all curves passing through the point $x \in M$. Intuitively, a linear subspace best approximates the manifold $M$ in a neighborhood of the point $x$.  It is equipped with an inner product $$g_{x}: T_{x}M \times T_{x}M  \longrightarrow \mathbb{R}$$
along with a norm metric $|| \cdot ||: T_{x}M \rightarrow \mathbb{R} $ 
defined by $||\nu||:=\sqrt{g_{x}(\nu,\nu)}$ for any $\nu \in T_{x}M$. 
\end{defn}

A well-known theorem states that for any given Riemannian manifold denoted as $M$, there exists an isometric embedding into the Euclidean space, given that the dimension of the Euclidean space is chosen to be sufficiently large. This theorem is famously attributed to the work of \textit{John Nash} \citep{Boothby1986} and remains a celebrated cornerstone in the field of differential geometry.
\begin{defn}[Riemannian manifold]\label{def:Rmanifold}
A Riemannian manifold, denoted as $M$, is a type of smooth manifold characterized by the presence of a metric, denoted as $g_x$, which smoothly varies and serves as an inner product on the tangent space $T_x M$ at each point $x$ within the manifold. This family of inner products, collectively referred to as a Riemannian metric, defines the geometric properties of the manifold.
\end{defn}

Distances within a manifold are crucial for establishing measures of centrality and dispersion in directional statistics. The \texttt{geodesic}, defined as the shortest curve connecting a pair of points on the manifold, is the fundamental concept in this context.
\begin{defn}[Geodesic curve]\label{def:Geodesic}
Given a couple of points on a Riemannian manifold $(p,q) \in M$ and the set of all curves $\gamma : [a,b]  \rightarrow M$ such that $\gamma(a)=p$ and  $\gamma(b)=x$
Suppose that a Riemannian manifold $M$ is pathwise connected, in the sense that for any two points $p,q$ there exists a smooth path $\gamma : [a,b]  \rightarrow M$ such that $\gamma(a)=p$ and  $\gamma(b)=q$, the \textit{geodesic curve} $\bar{\gamma}$ is the curve with the shortest total length from $p$ to $q$.
\end{defn}

\section{A review on extending of the PCA on torus}
\label{sm:reviewPCA}

Torus-specific PCA techniques have emerged from the need to analyze dihedral angles in proteins or RNA within bioinformatics, as demonstrated in previous research \citep{Mu2005, Altis2007, Nodehi2015, Eltzner2018, Nodehi2018, zoubouloglou2021, Mardia2021}. In our real-world application, we explore several modern torus data techniques that facilitate the practical use of these comparative methods. These techniques include:

\begin{itemize}

\item[1.] \textbf{Dihedral angles principal component analysis (dPCA)}: \\
\citet{Mu2005} and \citet{Altis2007} introduced dPCA as an extension of PCA designed for dihedral angles in proteins. The fundamental concept underlying dPCA involves applying PCA to dihedral angles after they have been transformed into sinusoidal and cosinusoidal representations. The transformation from the angular space to a linear metric coordinate space is achieved through the utilization of trigonometric functions.  However, the dPCA exhibits several limitations. Notably, it requires twice the number of variables, overlooks the identity $\cos^2\theta+\sin^2\theta=1$, and fails to account for the topological characteristics of angles. These constraints have motivated researchers to explore alternative extensions of dPCA.

\item[2.] \textbf{Angular PCA (aPCA)}: \\
\citet{Riccardi2009} introduced Angular PCA (aPCA) specifically tailored for toroidal data, centered on their circular means. This approach, as demonstrated by \citet{Riccardi2009}, yields free-energy landscapes that exhibit a striking similarity to those produced by dPCA.

\item[3.]\textbf{Complex dPCA}:\\
\citet{Altis2007} introduced Complex dPCA as an alternative extension to dPCA for toroidal data. Let $\theta_{j} \in [0,2\pi]$ represent an angle. One can transform $\theta_{j}$ into complex numbers using Euler's formula:
$$
z_{j}=e^{i\theta_{j}}~~ (j=1,\ldots,N).
$$
The covariance matrix corresponding to the complex variables $z_{j}$ is defined as:
$$
C_{kj}=\langle(z_{k}-\langle z_{k}\rangle)(z_{j}^\ast-\langle z_{j}^\ast\rangle)\rangle
$$
where $k,j=(1,\ldots,N)$, $z^{\ast}$ denotes the complex conjugate of $z$, and $\langle \ldots \rangle$ represents the average over all sampled conformations. The matrix $\matr{C}$ is Hermitian, possessing $N$ real-valued eigenvalues $\lambda_{j}$ and $N$ complex eigenvectors $\nu_{j}$,
$$\matr{C} {\boldsymbol\nu}_{j}= \lambda_{j} {\boldsymbol\nu}_{j}$$
The complex principal components are expressed as: 
$$
Y^{\text{Complex}}_{j}= {\boldsymbol\nu}_{j}^\top {\boldsymbol z}.
$$
In summary, Complex dPCA is an extension of dPCA that transforms dihedral angles into complex numbers using Euler's formula. The covariance matrix is defined using the complex variables, and the resulting complex eigenvectors and eigenvalues can be used to compute the complex principal components.\\

\item[4.]\textbf{PCA on torus (dPCA$+$)}:\\
Dihedral angles principal component analysis plus (dPCA$+$) is an extension of dPCA for torus data \citep{Sittel2017}. The method aims to minimize the projection error induced by periodicity by transforming the data such that the maximal gap of the sampling is shifted to the periodic boundary. The covariance matrix and its eigen-decomposition can then be computed in a standard manner. This method preserves the topological feature of the torus by defining the data points' correct neighborhoods. However, the main assumption underlying dPCA$+$ is that the data show a significant gap in their distribution, which is a limitation in general. Additionally, dPCA$+$ is relatively comparable to angular PCA (aPCA) \citep{Riccardi2009} in its practical application.

\item[5.]\textbf{Torus principal component analysis (T-PCA)}:\\
The T-PCA is an extension of PCA to torus data, as suggested by \citet{Eltzner2018}. This method deforms tori into spheres and then uses PNS \citep{Jung2012}. However, deforming tori into spheres creates singularities, which the authors circumvent by introducing a data-adaptive pre-clustering technique to avoid singularities. The authors proposed two data-driven orderings of variables, \textit{SI ordering} and \textit{SO ordering}, corresponding to sorting the variables in terms of descending and increasing amount of circular spread, respectively. However, \citet{zoubouloglou2021} showed that \textit{SI ordering} and \textit{SO ordering} yield significantly different outcomes that may affect subsequent analyses.

\item[6.]\textbf{Scaled torus principal component analysis (ST-PCA)}:\\
The ST-PCA was recently proposed by \citet{zoubouloglou2021} as a method for analyzing multivariate toroidal data. ST-PCA seeks a data-driven map from a torus to a sphere with the same size and radius as the torus. The method is based on the idea that spherical embeddings are more appropriate than Euclidean ones when considering transformations from $d$-torus, $\mathbb{T}^d$, as proposed by \citet{Eltzner2018}. The analyses are done on spheres, and after finding the best fit of data, it can be inverted back to the torus. To minimize the disparity between pairwise geodesic distances in both spaces, the map is built via multidimensional scaling \citep{CoxCox1991,CoxCox2008}.
\end{itemize}

\section{A review on TPPCA}
\label{SecSM-5}

In this section, we provide a more detailed explanation of TPPCA, as referenced in the main paper.

\subsection{Step 0: Initial values}
\label{sec:initial}

In this section, we outline the procedure for selecting suitable initial values for the parameters ($\vect{\mu}$, $\matr{W}$, and $\sigma^2$) and impute the missing values $\sample{K}$ in the model. The initial values for the $\EM$ algorithm are obtained through a two-step process:
\begin{itemize}
    \item[1.] Estimation of initial values for $\sample{K}$: In this step, an unstructured variance-covariance matrix $\matr{\Sigma}$ is assumed, such that $\vect{Y} \sim WN_{D}(\vect{\mu}, \matr{\Sigma})$. Initial values $\hat{\sample{K}}_0 = (\hat{\vect{k}}_1, \ldots, \hat{\vect{k}}_N)$ are obtained using the Classification Expectation-Maximization (CEM) algorithm proposed by \citet{Nodehi2018}. 
    
    \item[2.] Estimation of initial values for the remaining parameters: a strategy similar to that used by \citet{TippingBishop1999} is employed to maximize the log-likelihood $\ell(\vect{\mu}, \matr{W}, \sigma^2, \hat{\sample{K}}_0)$ with respect to the parameters, yielding the initial estimates $(\hat{\vect{\mu}}_0, \hat{\matr{W}}_0, \hat{\sigma}^2_0)$. These estimates are then used to initiate the iterative algorithm (Steps 1 and 2). For more details on PPCA, please refer to SM-1. 
\end{itemize}

\subsection{Step 1}
\label{Sub:step1}
The update of $\vect{\mu}$ and $\sample{K}$ can be performed as outlined in \citep{Nodehi2018}, using the CEM algorithm. The CEM algorithm, when applied to $\Gamma(\vect{\mu}, \hat{\matr{W}}, \hat{\sigma}^2, \sample{K}|\sample{Y})$, is an iterative classification algorithm that performs the estimation of the parameter ($\vect{\mu}$) and the imputation of missing classification ($\sample{K}$).
\begin{itemize}
\item {\bf E step (Expectation)}: The expectation of log-likelihood over the latent variables based on the data samples and initial values as shown in $\Gamma(\vect{\mu}, \hat{\matr{W}}, \hat{\sigma}^2, \sample{K} | \sample{Y})$. 
\item {\bf C step (Classification)}: Update $\hat{\sample{K}}$ by
\begin{align*}
 \hat{\sample{K}} & = \arg\max_{\sample{K} \in \mathbb{Z}^D} \sum_{j=1}^{N} \Gamma(\hat{\vect{\mu}}, \hat{\matr{W}}, \hat{\sigma}^2, \sample{K} | \sample{Y})\\
 & = \arg\max_{\vect{k}_j \in \mathbb{Z}^D, j=1,\ldots,N} \sum_{j=1}^{N} \Gamma(\hat{\vect{\mu}}, \hat{\matr{W}}, \hat{\sigma}^2, \vect{k}_j |\vect{y}_j) \ , \qquad \
\end{align*}
which is equivalent to maximizing component by component $\Gamma(\hat{\vect{\mu}}, \hat{\matr{W}}, \hat{\sigma}^2, \vect{k}_j |\vect{y}_j)$,  i.e.
\begin{equation*}
\hat{\vect{k}}_j = \arg\max_{\vect{k}_j \in \mathbb{Z}^D} \Gamma(\hat{\vect{\mu}}, \hat{\matr{W}}, \hat{\sigma}^2, \vect{k}_j |\vect{y}_j) \ , \qquad j=1, \ldots, N.
\end{equation*}
\item {\bf M step (Maximization)}:
Using equation (2) in the paper, and considering $\Gamma(\vect{\mu}, \hat{\matr{W}}, \hat{\sigma}^2, \hat{\sample{K}} | \sample{Y})$ we update $\hat{\vect{\mu}}$ by evaluating
\begin{align*}
& \frac{\partial{\Gamma(\vect{\mu}, \hat{\matr{W}}, \hat{\sigma}^2, \hat{\sample{K}} | \sample{Y})}}{\partial{\vect{\mu}}} = \\
& \sum_{j=1}^{N} \left[ \frac{(\vect{y}_j + 2\pi \hat{\vect{k}}_j - \vect{\mu})^\top }{\sigma^2} - \frac{\E(\matr{Z}_j^\top |\vect{y}_j + 2\pi \hat{\vect{k}}_j) \matr{W}^\top }{\sigma^2} \right] =\\
& \sigma^{-2} \sum_{j=1}^{N} (\vect{y}_j +2\pi \hat{\vect{k}}_j -\vect{\mu})^\top - \sum_{j=1}^{N} (\vect{y}_j + 2\pi \hat{\vect{k}}_j - \vect{\mu})^\top \matr{W} \matr{M}^{-1} \matr{W}^\top ,
\end{align*}
equating it to $\vect{0}$ and solving for $\vect{\mu}$ leads to
\begin{equation*}
N \vect{\mu}^\top (\matr{I}_D - \matr{W} \matr{M}^{-1} \matr{W}^\top ) = \sum_{j=1}^{N} (\vect{y}_j + 2\pi \hat{\vect{k}}_j )^\top (\matr{I}_D - \matr{W} \matr{M}^{-1} \matr{W}^\top)
\end{equation*}
that is
\begin{equation*}
\hat{\vect{\mu}} = \frac{\sum_{j=1}^{N} (\vect{y}_j +2\pi \hat{\vect{k}}_j )}{N}= \frac{\sum_{j=1}^{N} \hat{\vect{x}}_j }{N}
\end{equation*}
where $\hat{\vect{x}}_j = \vect{y}_j + 2\pi \hat{\vect{k}}_j$ is an estimate of the unobserved $\vect{x}_j$ ($j=1, \ldots, N$).
\end{itemize}

\subsection{Step 2}
\label{Sub:step2}

Considering the derivative of the function $\Gamma(\hat{\vect{\mu}}, \matr{W}, \sigma^2, \hat{\sample{K}} | \sample{Y})$ with respect to the variables $\matr{W}$ and $\sigma^2$, this could be formulated as follows:
\begin{align*}
&\frac{\partial{\Gamma(\hat{\vect{\mu}}, \matr{W}, \sigma^2, \hat{\sample{K}} | \sample{Y})}}{\partial{\matr{W}}} = \\
& \sum_{j=1}^{N} \left[ -\frac{\matr{W} \mathbbm{E}(\matr{Z}_j \matr{Z}_j ^\top | \hat{\vect{x}}_j)}{2 \sigma^2} + \sigma^{-2} (\hat{\vect{x}}_j -\hat{\vect{\mu}}) \E(\matr{Z}_j^\top | \hat{\vect{x}}_j ) \right],
\end{align*}
we obtain
\begin{align*}
 \hat{\matr{W}}  =& \left[ \sum_{j=1}^{N} (\hat{\vect{x}}_j -\hat{\vect{\mu}}) \E(\matr{Z}_j^\top | \hat{\vect{x}}_j ) \right] \left[ \sum_{j=1}^{N} \E(\matr{Z}_j \matr{Z}_j^\top | \hat{\vect{x}}_j ) \right]^{-1} \\
& = \left[ \sum_{j=1}^{N} (\hat{\vect{x}}_j -\hat{\vect{\mu}}) (\hat{\vect{x}}_j -\hat{\vect{\mu}})^\top \matr{W} \matr{M}^{-1} \right] \times \\
& \left[ N \sigma^2 \matr{M}^{-1} + \sum_{j=1}^{N} \matr{M}^{-1} \matr{W}^\top (\hat{\vect{x}}_j -\hat{\vect{\mu}}) (\hat{\vect{x}}_j -\hat{\vect{\mu}})^\top \matr{W} \matr{M}^{-1} \right]^{-1} \\
& = \matr{S} \matr{W} \matr{M}^{-1} \{\sigma^2 \matr{M}^{-1} + (\matr{M}^{-1} \matr{W}^\top \matr{S} \matr{W} \matr{M}^{-1})\}^{-1} \\
& = \matr{S} \matr{W} \matr{M}^{-1} \matr{M} (\sigma^2 \matr{I}_d + \matr{M}^{-1} \matr{W}^\top \matr{S} \matr{W})^{-1} \\
& = \matr{S} \matr{W} (\sigma^2 \matr{I}_d + \matr{M}^{-1} \matr{W}^\top \matr{S} \matr{W})^{-1}
\end{align*}
where $\matr{S}= \frac{1}{N}\sum_{j=1}^{N} (\hat{\vect{x}}_j - \hat{\vect{\mu}}) (\hat{\vect{x}}_j - \hat{\vect{\mu}})^\top$ has the same form as the usual maximum likelihood  estimates of the variance-covariance matrix for the (estimated) sample $\hat{\vect{x}}_1, \ldots, \hat{\vect{x}}_N$ with estimated mean vector $\hat{\vect{\mu}}$ and $\matr{M} = \matr{W}^\top \matr{W} + \sigma^2 \matr{I}_d$.

After substituting $\hat{\matr{W}}$ in $\Gamma(\hat{\vect{\mu}}, \hat{\matr{W}}, \sigma^2, \hat{\sample{K}} | \sample{Y})$ and using Equation (2) in the paper, the estimating equation of $\sigma^2$ is
\begin{align*}
& \frac{\partial{\Gamma(\hat{\vect{\mu}},\hat{\matr{W}},\sigma^2,\hat{\sample{K}},|\sample{Y})}}{\partial{\sigma^2}} = \sum_{j=1}^{N} -\frac{D}{\sigma^2} - \frac{1}{\sigma^4}\left[ \trace[(\hat{\vect{x}}_j -\hat{\vect{\mu}})(\hat{\vect{x}}_j -\hat{\vect{\mu}})^\top] \right. \\
& - \left.\trace(\hat{\matr{W}} \E(\matr{Z}_j \matr{Z}_j^\top | \hat{\vect{x}}_j) \hat{\matr{W}}^\top ) \right] - \frac{2}{\sigma^4} \left[(\hat{\vect{x}}_j - \hat{\vect{\mu}}) \E(\matr{Z}_j^\top | \hat{\vect{x}}_j) \hat{\matr{W}}^\top \right] = 0,
\end{align*}
and by calling $\hat{\matr{M}}= \hat{\matr{W}}^\top \hat{\matr{W}} + \hat{\sigma}^2 \matr{I}_d$, we obtain 
\begin{align*}
& \hat{\sigma}^2  = \frac{1}{N D} \sum_{j=1}^{N} \big\{ \trace[(\hat{\vect{x}}_j - \hat{\vect{\mu}})(\hat{\vect{x}}_j - \hat{\vect{\mu}})^\top] - \\
&  2 (\hat{\vect{x}}_j - \hat{\vect{\mu}}) \E(\matr{Z}_j^\top | \hat{\vect{x}}_j ) \hat{\matr{W}}^\top   - \trace(\hat{\matr{W}} \E(\matr{Z}_j \matr{Z}_j^\top | \hat{\vect{x}}_j )) \hat{\matr{W}}^\top \big\} \\
& = \frac{1}{D} \trace[ \matr{S} - 2 \matr{S} \hat{\matr{W}} \hat{\matr{M}}^{-1} \hat{\matr{W}}^\top + \hat{\matr{W}} (\sigma^2 \hat{\matr{M}}^{-1} + \hat{\matr{M}}^{-1} \hat{\matr{W}}^\top \matr{S} \hat{\matr{W}} \hat{\matr{M}}^{-1})\hat{\matr{W}}^\top ] \\
& = \frac{1}{D} \trace[ \matr{S}(\vect{I}-\hat{\matr{W}} \hat{\matr{M}}^{-1} \hat{\matr{W}}^\top]\\
& = \frac{1}{D} \trace[ \matr{S}(\vect{I}-\hat{\matr{W}} (\hat{\matr{W}}^\top \hat{\matr{W}} + \hat{\sigma}^2 \matr{I}_d)^{-1} \hat{\matr{W}}^\top].
\end{align*}

\section{Number of Components: Cross-Validation}
\label{sm:sec:cv}
Determining the optimal number of components in PCA is a well-known challenge, as discussed in Section 2. In this section, we will elaborate on the utilization of a Cross-Validation procedure to address this issue.
Let's consider a data matrix $\matr{X}$ of dimensions $n \times D$, containing $D$ variables, $X_1, X_2, \ldots, X_D$, which we assume to be mean-centered. Using the Singular Value Decomposition (SVD) of $X$, we can represent $X$ as:
\begin{eqnarray}
\matr{X} = \matr{U} \matr{\Lambda} \matr{V}^{\top}
\label{svd}
\end{eqnarray}
where $\matr{U}^{\top}\matr{U} = \matr{I}_{D}$, $\matr{V}^{\top}\matr{V} = \matr{V}\matr{V}^{\top} = \matr{I}_{D}$ and $\matr{\Lambda} = (\lambda_{1},\cdots,\lambda_{D})$ with $\lambda_{1} \geq \cdots \geq \lambda_{D} \geq 0$. 
If $\matr{X}$ has rank $D$ and all the $\lambda_{i}$ ($i=1,\cdots,D$) are distinct, then decomposition (\ref{svd}) is unique apart from corresponding sign changes in $\matr{U}$ and $\matr{V}$. If we consider $x_{ij}$ and $u_{ij}$  as the $(i,j)-$th elements of the matrices $\matr{X}$ and $\matr{U}$, respectively, decomposition (\ref{svd}) has its elementwise representation as 
\begin{equation*}
x_{ij}=\sum_{t=1}^{D} u_{it} \lambda_{t} v_{tj}.  
\end{equation*}
Hence, according to the PCA, if we consider a lower dimension of $m$ dimensions, the variation in the remaining $(D-m)$ dimensions can be regarded as random noise. We can represent the data with $m$ components as:
\begin{equation*}
x_{ij}=\sum_{t=1}^{m} u_{it} \lambda_{t} v_{tj}+\epsilon_{ij}.  
\label{Xm}
\end{equation*}
where $\epsilon_{ij}$ is a residual term.

To do Cross-Validation, at first, by denoting $\matr{X}_{-j}$ as the result of deleting the $j-$th column and $\matr{X}_{-i}$ as the result of deleting the $i-$th row of $\matr{X}$, there are the following equations: 
\begin{eqnarray*}
\matr{X}_{-i}=\matr{\bar{U}}\bar{\matr{\Lambda}}\bar{\matr{V}}^{\top}
\end{eqnarray*}
where $\bar{\matr{U}}=\bar{u}_{st}$, $\bar{\matr{V}}=\bar{v}_{st}$ and $\bar{\matr{\Lambda}}=\operatorname{diag}(\bar{\lambda}_{1}, \cdots, \bar{\lambda}_{D})$ and  
\begin{eqnarray*}
X_{-j}=\tilde{U}\tilde{\Lambda}\tilde{V}^{\top}
\end{eqnarray*}
with $\tilde{U}=\tilde{u}_{st}$, $\tilde{V}=\tilde{v}_{st}$ and $\tilde{\Lambda}=\operatorname{diag}(\tilde{\lambda}_{1}, \cdots, \tilde{\lambda}_{D-1})$. After that, the predictor can be computed as   
\begin{eqnarray}
\hat{x}_{ij}(m)=\sum_{t=1}^{m} (\tilde{u}_{it} \sqrt{\tilde{\lambda}_{t}}) (\bar{u}_{tj} \sqrt{\bar{\lambda}_{t}}). 
\label{svdxm0}
\end{eqnarray}
Each element on the right-hand side of this equation is derived from the SVD of $\matr{X}$ after the removal of either the $i$-th row or the $j$-th column. It's important to note that the algorithms developed by \citet{BunchNielsen} and \citet{Bunchetal} efficiently compute $\tilde{\matr{U}}$, $\tilde{\matr{V}}$, $\bar{\matr{U}}$, and $\bar{\matr{V}}$, which are needed for \eqref{svdxm0}.
Finally, based on $\hat{x}_{ij}(m)$ for a given number $m$ of components, we can calculate the average squared discrepancy between the actual and predicted values, defined as:
\begin{eqnarray*}
PRESS(m)= \frac{1}{nD} \sum_{i=1}^{n} \sum_{j=1}^{D} (\hat{x}_{ij}(m)-x_{ij})^{2}. 
\end{eqnarray*}

Consider fitting components sequentially in (\ref{svdxm0}). Define 
\begin{equation*}
W_{m}= \frac{PRESS(m-1)-PRESS(m)}{\sample{D}_{m}} \div \frac{PRESS(m)}{\sample{D}_{r}}
\end{equation*}
In this equation, $\sample{D}_m$ stands for the degrees of freedom required to fit the $m$-th component, and $\sample{D}_r$ represents the remaining degrees of freedom after fitting this component. The determination of $\sample{D}_m$ involves considering the number of parameters that need to be estimated as well as the constraints applied to the eigenvectors at each stage. It can be demonstrated that $\sample{D}_m = n + D - 2m$. Consequently, $\sample{D}_r$ can be derived through successive subtraction, given the presence of $(n-1)D$ degrees of freedom in the mean-centered matrix $X$ \citep{Wold1978}.
Following the guidance of \citet{Krzanowski1983}, it is recommended to choose the optimal number of components as the highest value of $m$ for which $W_m$ exceeds $0.9$.

\section{Further numerical experiments}
\label{sm:sec:simulation}

\begin{figure}
\begin{center}
\includegraphics[width=0.43\textwidth]{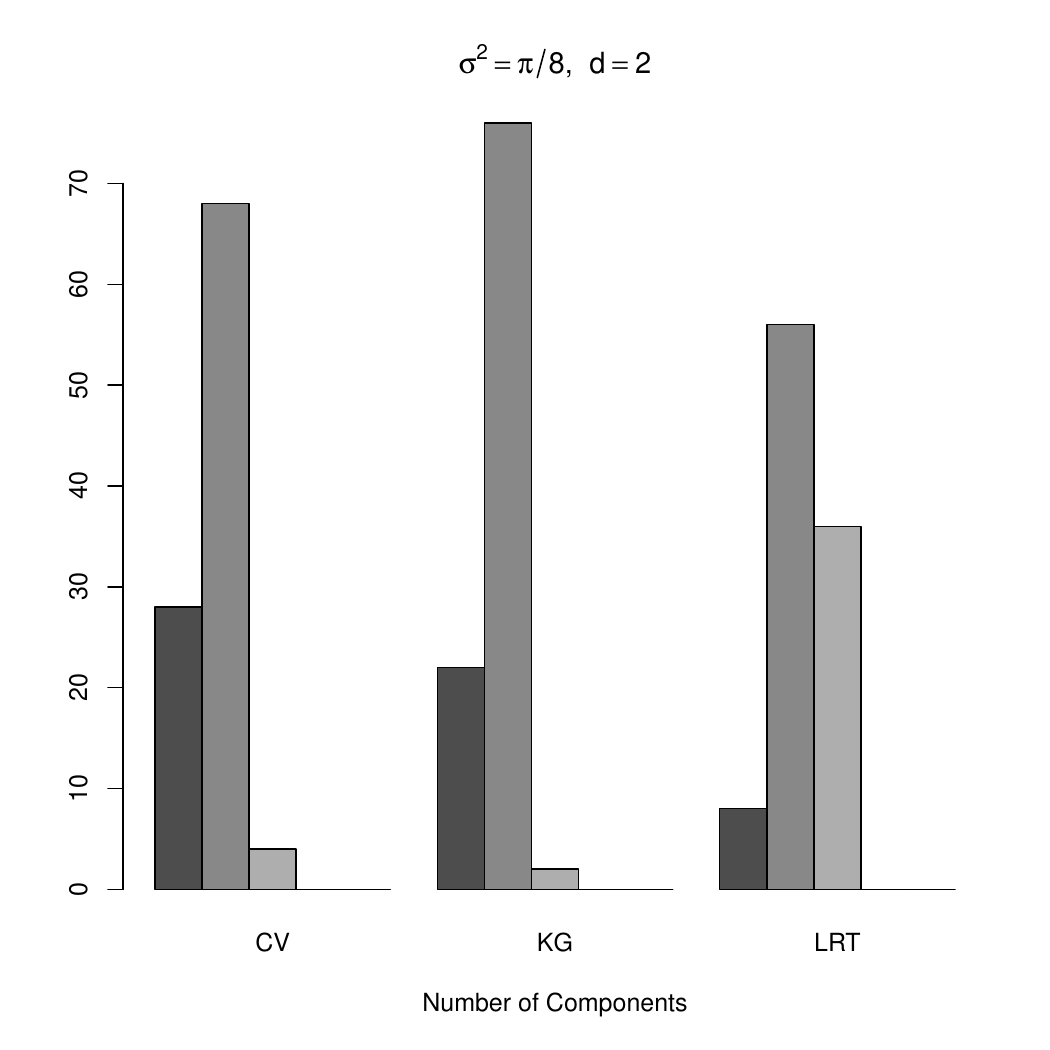}
\includegraphics[width=0.43\textwidth]{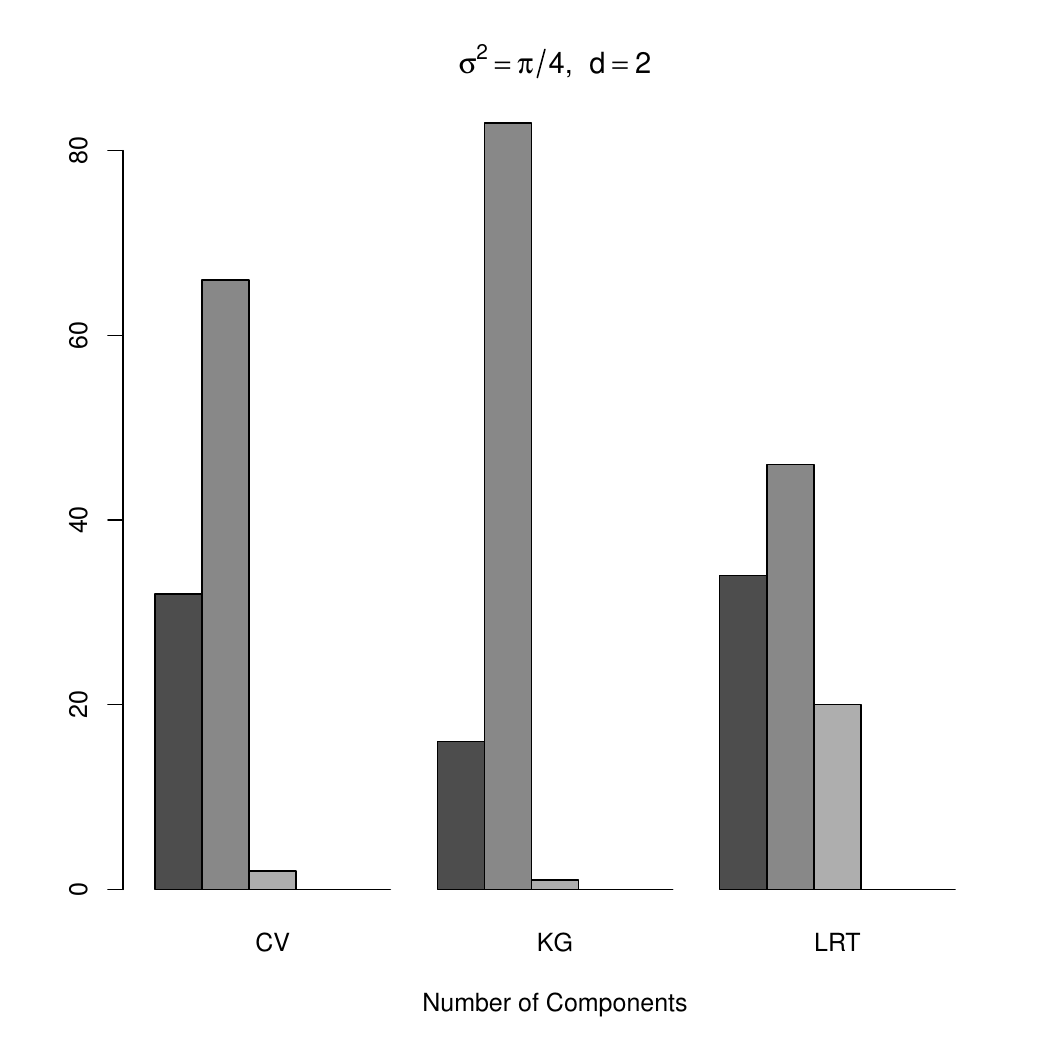} \\
\includegraphics[width=0.43\textwidth]{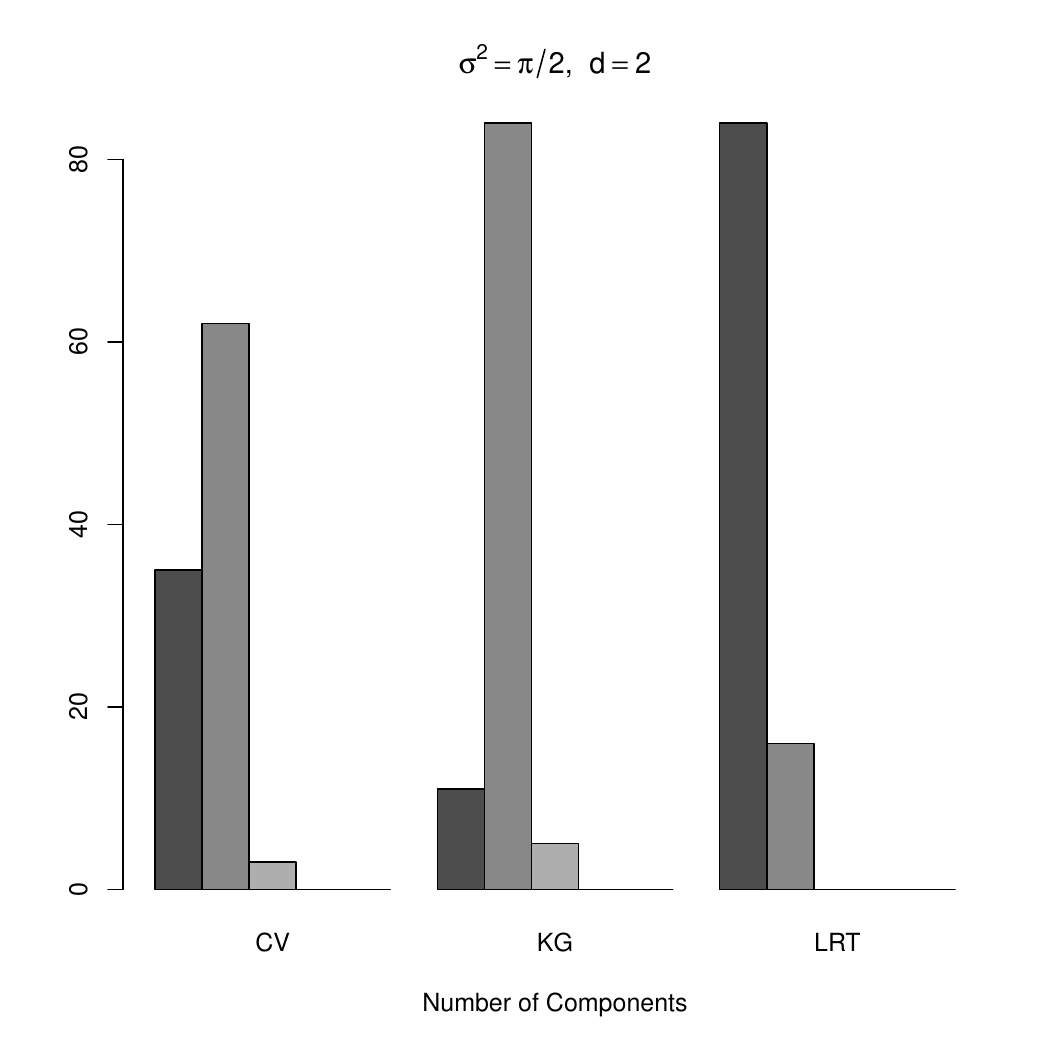}
\includegraphics[width=0.43\textwidth]{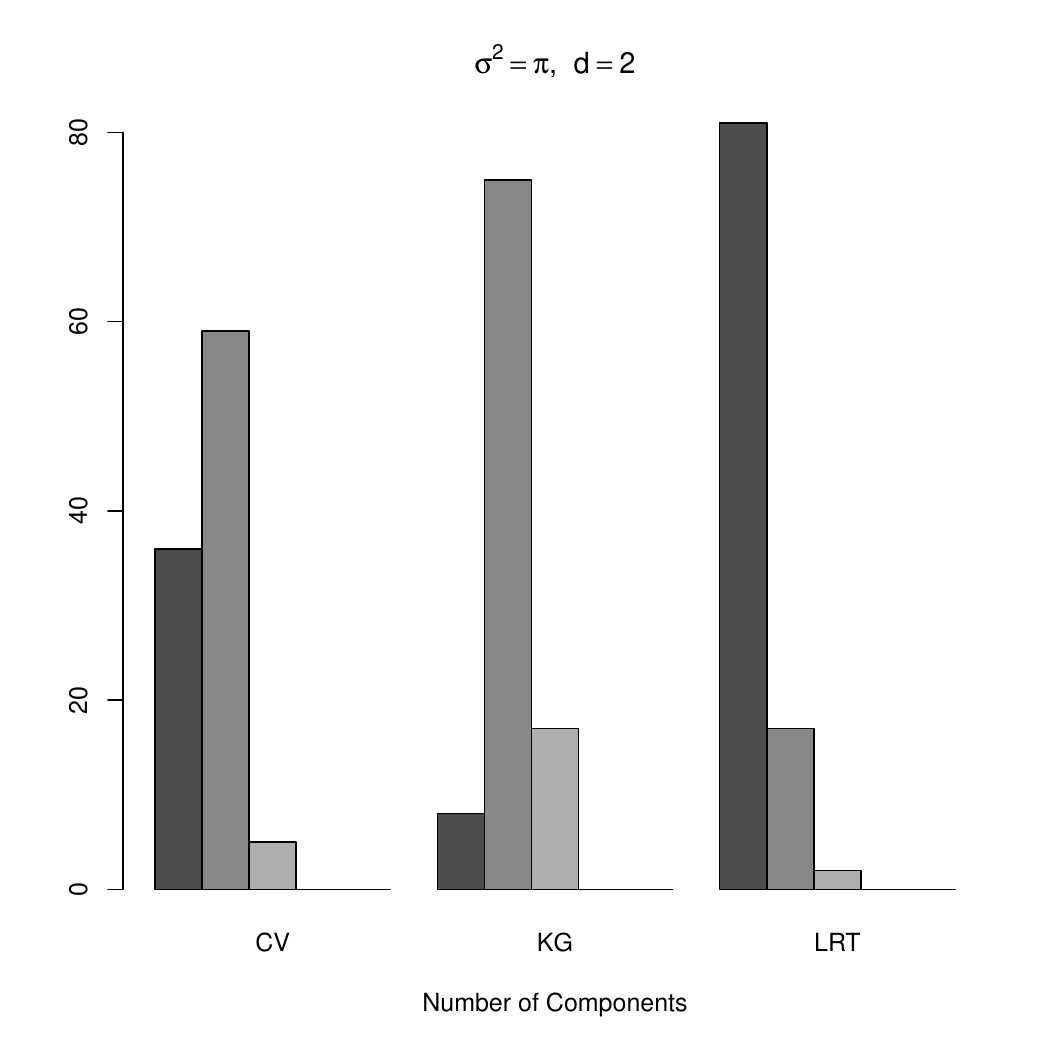} \\
\includegraphics[width=0.43\textwidth]{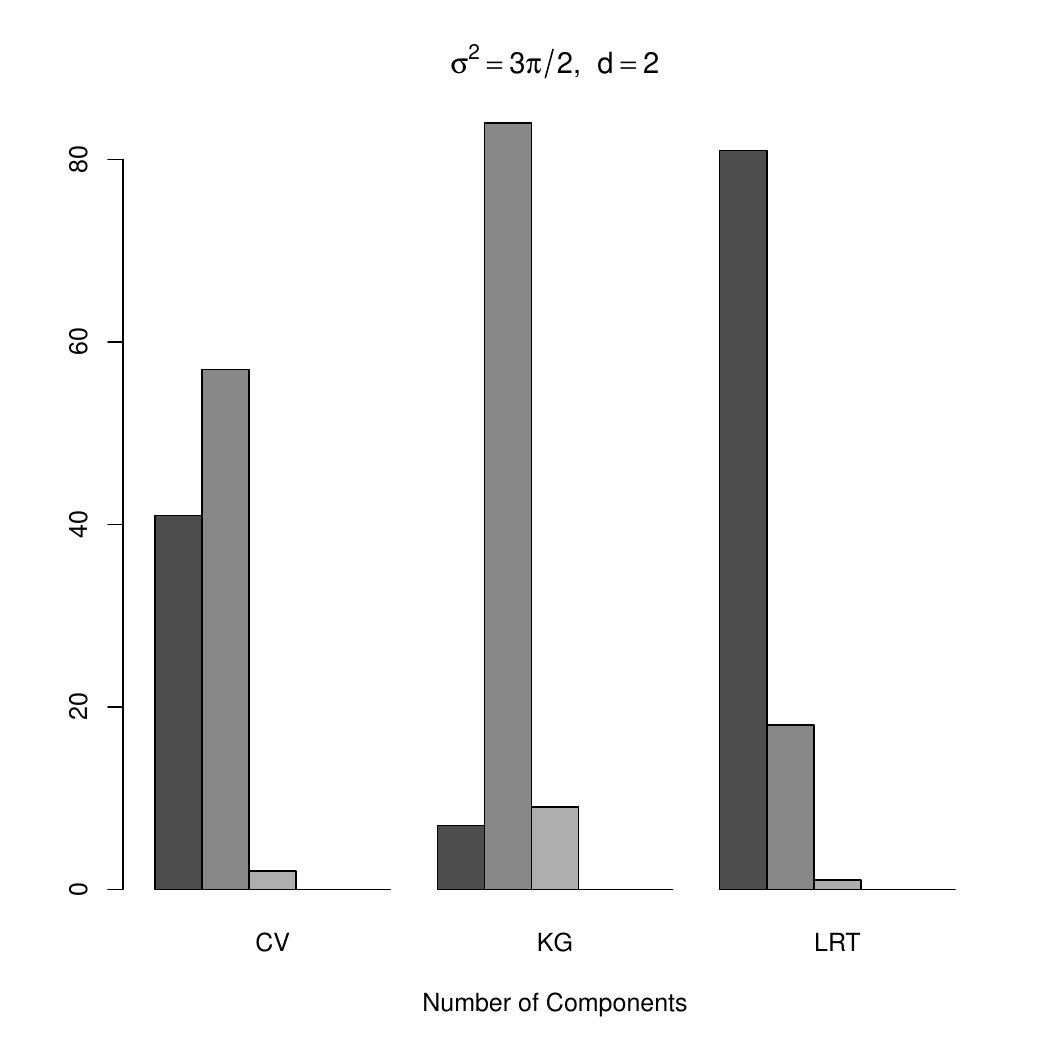}
\includegraphics[width=0.43\textwidth]{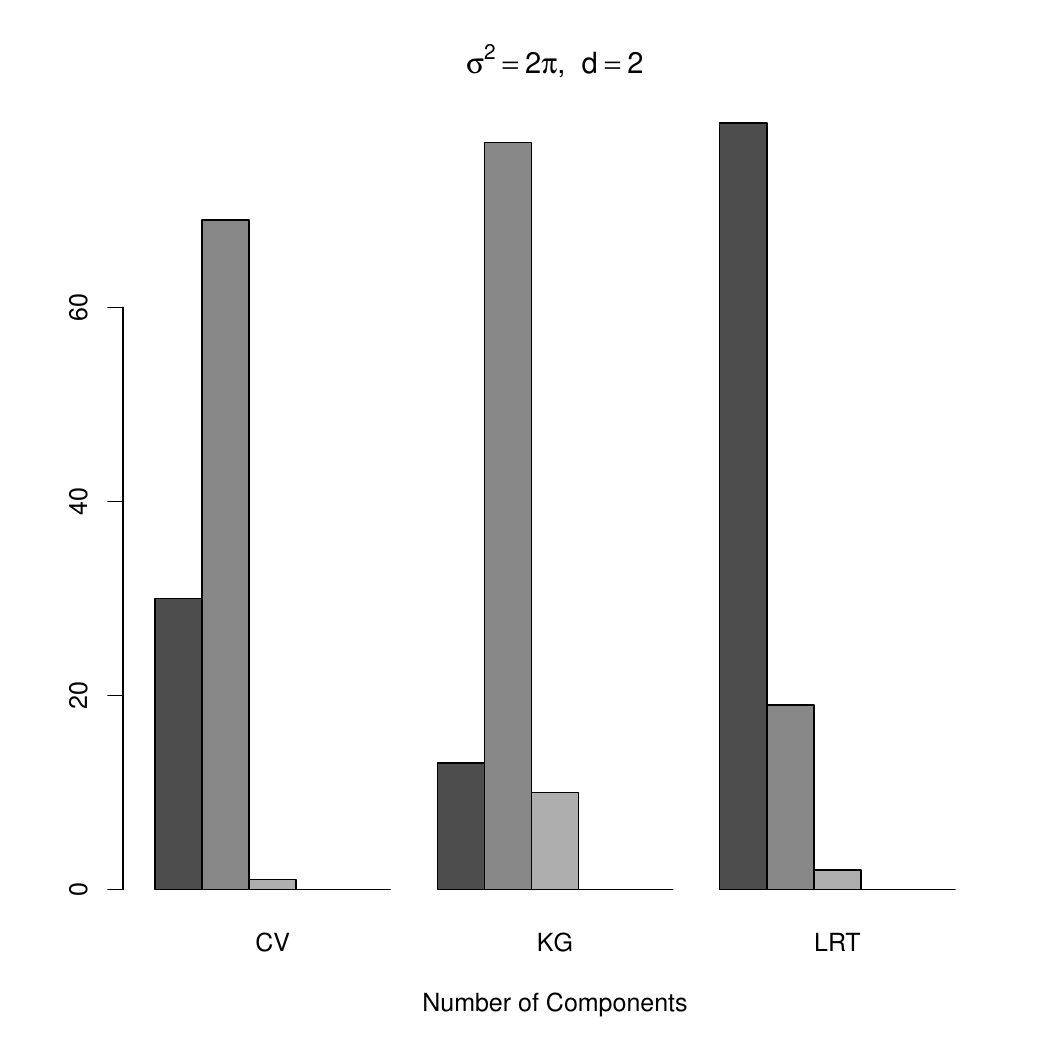}
\end{center}
\caption{Monte Carlo experiment. Frequencies of selection of dimension for CV, KG and LRT. True value is $d=2$, sample size is $50$.}
\label{sm:fig:sim:2:50}
\end{figure}

\begin{figure}
\begin{center}
\includegraphics[width=0.43\textwidth]{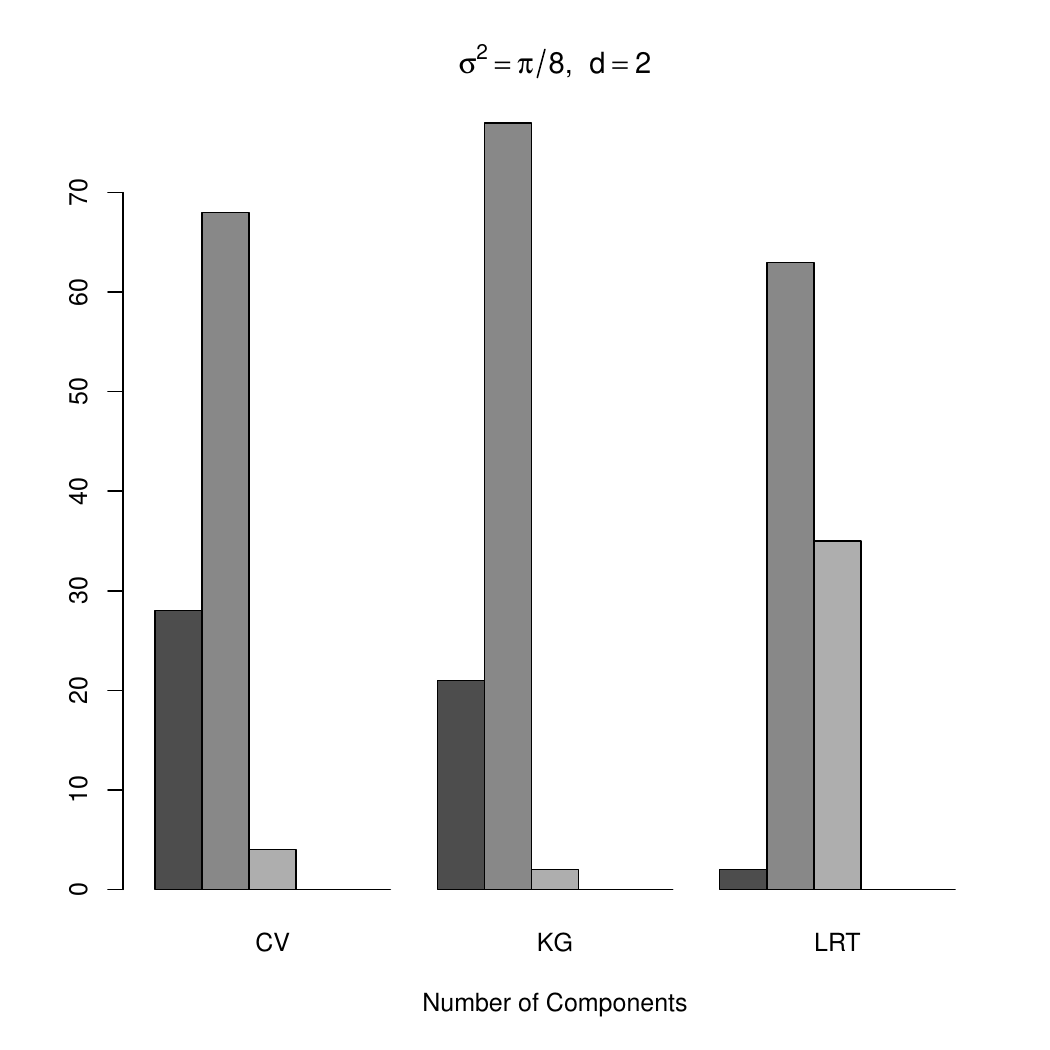}
\includegraphics[width=0.43\textwidth]{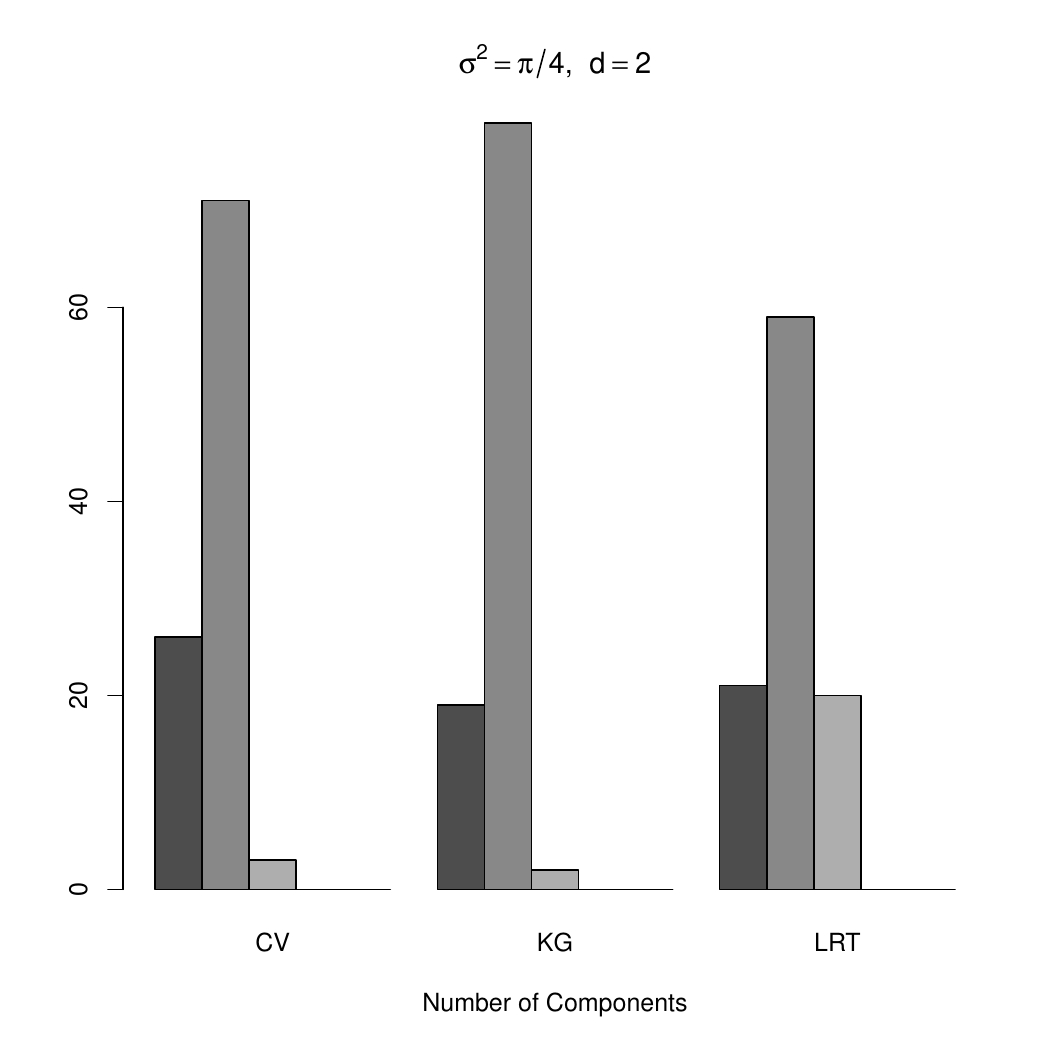} \\
\includegraphics[width=0.43\textwidth]{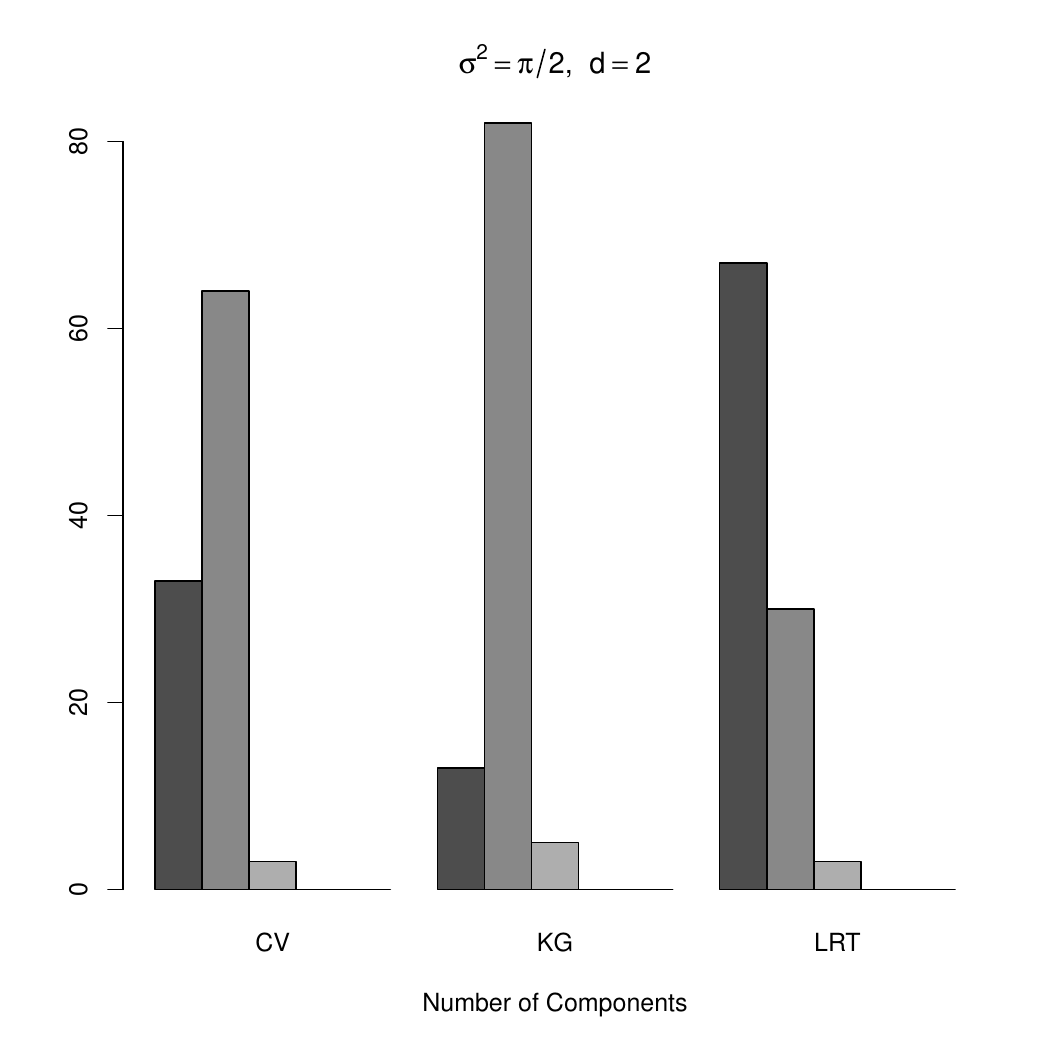}
\includegraphics[width=0.43\textwidth]{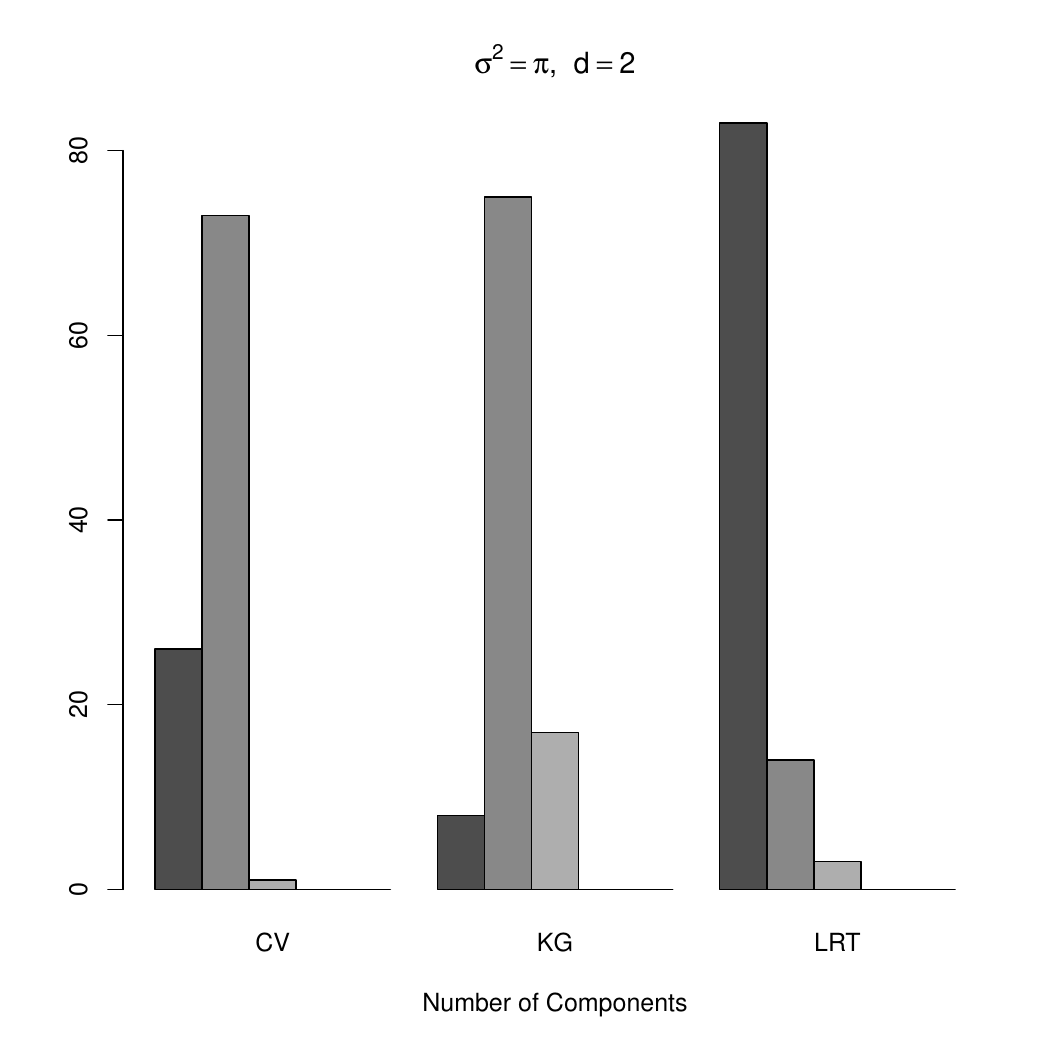} \\
\includegraphics[width=0.43\textwidth]{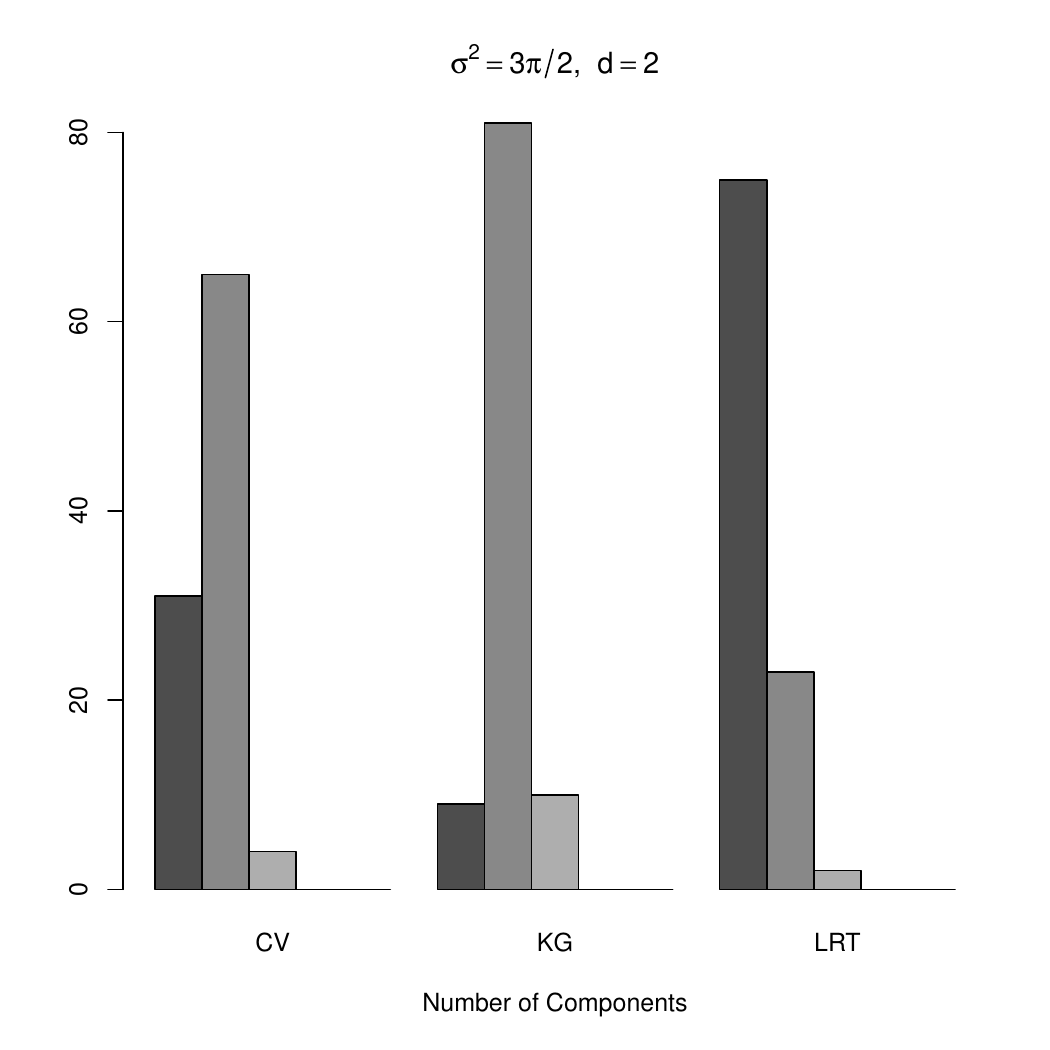}
\includegraphics[width=0.43\textwidth]{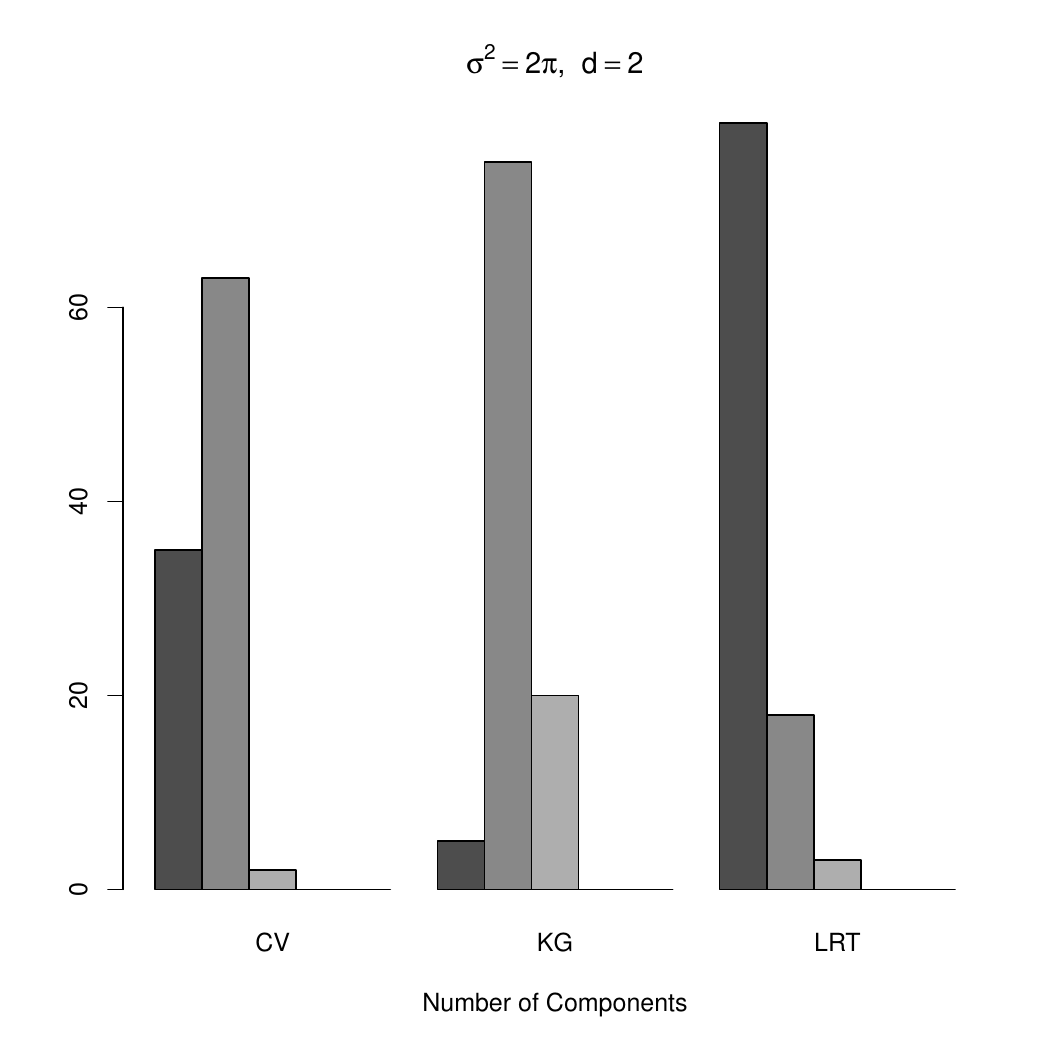}
\end{center}
\caption{Monte Carlo experiment. Frequencies of selection of dimension for CV, KG and LRT. True value is $d=2$, sample size is $100$.}
\label{sm:fig:sim:2:100}
\end{figure}

\begin{figure}
\begin{center}
\includegraphics[width=0.43\textwidth]{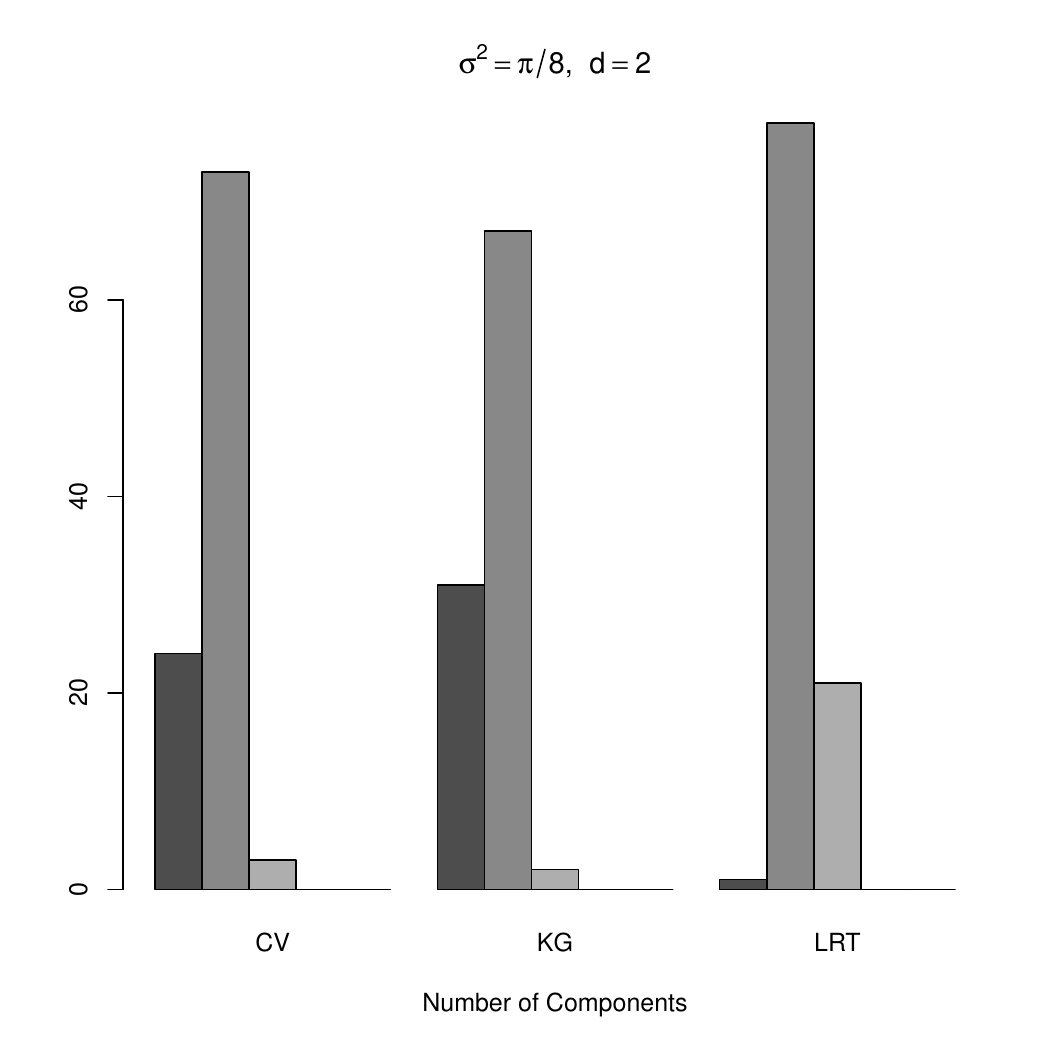}
\includegraphics[width=0.43\textwidth]{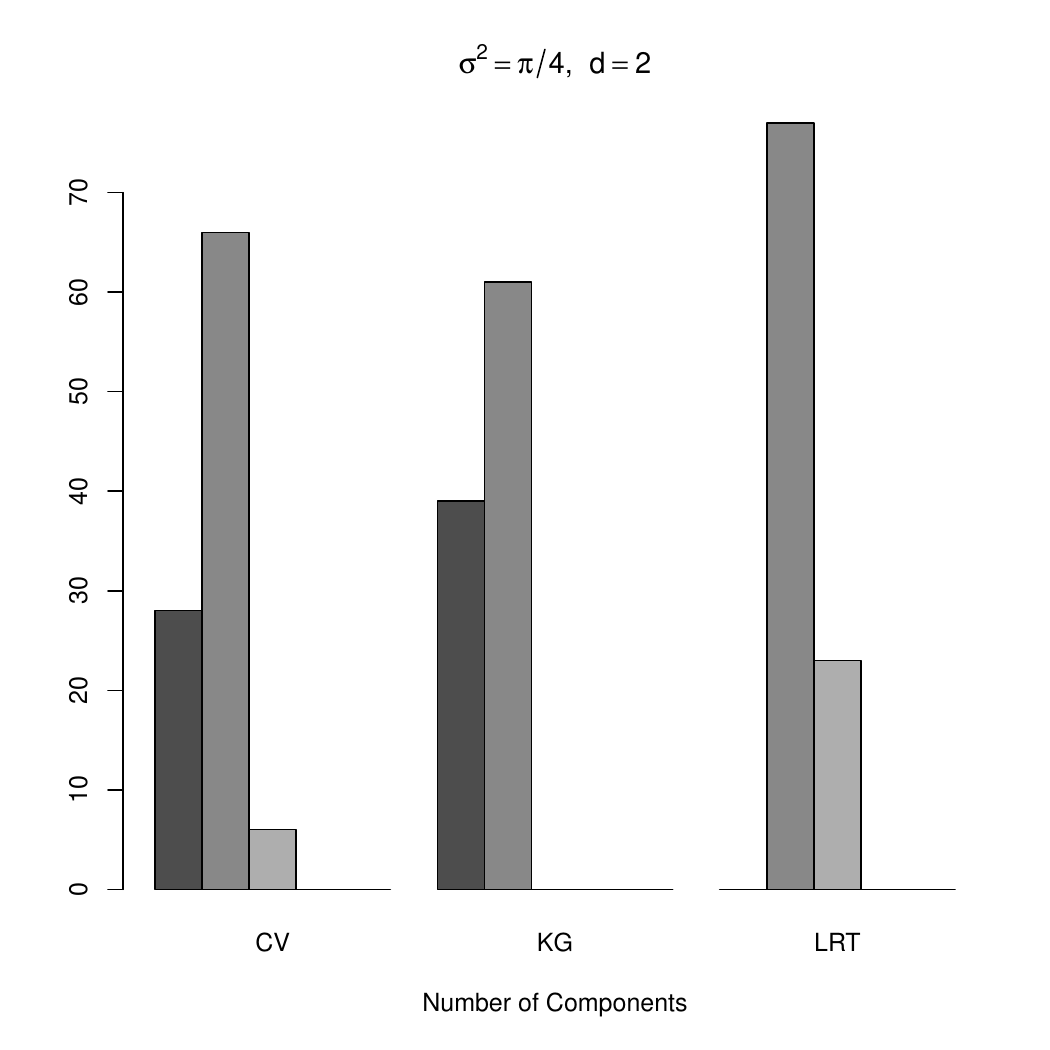} \\
\includegraphics[width=0.43\textwidth]{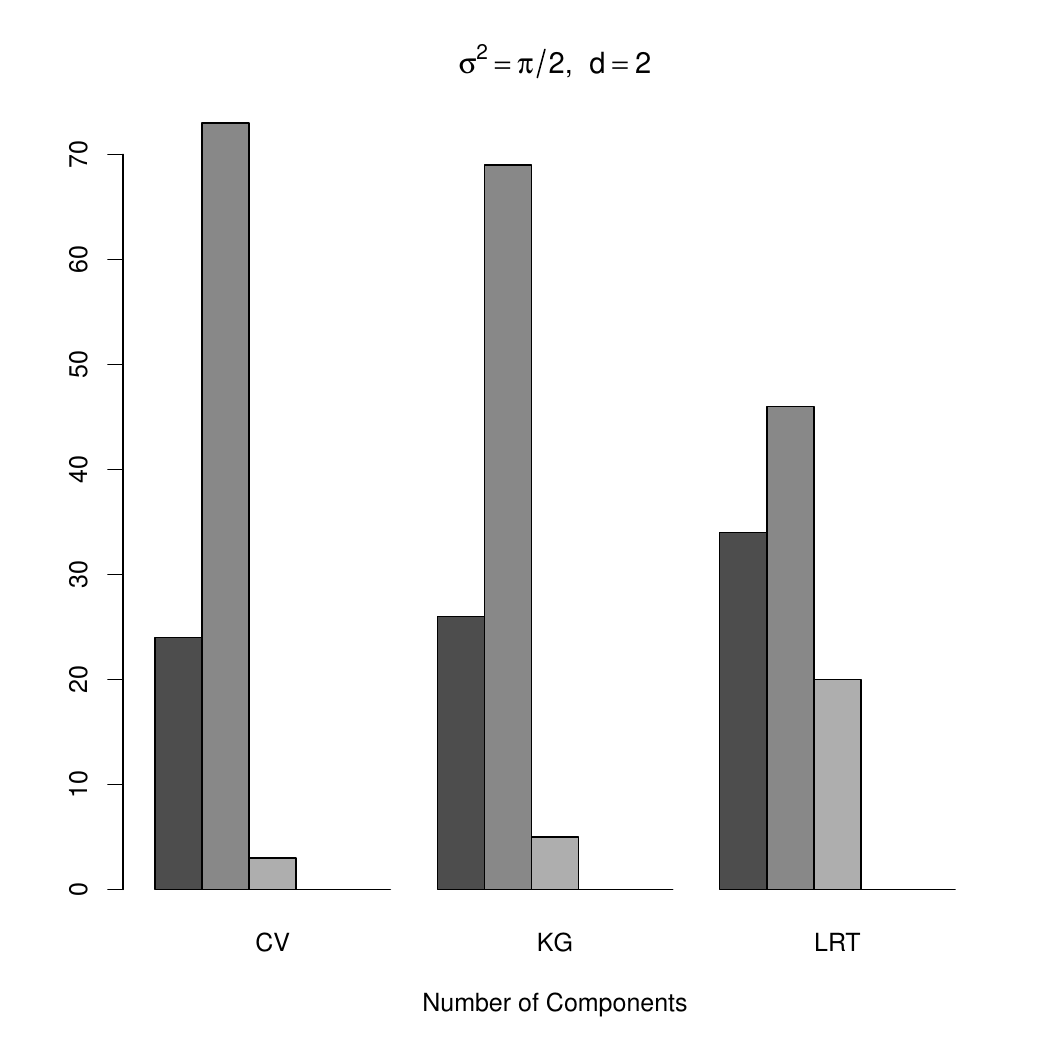}
\includegraphics[width=0.43\textwidth]{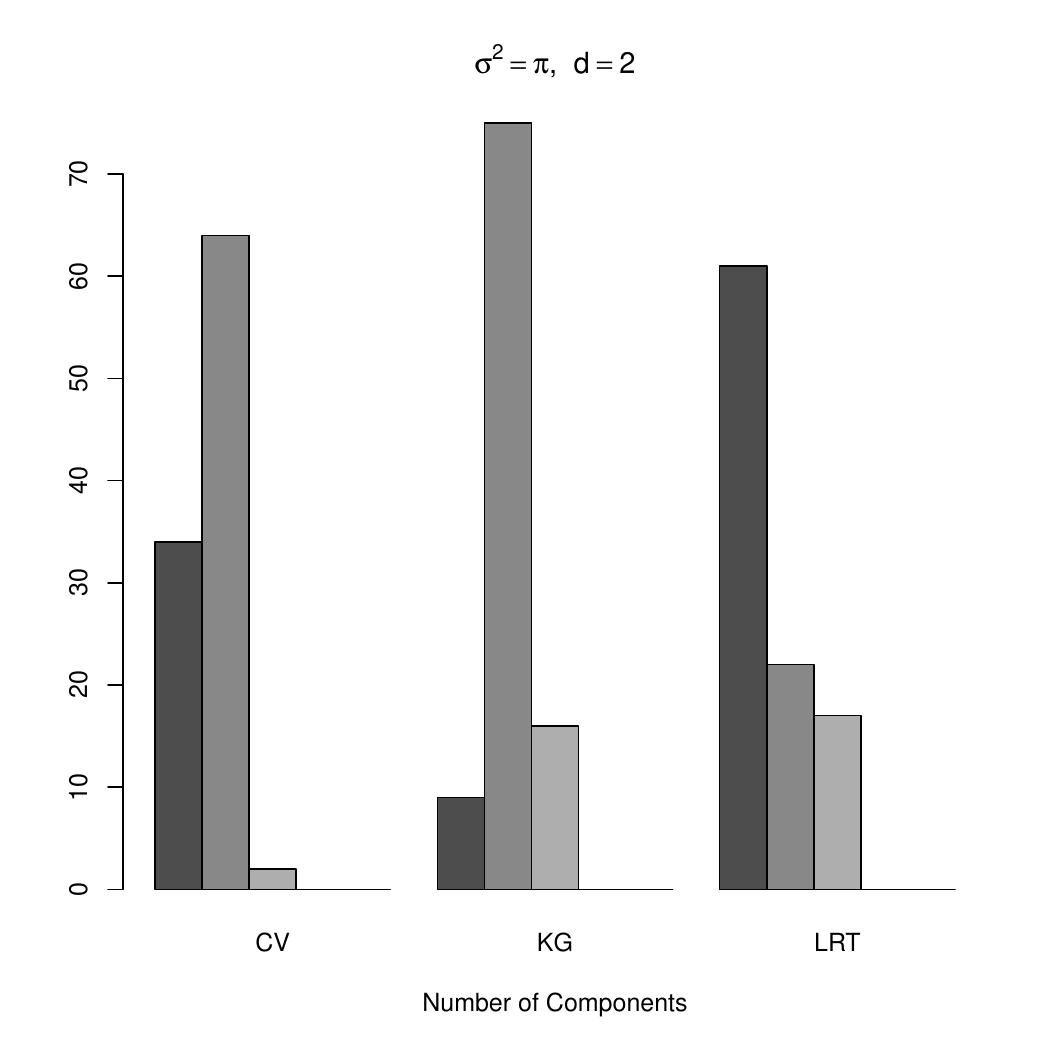} \\
\includegraphics[width=0.43\textwidth]{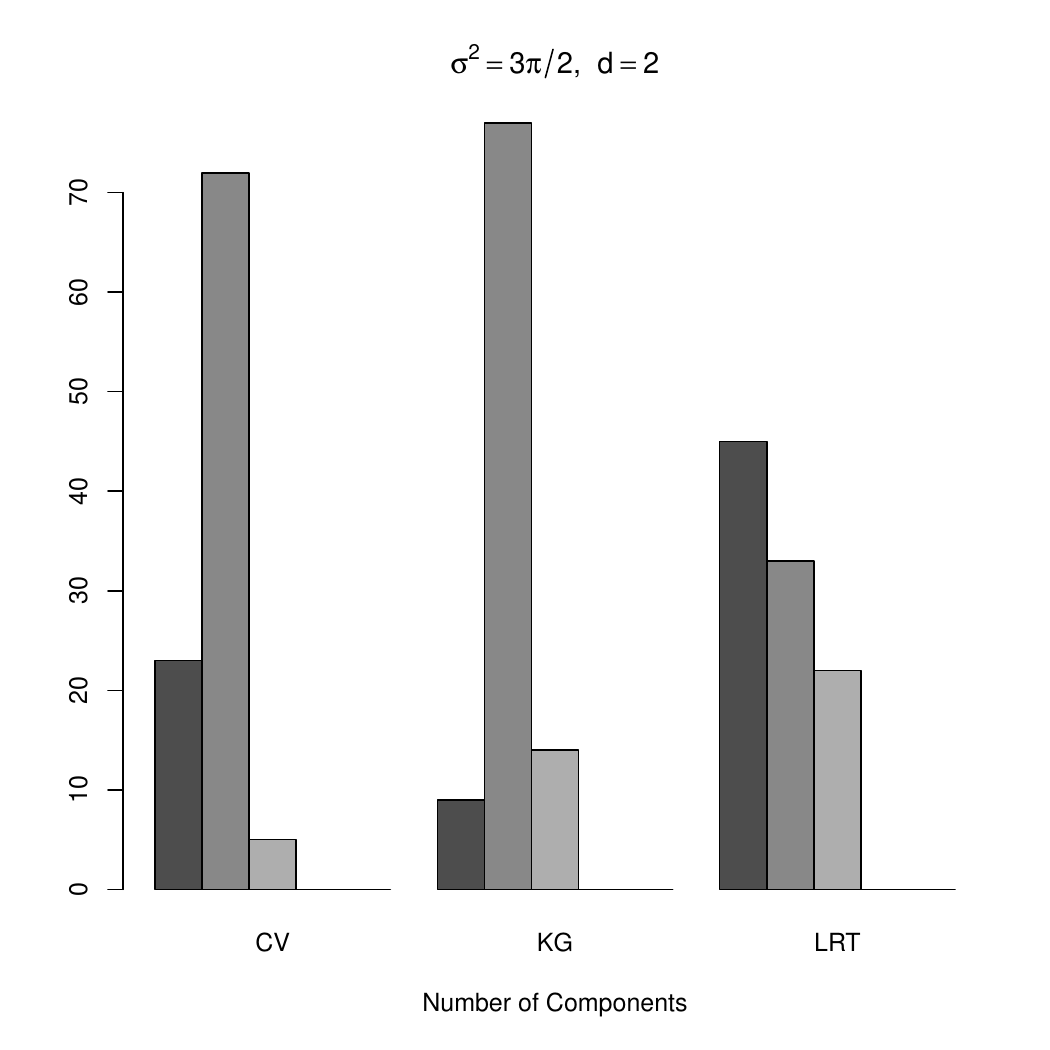}
\includegraphics[width=0.43\textwidth]{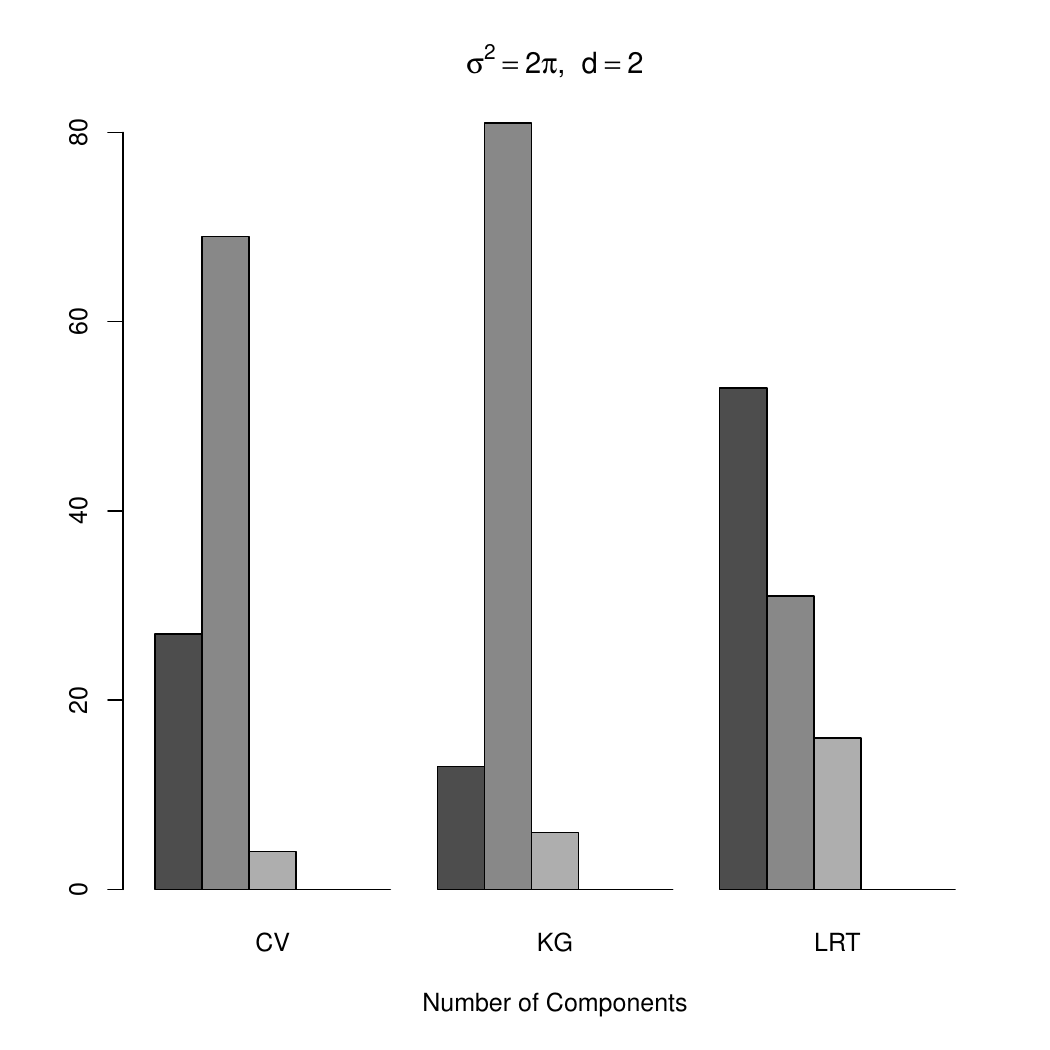}
\end{center}
\caption{Monte Carlo experiment. Frequencies of selection of dimension for CV, KG and LRT. True value is $d=2$, sample size is $500$.}
\label{sm:fig:sim:2:500}
\end{figure}

\begin{figure}
\begin{center}
\includegraphics[width=0.43\textwidth]{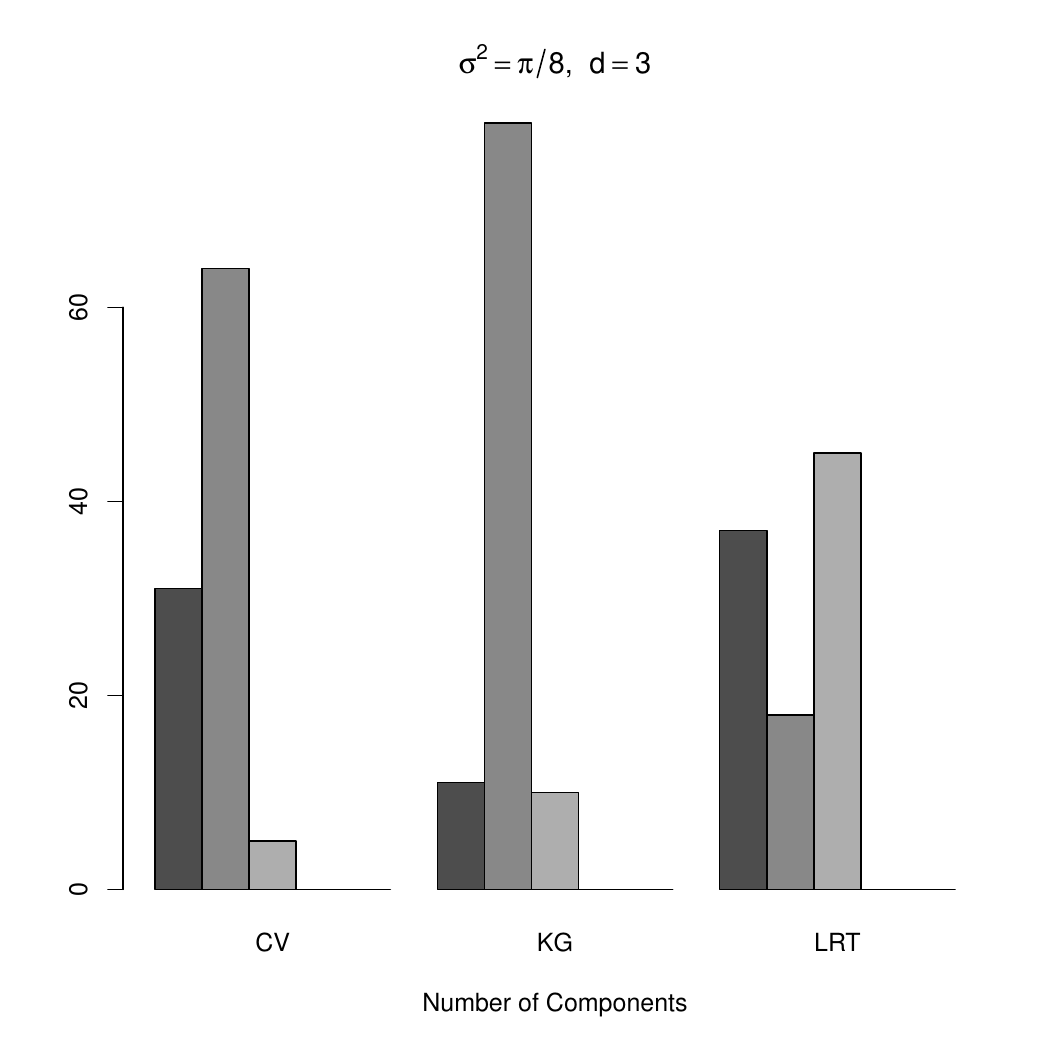}
\includegraphics[width=0.43\textwidth]{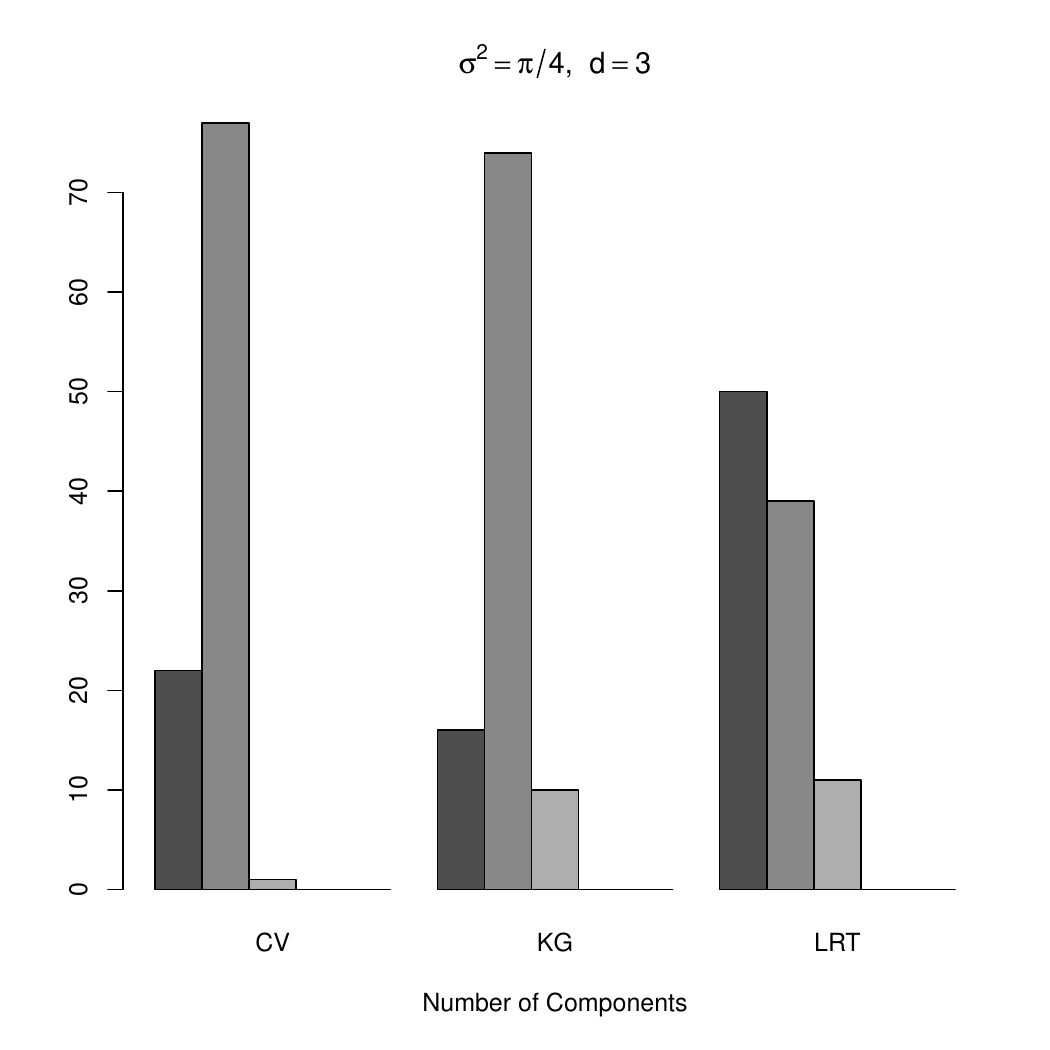} \\
\includegraphics[width=0.43\textwidth]{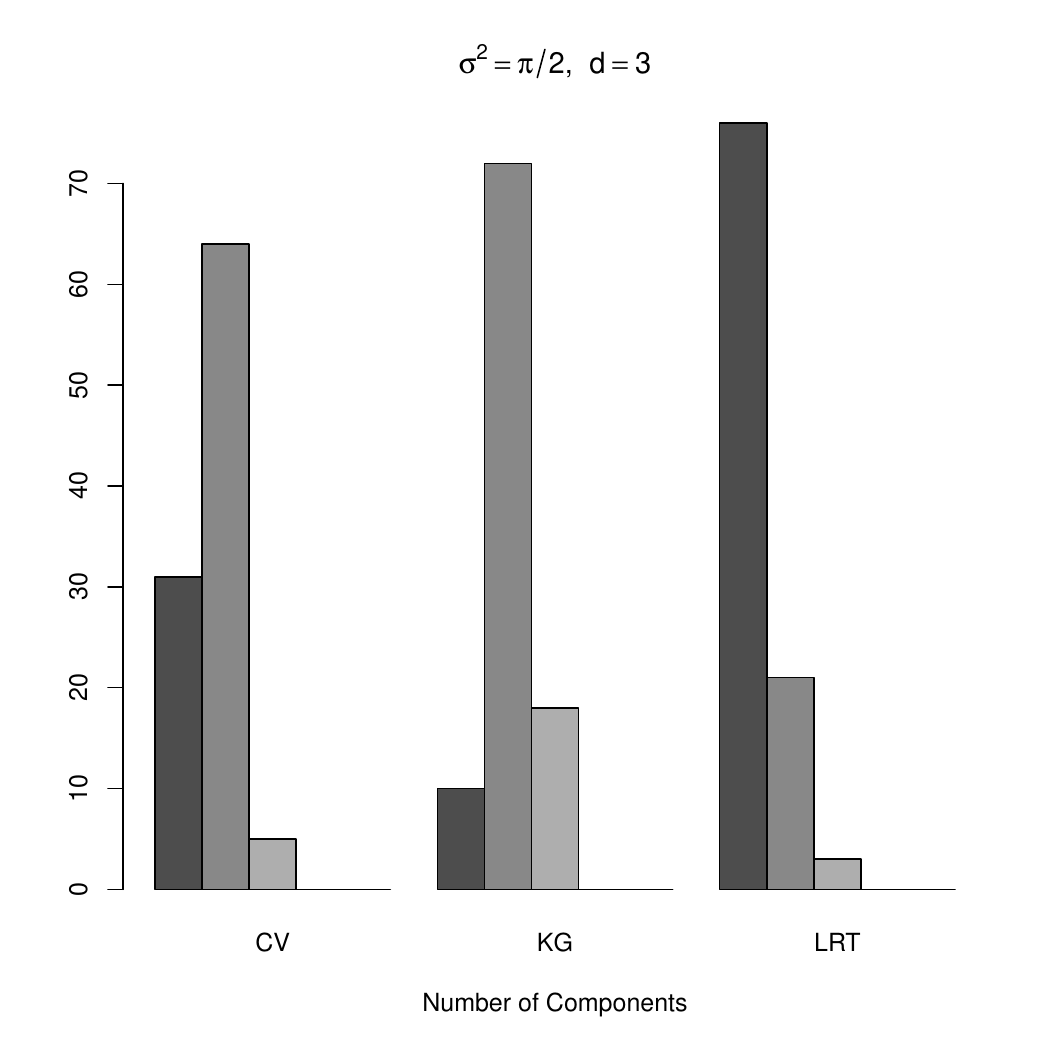}
\includegraphics[width=0.43\textwidth]{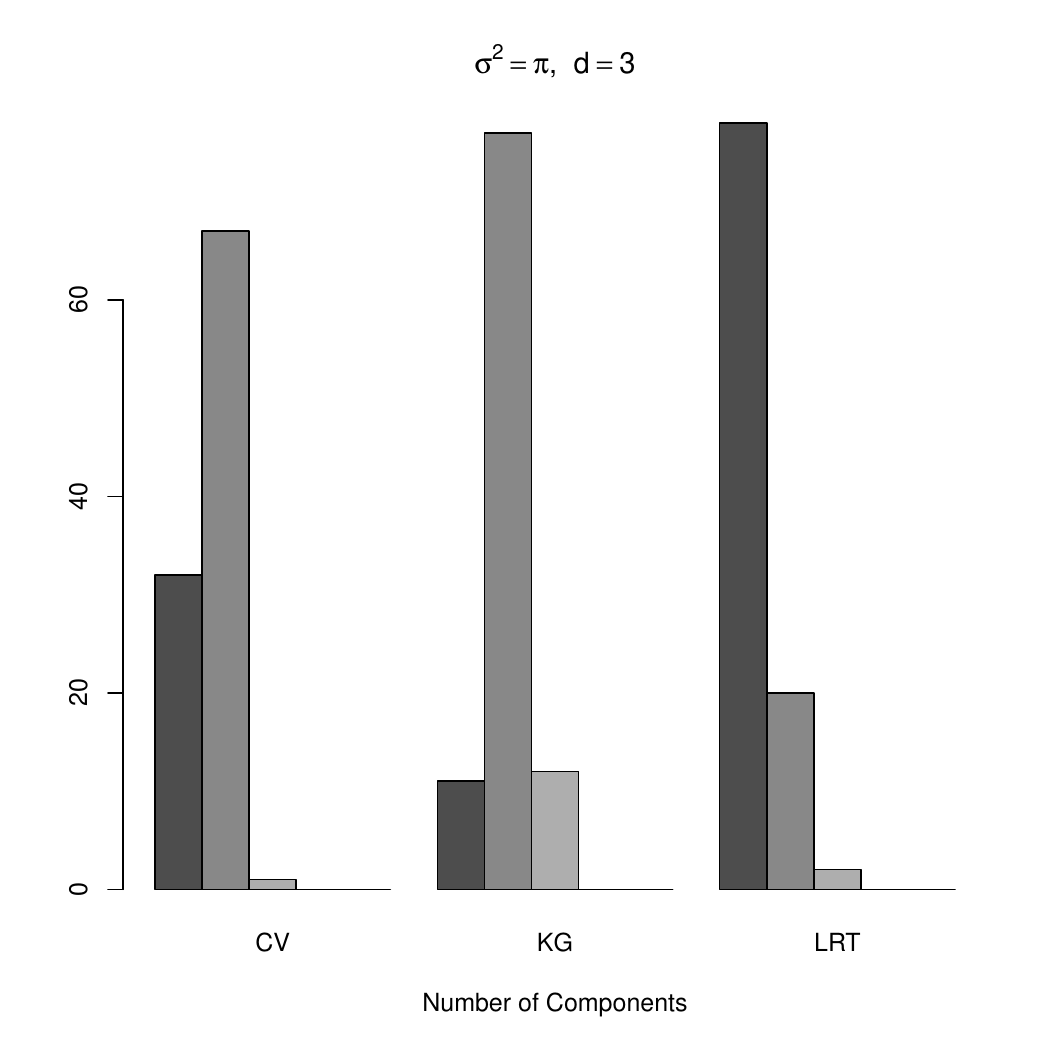} \\
\includegraphics[width=0.43\textwidth]{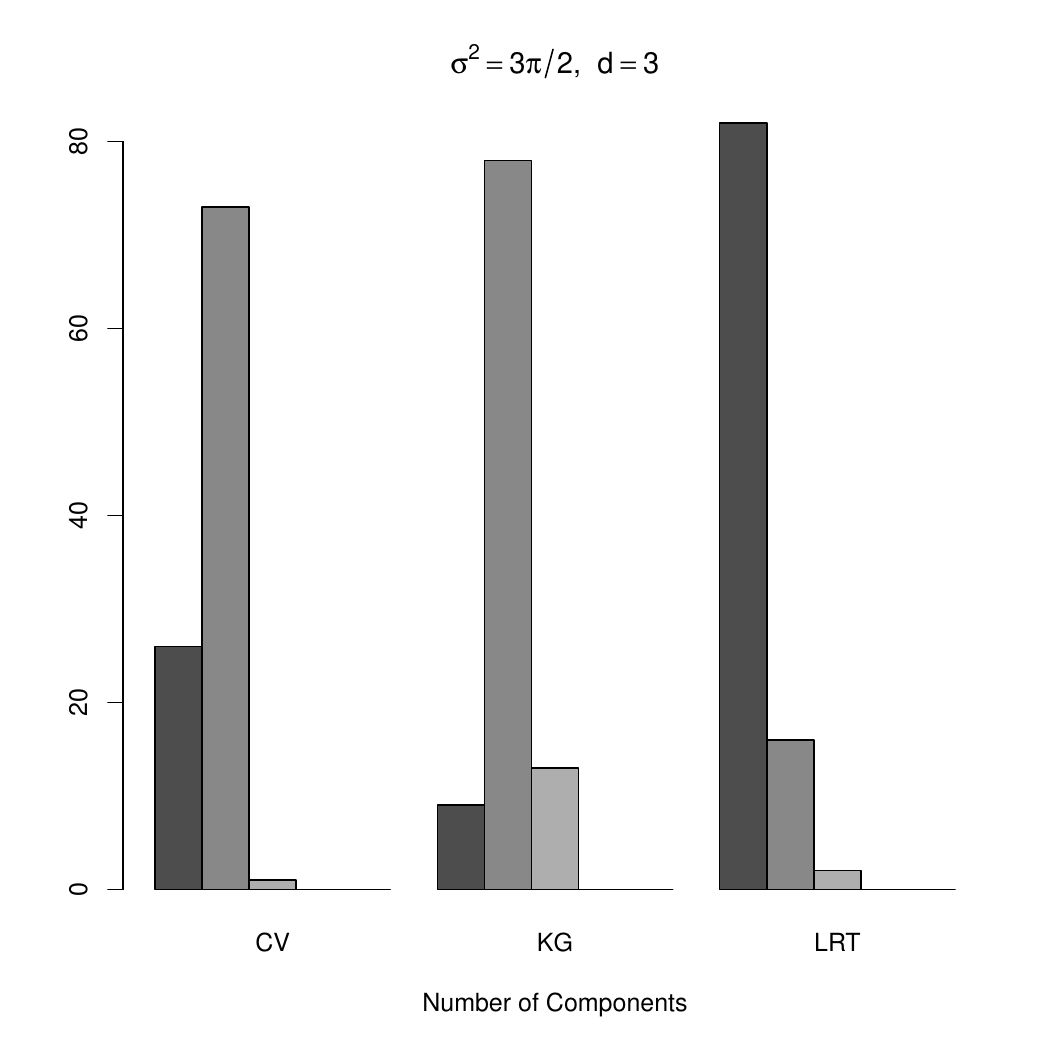}
\includegraphics[width=0.43\textwidth]{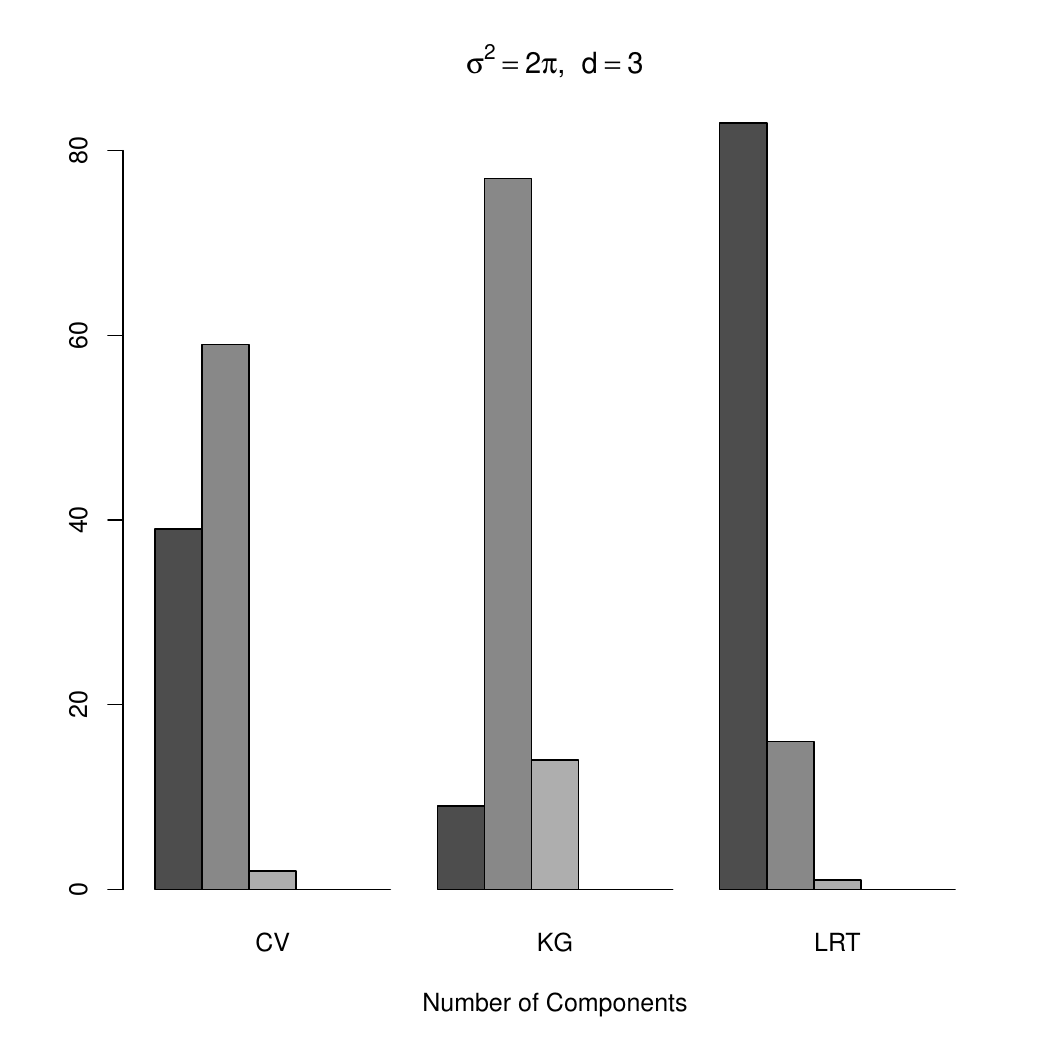}
\end{center}
\caption{Monte Carlo experiment. Frequencies of selection of dimension for CV, KG and LRT. True value is $d=3$, sample size is $50$.}
\label{sm:fig:sim:3:50}
\end{figure}

\begin{figure}
\begin{center}
\includegraphics[width=0.43\textwidth]{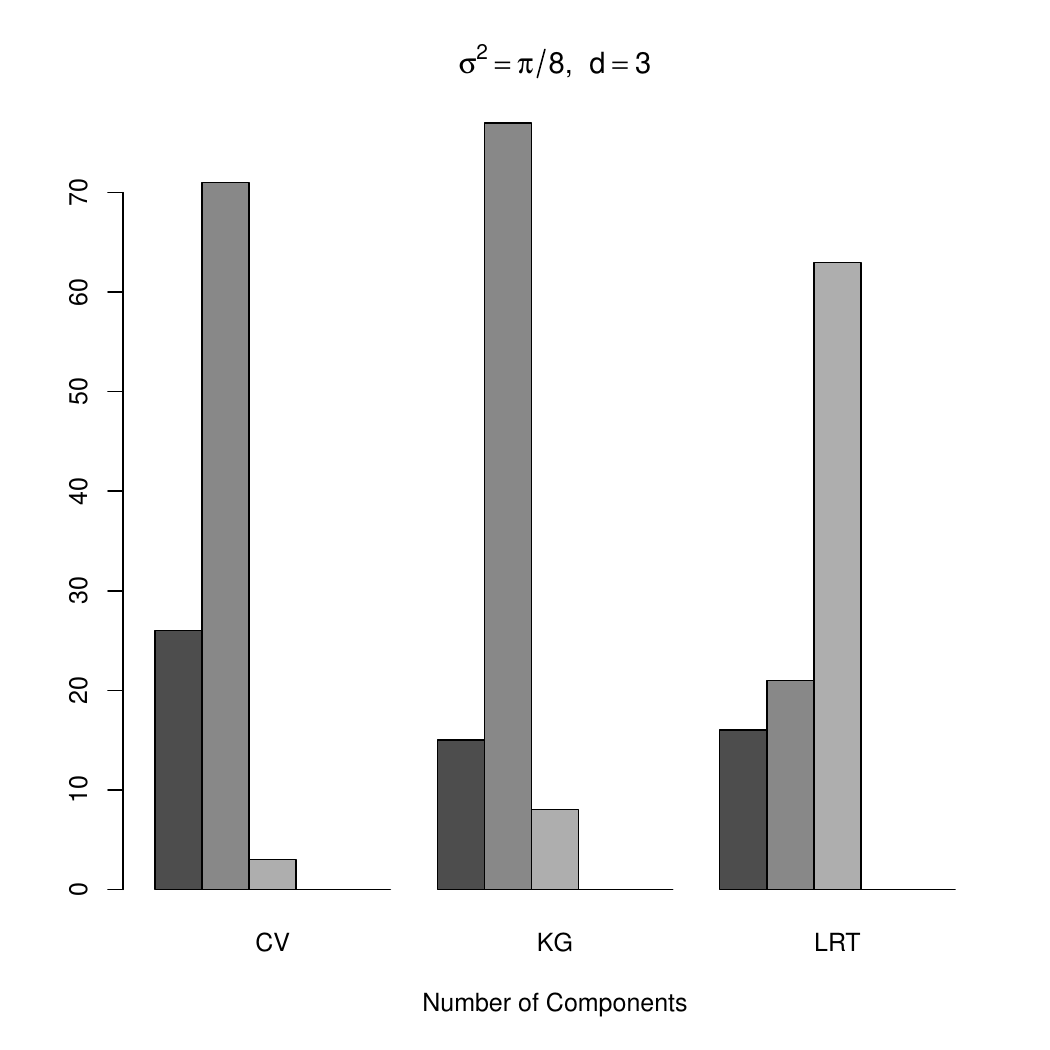}
\includegraphics[width=0.43\textwidth]{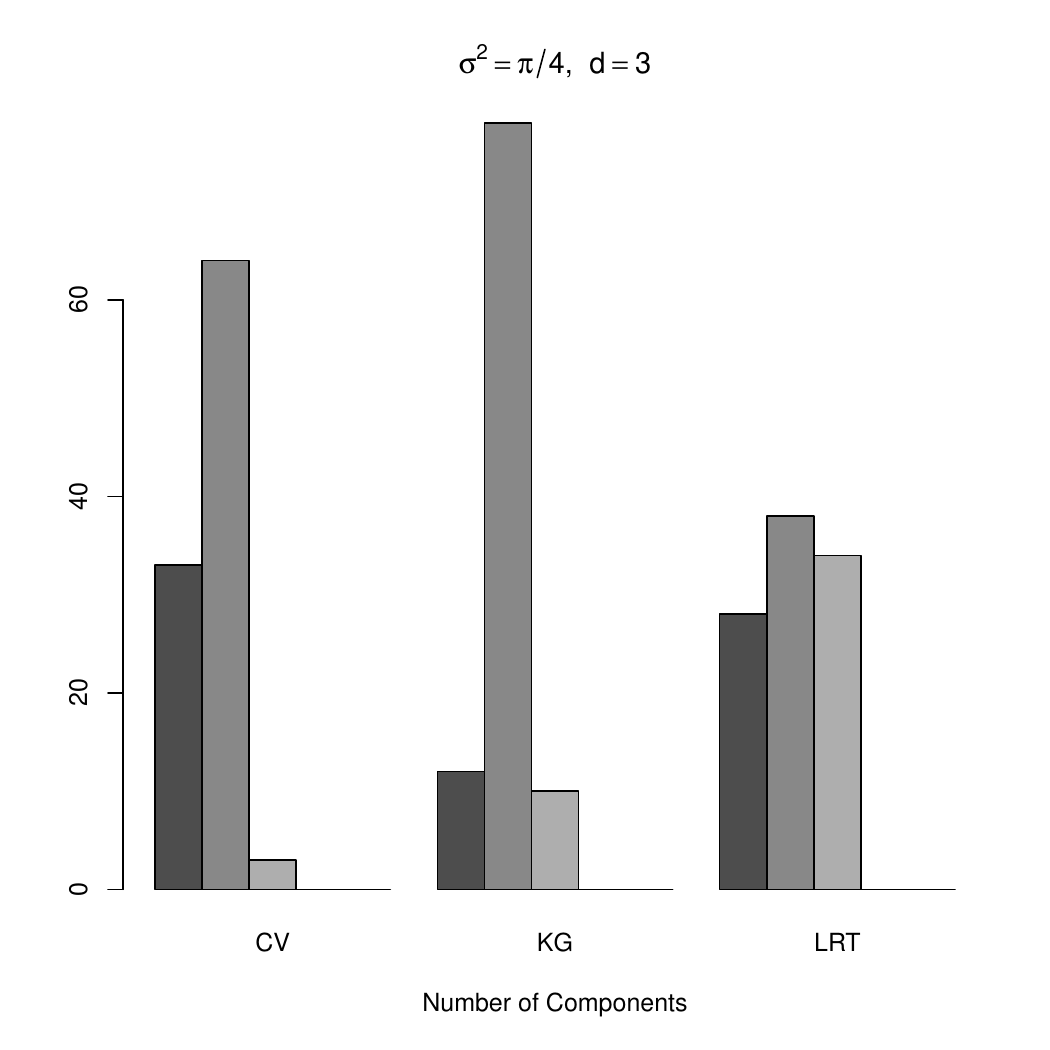} \\
\includegraphics[width=0.43\textwidth]{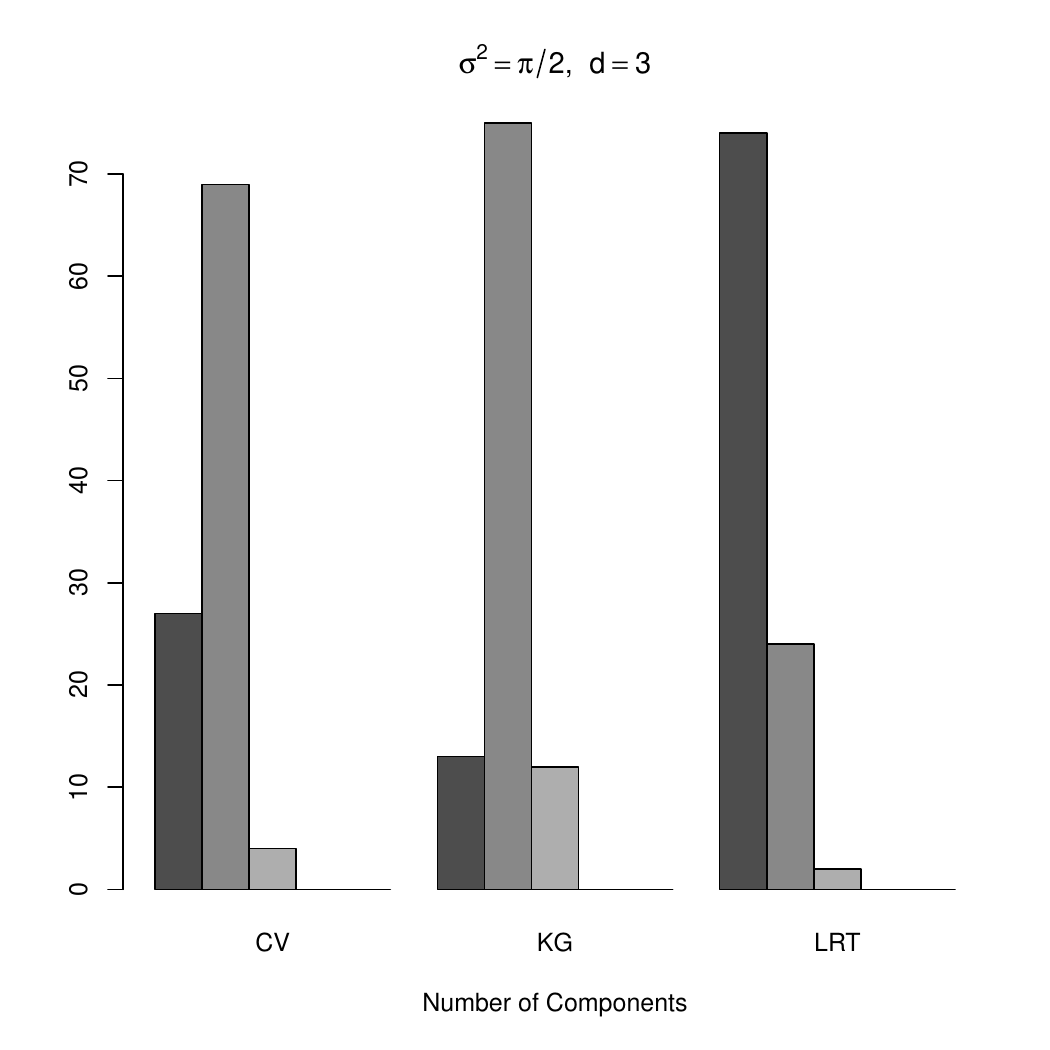}
\includegraphics[width=0.43\textwidth]{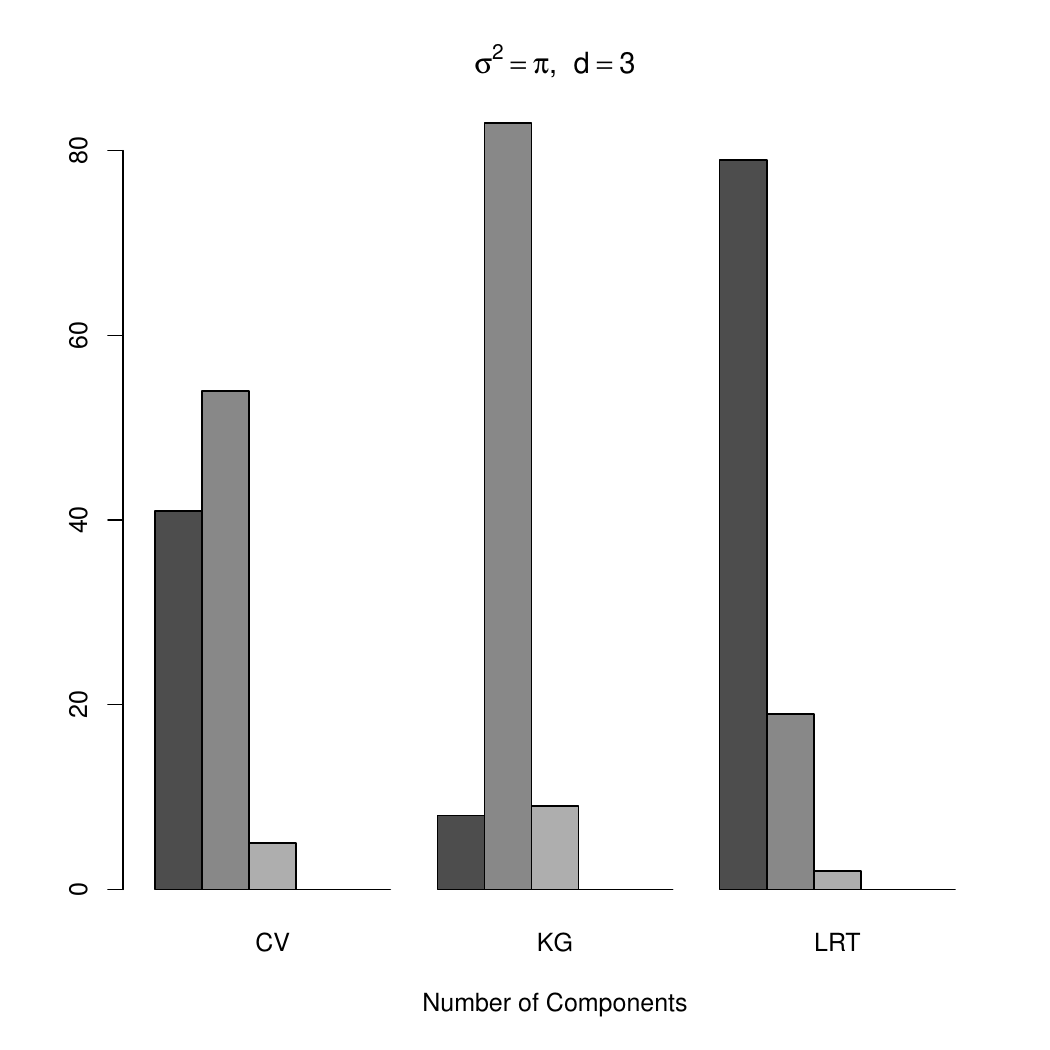} \\
\includegraphics[width=0.43\textwidth]{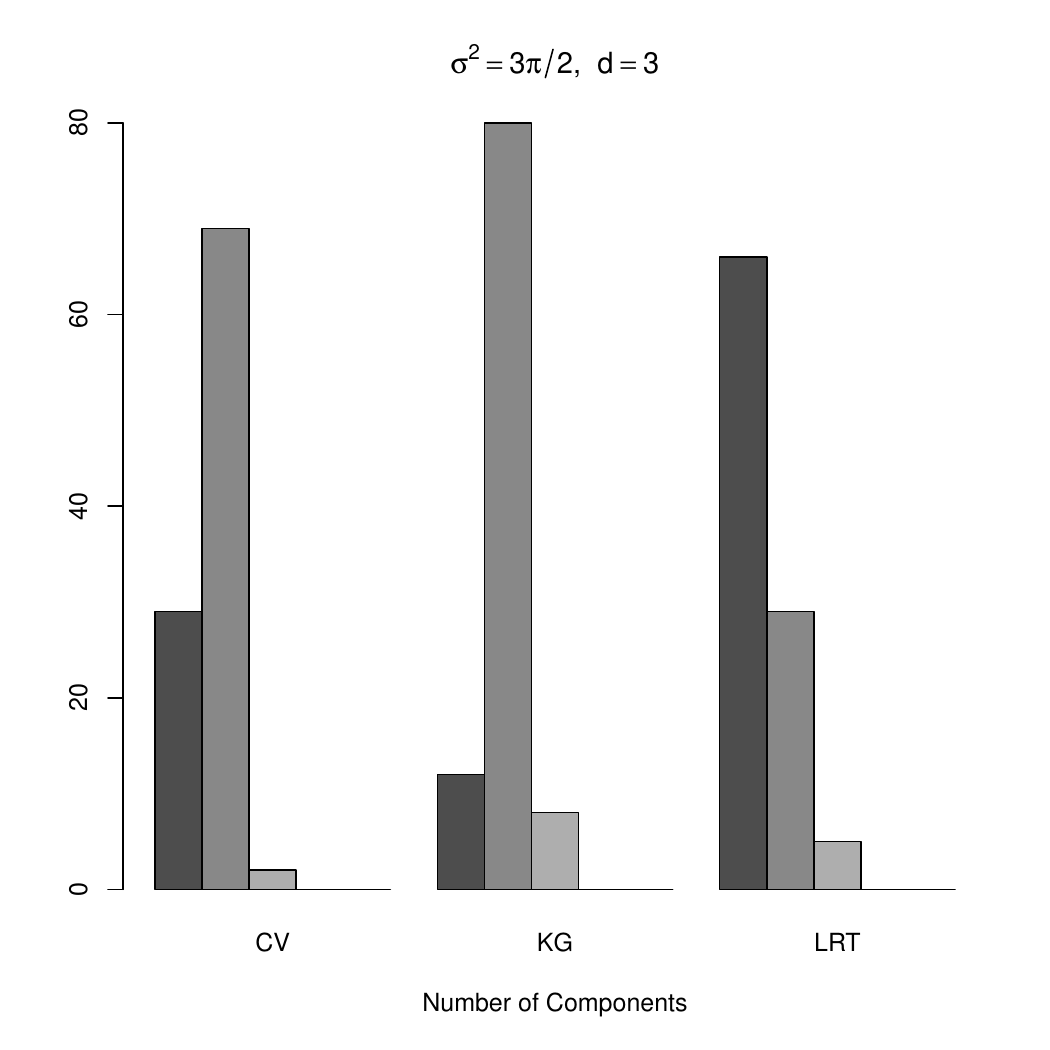}
\includegraphics[width=0.43\textwidth]{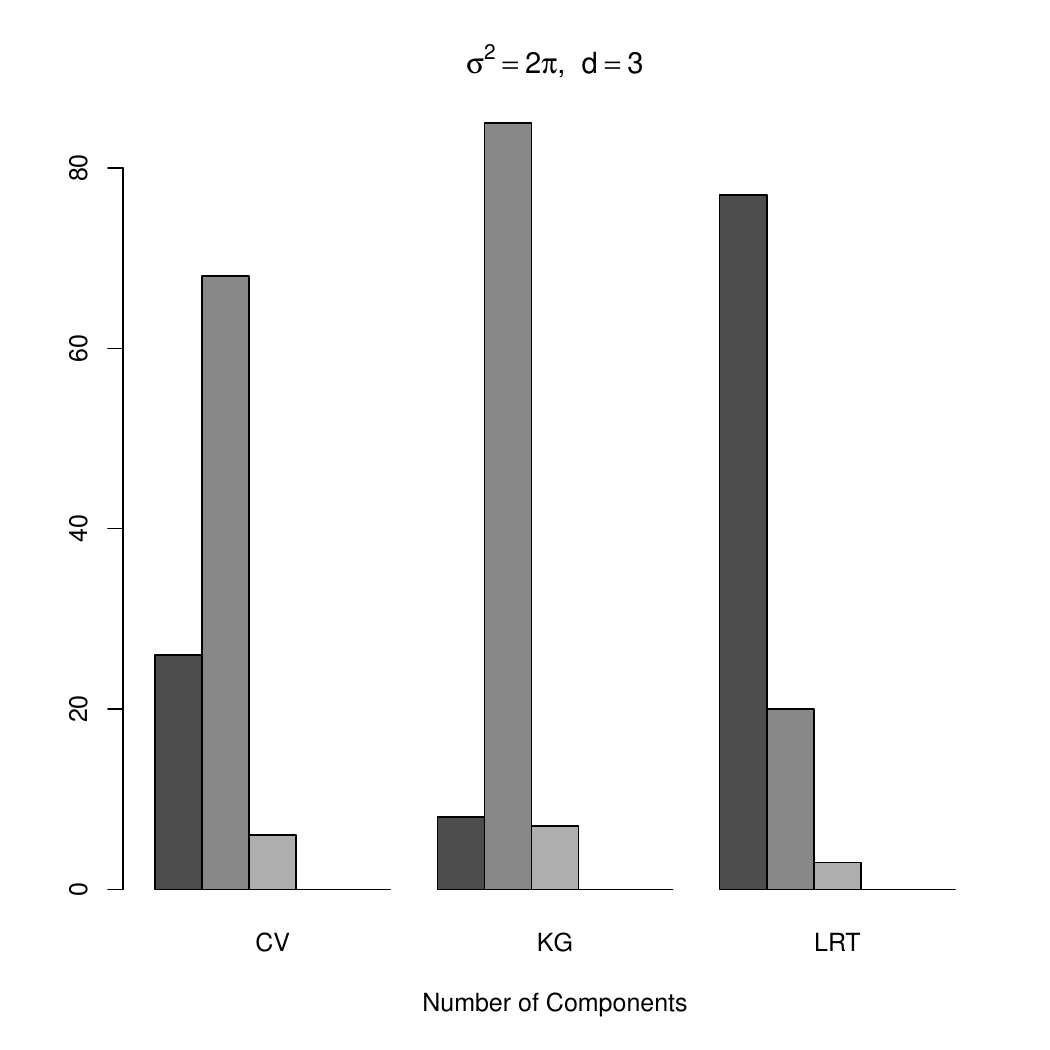}
\end{center}
\caption{Monte Carlo experiment. Frequencies of selection of dimension for CV, KG and LRT. True value is $d=3$, sample size is $100$.}
\label{sm:fig:sim:3:100}
\end{figure}

\begin{figure}
\begin{center}
\includegraphics[width=0.43\textwidth]{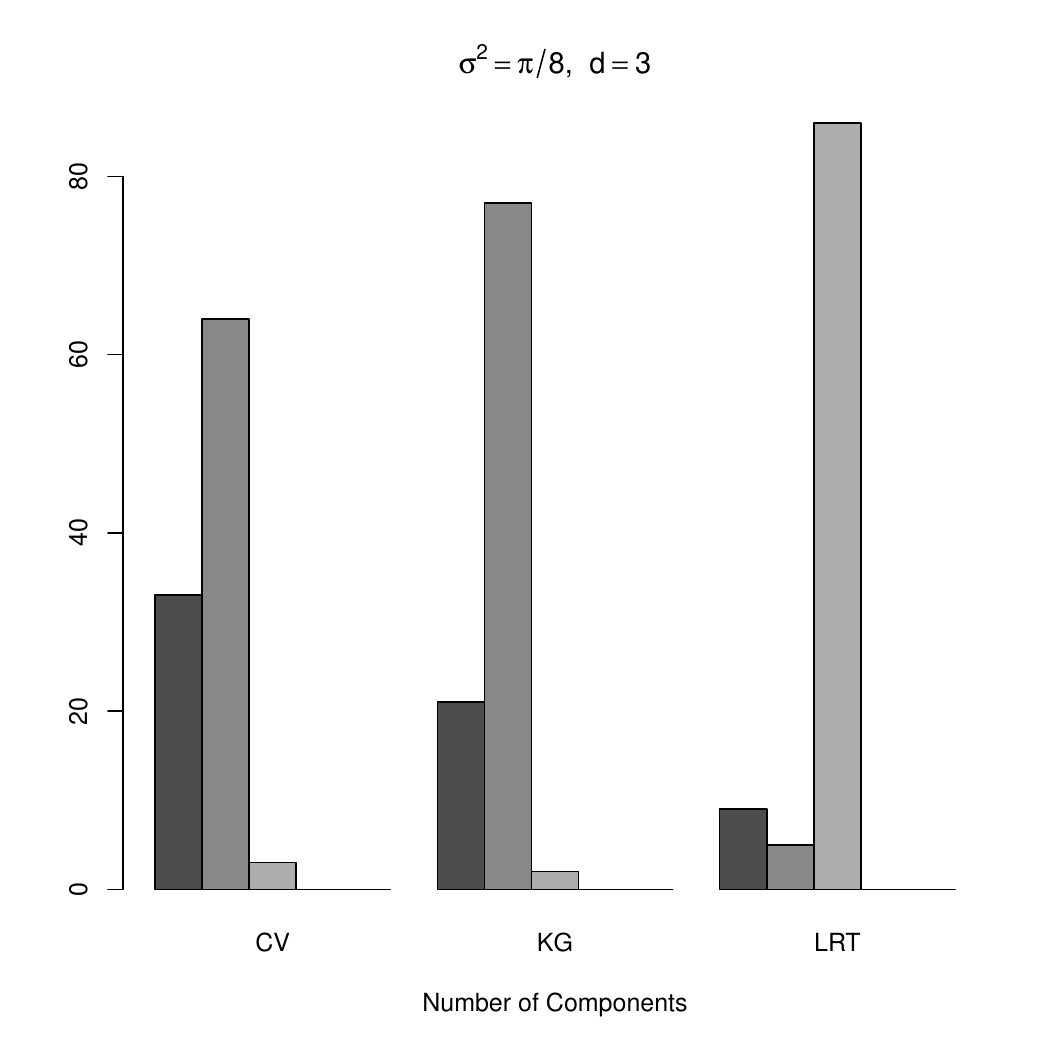}
\includegraphics[width=0.43\textwidth]{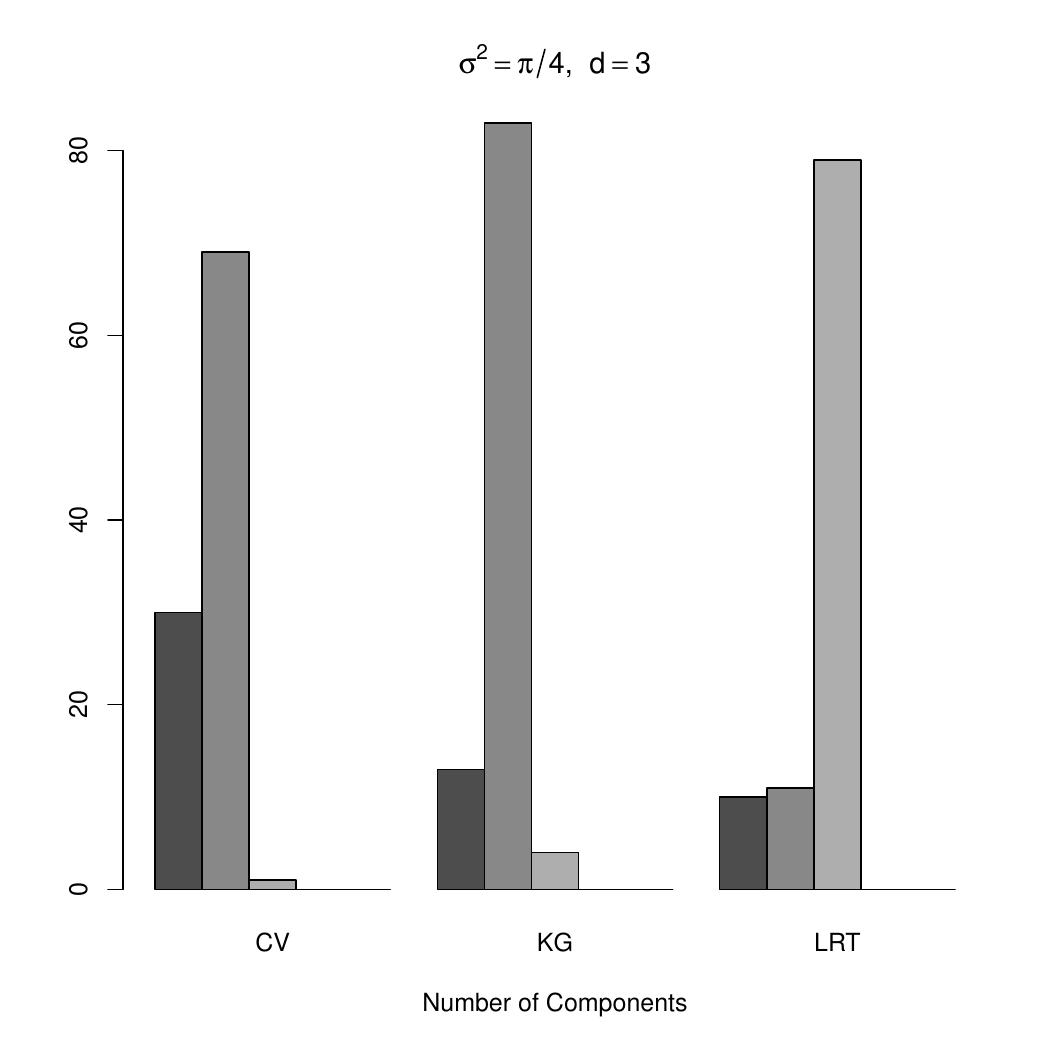} \\
\includegraphics[width=0.43\textwidth]{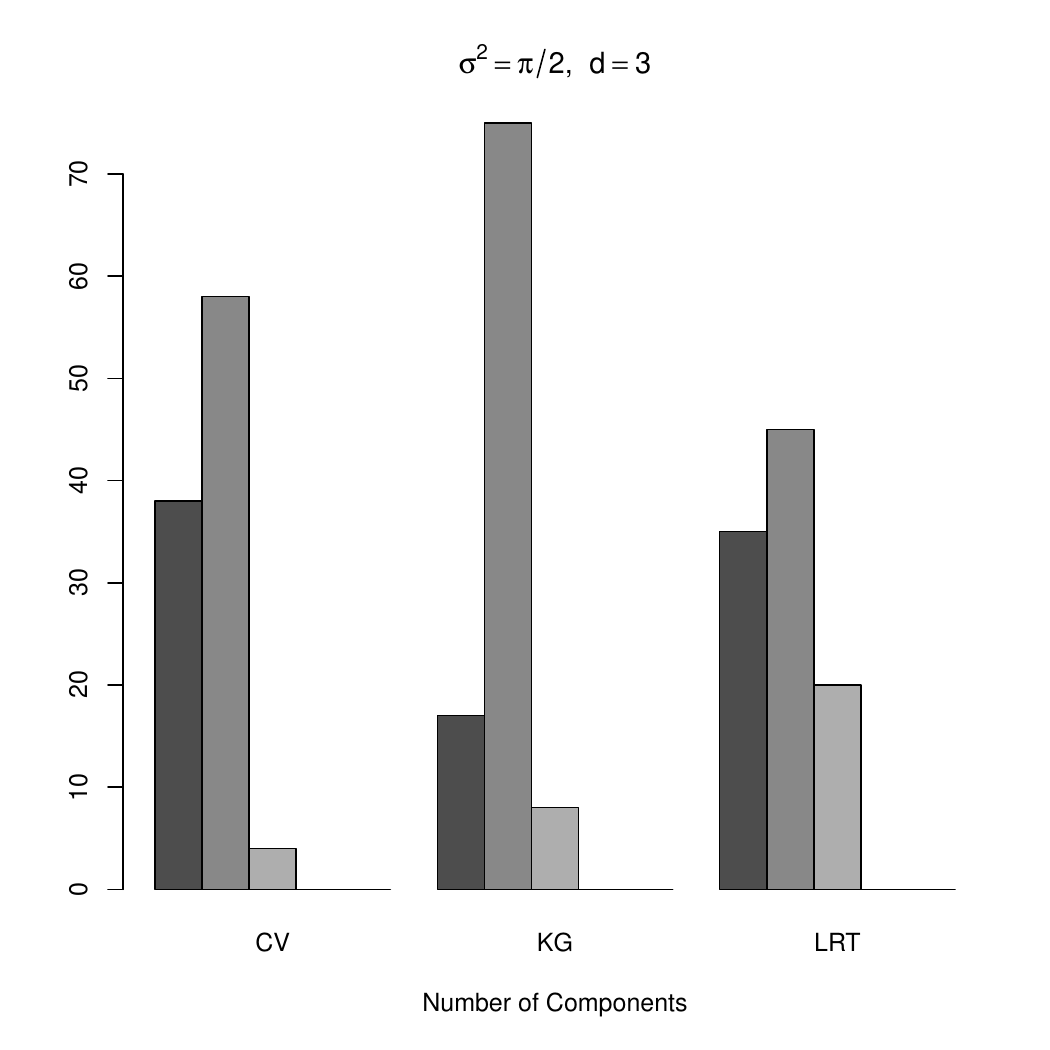}
\includegraphics[width=0.43\textwidth]{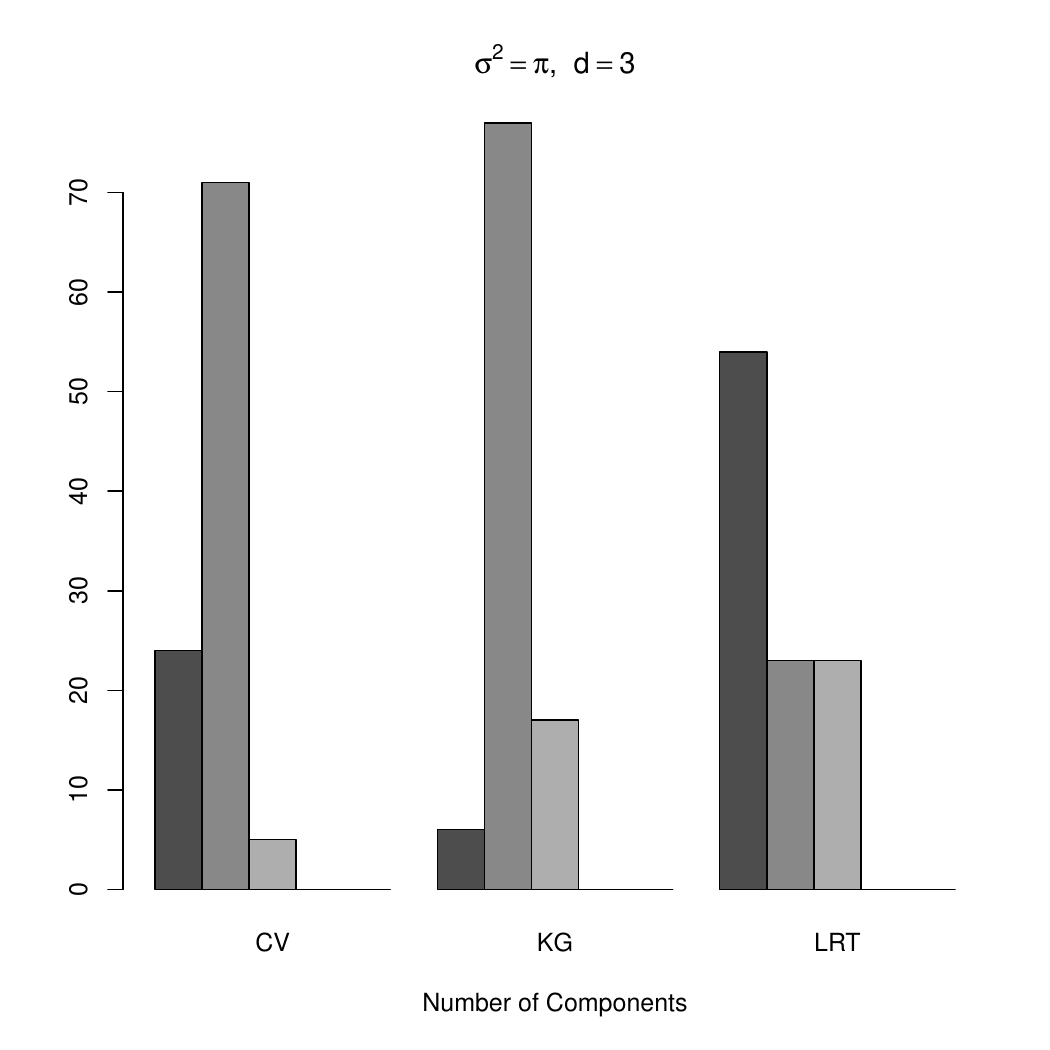} \\
\includegraphics[width=0.43\textwidth]{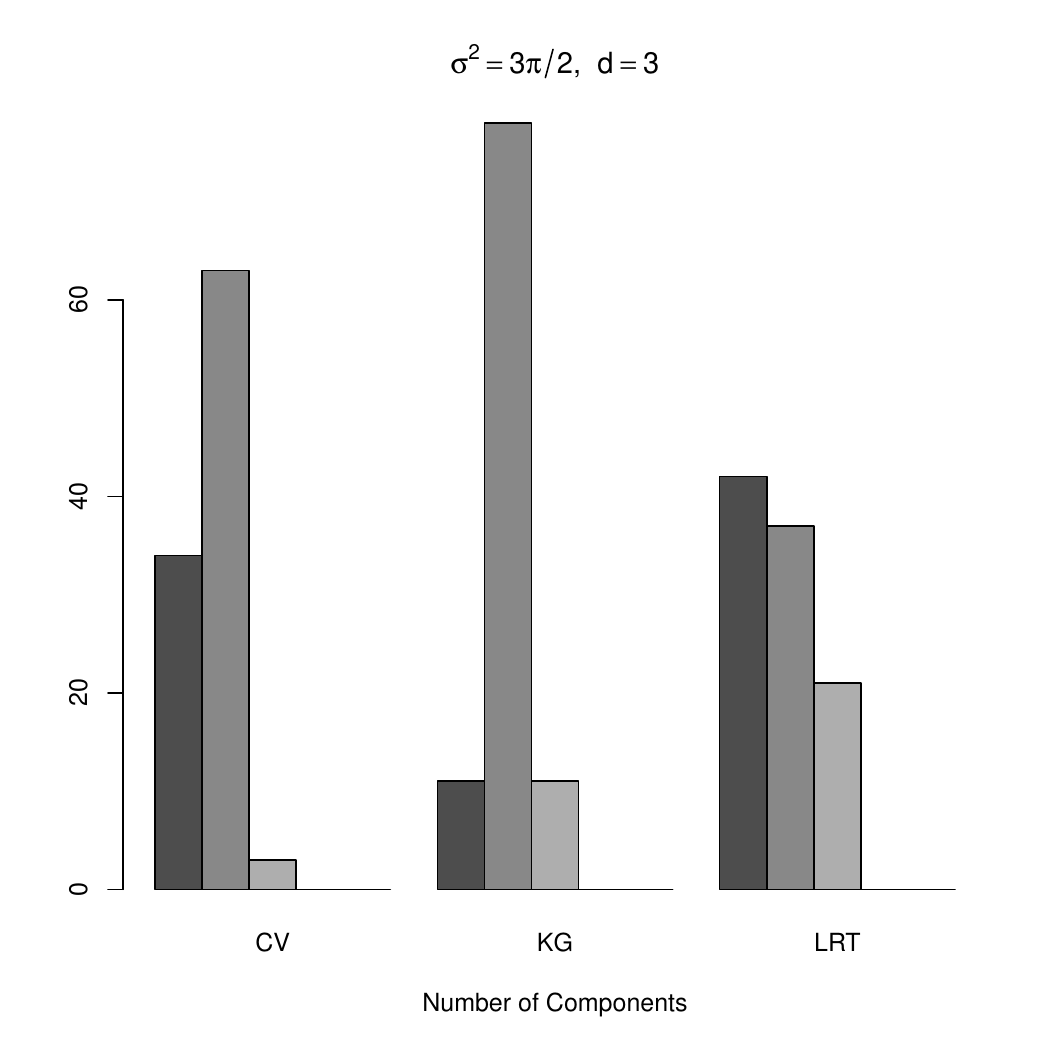}
\includegraphics[width=0.43\textwidth]{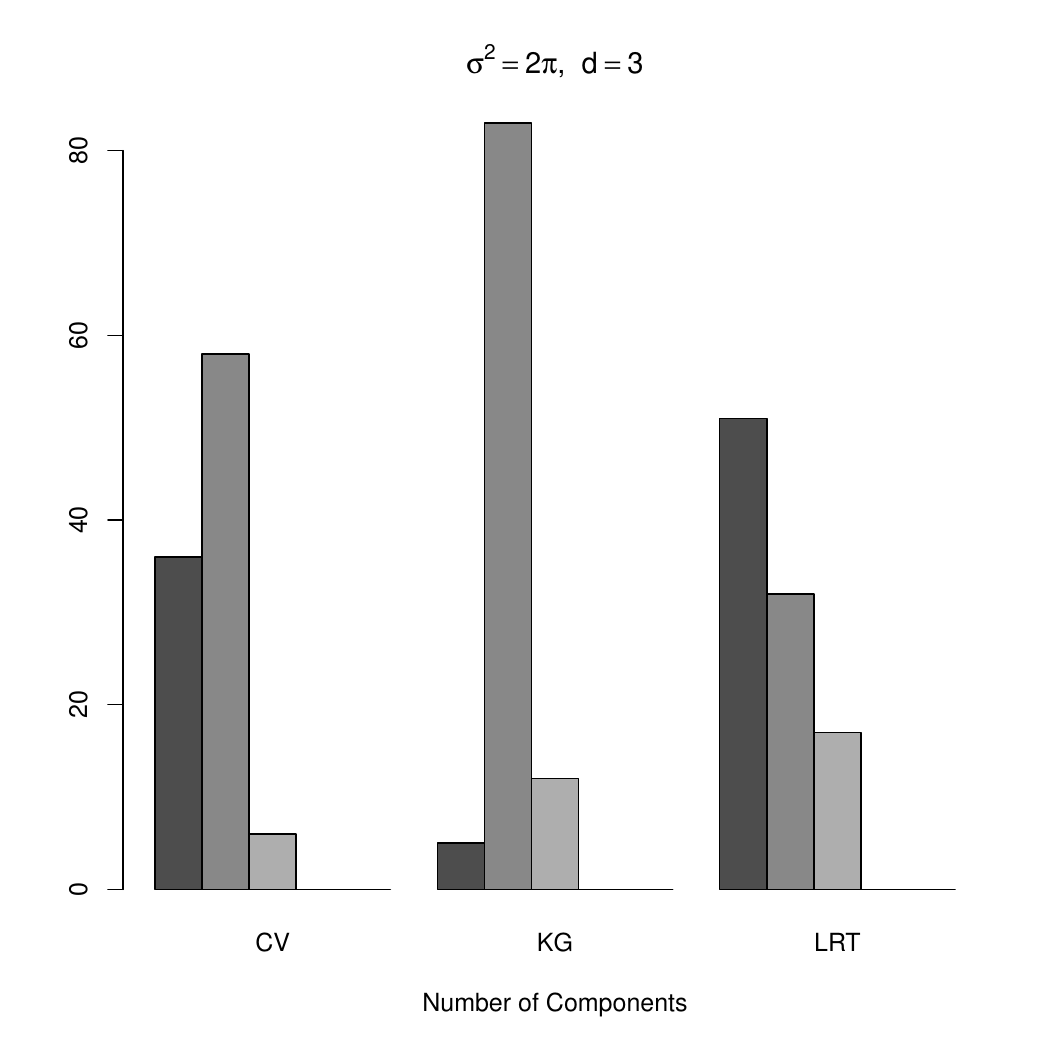}
\end{center}
\caption{Simulation study. Frequencies of selection of dimension for CV, KG, and LRT. The true value is $d=3$ and the sample size is $500$.}
\label{sm:fig:sim:3:500}
\end{figure}

In this section, we provide additional outputs from the simulation studies. Figures \ref{sm:fig:sim:2:50}--\ref{sm:fig:sim:2:500} and Figures \ref{sm:fig:sim:3:50}--\ref{sm:fig:sim:3:500} show the frequency of dimension selection when the true dimension is $d=2$ and $d=3$, respectively. Each figure corresponds to a different sample size ($N=50$, $100$, $500$), and each panel within the figures represents a different value of $\sigma^2$.

\begin{figure}
\begin{center}
\includegraphics[width=0.87\textwidth]{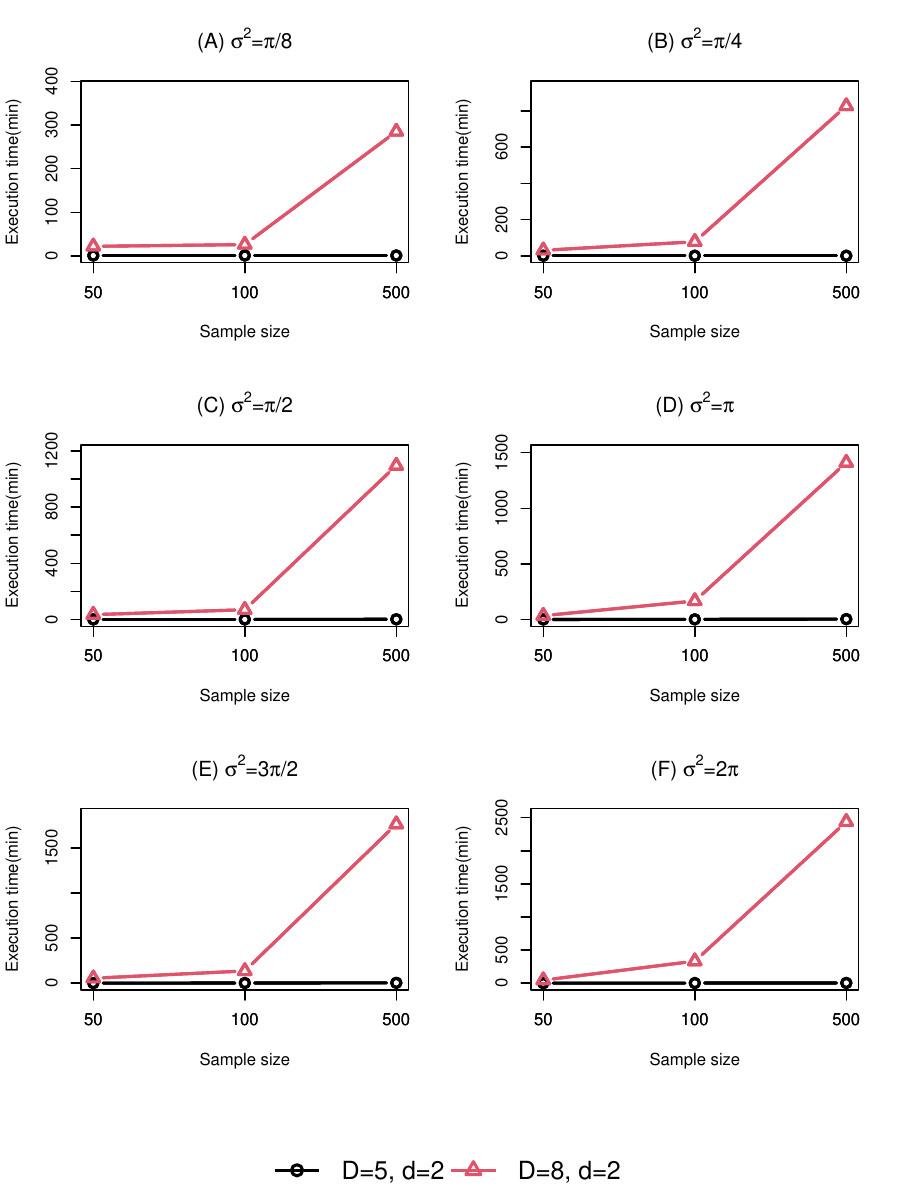} 
\end{center}
\caption{Mean execution times for N = (50,100,500), and $\sigma^2=(\pi/8, \pi/4, \pi/2, \pi, 3/2\pi, 2\pi)$ and $d=2$. Black line: $D=5$, red line: $D=8$}
\label{sm:fig:time:d2}
\end{figure}

\begin{figure}
\begin{center}
\includegraphics[width=0.87\textwidth]{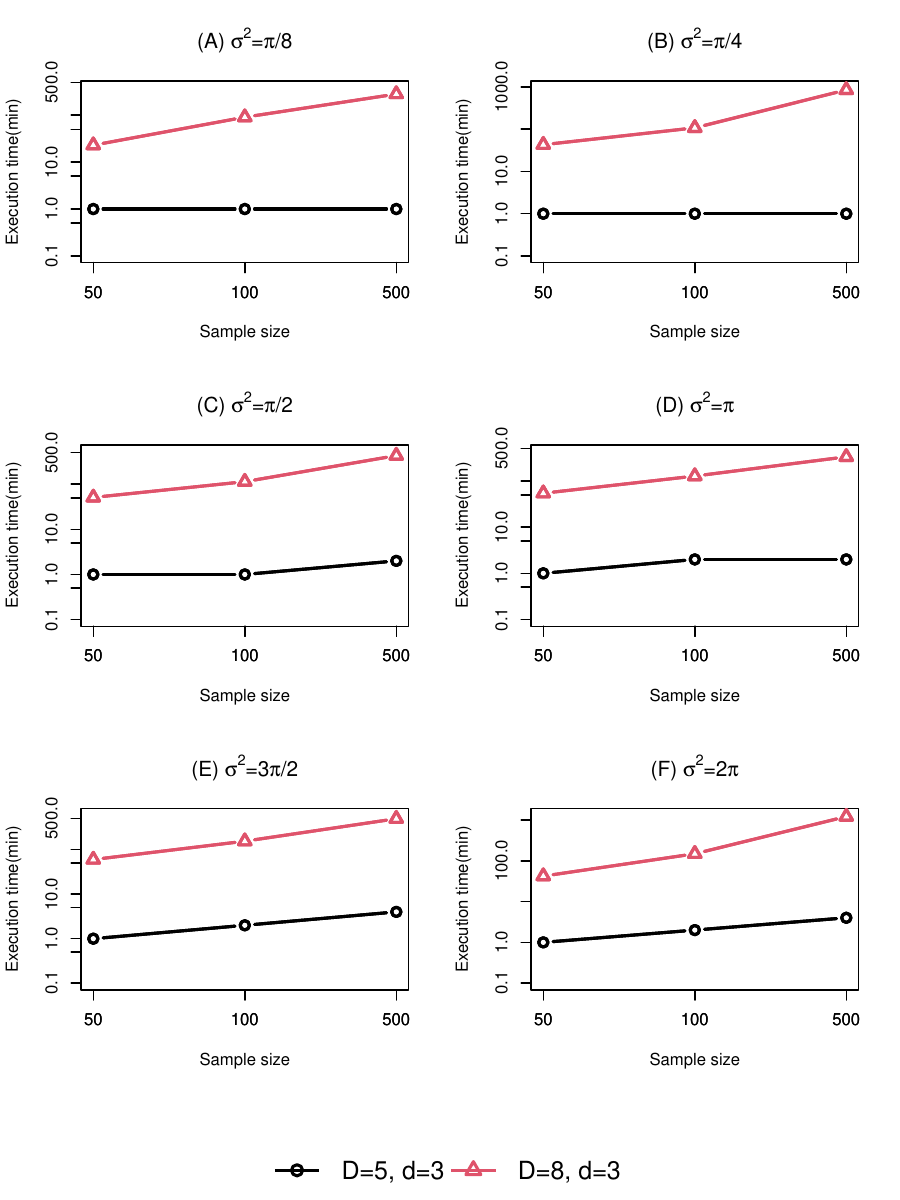} 
\end{center}
\caption{Mean execution times for N = (50, 100, 500), and $\sigma^2=(\pi/8, \pi/4, \pi/2, \pi, 3/2\pi, 2\pi)$ and $d=3$. Black line: $D=5$, red line: $D=8$}
\label{sm:fig:time:d3}
\end{figure}

\clearpage

\section{Large RNA data set}
\label{sm:sec:rna}
In this section, we present the analysis results from the large RNA (\textit{bigRNA}) dataset. Table \ref{sm:tab:rna:ncomp} shows the cumulative percentage of explained variance for each of the $23$ clusters after applying the TPPCA algorithm. The number of components was selected based on the LRT, KG, and CV criteria. Consistent with our simulation study findings, the LRT criterion consistently outperformed the others in this real-world scenario.

\begin{table}
\begin{center}
\begin{tabular}{|rr|rrr|rrr|}
\hline
\multicolumn{2}{|c}{} & \multicolumn{3}{|c}{TPPCA} &  \multicolumn{3}{|c|}{N. comp.} \\ 
Cluster \# & \# points & CV & KG & LRT & CV & KG & LRT \\
  \hline
\hline
  1 & 4917 & 29.85 & 83.37 & 93.74 &  1 & 4 & 5 \\ 
  2 & 477 & 51.73 & 74.53 & 96.77 &  1 & 2 & 5 \\ 
  3 & 232 & 55.21 & 73.72 & 95.09 &  2 & 3 & 5 \\ 
  4 & 211 & 47.82 & 76.59 & 96.72 &  1 & 2 & 5 \\ 
  5 & 140 & 69.27 & 84.29 & 97.44 &  1 & 2 & 5 \\ 
  6 & 139 &  46.19 & 73.12 & 97.09 &   1 & 2 & 5 \\ 
  7 & 139 & 32.01 & 77.93 & 93.66  &  1 & 3 & 5 \\ 
  8 & 138 & 83.71 & 83.71 & 98.57  &   2 & 2 & 5 \\ 
  9 & 128 & 27.53 & 83.81 & 92.14 &  1 & 4 & 5 \\ 
  10 & 122 & 79.49 & 66.27 & 97.90 &  2 & 1 & 5 \\ 
  11 & 85 & 57.67 & 72.89 & 92.57 & 2 & 3 & 5 \\ 
  12 & 84 & 76.75 & 76.75 & 96.54 &  2 & 2 & 5 \\ 
  13 & 79 & 68.52 & 83.87 & 95.51 &  2 & 3 & 5 \\ 
  14 & 72 & 71.80 & 71.80 & 97.04 &   2 & 2 & 5 \\ 
  15 & 60 & 70.99 & 70.99 & 96.77 &  2 & 2 & 5 \\ 
  16 & 60 & 99.45 & 98.59 & 99.75 &  2 & 1 & 3 \\ 
  17 & 59 & 68.00 & 68.00 & 95.77 &  2 & 2 & 5 \\ 
  18 & 54 & 65.62 & 80.77 & 80.77 &   2 & 3 & 3 \\ 
  19 & 52 & 82.17 & 82.17 & 97.71 &  2 & 2 & 5 \\ 
  20 & 46 & 96.63 & 94.11 & 99.56 &  2 & 1 & 4 \\ 
  21 & 35 & 65.49 & 81.16 & 95.97 &  2 & 3 & 5 \\ 
  22 & 33 & 87.17 & 87.17 & 96.61 &  2 & 2 & 4 \\ 
  23 & 28 & 64.27 & 64.27 & 64.27 &  2 & 2 & 2 \\ 
 \hline
\end{tabular}
\end{center}
\caption{RNA data set. Cumulative of percentage of variance explained by selected components (using CV, KG and LRT) of TPPCA.}
\label{sm:tab:rna:ncomp}
\end{table}

\newpage
\bibliography{tppca}   

\begin{thebibliography}{65}
\providecommand{\natexlab}[1]{#1}
\providecommand{\url}[1]{\texttt{#1}}
\expandafter\ifx\csname urlstyle\endcsname\relax
  \providecommand{\doi}[1]{doi: #1}\else
  \providecommand{\doi}{doi: \begingroup \urlstyle{rm}\Url}\fi

\bibitem[Agrawal et~al.(2020)Agrawal, Chiu, Le, Halperin, and
  Sankararaman]{AgrawalETAL:2020}
A.~Agrawal, A.~M. Chiu, M.~Le, E.~Halperin, and S.~Sankararaman.
\newblock Scalable probabilistic pca for large-scale genetic variation data.
\newblock \emph{PLoS genetics}, 16\penalty0 (5):\penalty0 e1008773, 2020.

\bibitem[Akaike(1987)]{Akaike1987}
H.~Akaike.
\newblock Factor analysis and {AIC}.
\newblock \emph{Psychometrika}, 52:\penalty0 317--332, 1987.

\bibitem[Altis et~al.(2007)Altis, Nguyen, Hegger, and Stock]{Altis2007}
A.~Altis, P.~Nguyen, R.~Hegger, and G.~Stock.
\newblock Dihedral angles principal component analysis of molecular dynamics
  simulations.
\newblock \emph{Journal of Chemical Physics}, 26:\penalty0 244111.1--244111,
  2007.

\bibitem[Bartlett(1950)]{Bartlett1950}
M.S. Bartlett.
\newblock Tests of significance in factor analysis.
\newblock \emph{British Journal of Psychology (Statistics Section)},
  3:\penalty0 77--85, 1950.

\bibitem[Basilevsky(1994)]{Basilevsky1994}
A.~Basilevsky.
\newblock \emph{Statistical Factor Analysis and Related Methods}.
\newblock Wiley, New York, 1994.

\bibitem[Boothby(1986)]{Boothby1986}
W.~M. Boothby.
\newblock \emph{An Introduction to Differentiable Manifolds and Riemannian
  Geometry}.
\newblock Academic Press, New York, 1986.

\bibitem[Bozdogan(1994)]{Bozdogan1994}
H.~Bozdogan.
\newblock On the frontiers of statistical modeling: An informational approach.
\newblock \emph{Proceedings of the first US/Japan conference}, 1994.

\bibitem[Brock et~al.(2008)Brock, Pihur, Datta, and Datta]{Brock2008}
G.~Brock, V.~Pihur, S.~Datta, and S.~Datta.
\newblock clvalid: An r package for cluster validation.
\newblock \emph{Journal of statistical Software}, 25:\penalty0 1--22, 2008.

\bibitem[Bunch and Nielsen(1978)]{BunchNielsen}
J.~R. Bunch and C.~P. Nielsen.
\newblock Updating the singular value decomposition.
\newblock \emph{Numerische Mathematik}, 31:\penalty0 111--129, 1978.

\bibitem[Bunch et~al.(1978)Bunch, Nielsen, and Sorensen]{Bunchetal}
J.~R. Bunch, C.~P. Nielsen, and D.~C. Sorensen.
\newblock Rank one modification of the symmetric eigenproblem.
\newblock \emph{Numerische Mathematik}, 31:\penalty0 31--48, 1978.

\bibitem[Charrad et~al.(2014)Charrad, Ghazzali, Boiteau, and
  Niknafs]{charrad2014}
M.~Charrad, N.~Ghazzali, V.~Boiteau, and A.~Niknafs.
\newblock Nbclust: an r package for determining the relevant number of clusters
  in a data set.
\newblock \emph{Journal of statistical software}, 61:\penalty0 1--36, 2014.

\bibitem[Cox and Cox(2008)]{CoxCox2008}
M.~A.~A. Cox and T.~F. Cox.
\newblock Multidimensional scaling.
\newblock In \emph{Handbook of data visualization}, pages 315--347. Springer,
  2008.

\bibitem[Cox and Cox(1991)]{CoxCox1991}
T.~F. Cox and M.~A.~A. Cox.
\newblock Multidimensional scaling on a sphere.
\newblock \emph{Communications in Statistics-Theory and Methods}, 20\penalty0
  (9):\penalty0 2943--2953, 1991.

\bibitem[Duarte and Pyle(1998)]{Duarte1998}
C.~M. Duarte and A.~M. Pyle.
\newblock Stepping through an rna structure: a novel approach to conformational
  analysis.
\newblock \emph{Journal of molecular biology}, 284\penalty0 (5):\penalty0
  1465--1478, 1998.

\bibitem[Eltzner et~al.(2018)Eltzner, Huckemann, and Mardia]{Eltzner2018}
B.~Eltzner, S.~Huckemann, and K.V. Mardia.
\newblock Torus principal component analysis with applications to {RNA}
  structure.
\newblock \emph{Annals of Applied Statistics}, 12\penalty0 (2):\penalty0
  1332--1359, 2018.

\bibitem[Fletcher et~al.(2004)Fletcher, Lu, Pizar, and Joshi]{Fletcher2004}
P.~T. Fletcher, C.P. Lu, S.~Pizar, and S.~C. Joshi.
\newblock Principal geodesic analysis for the study of nonlinear statistics of
  shape.
\newblock \emph{IEEE Trans. Med. Imaging}, 23:\penalty0 995--1005, 2004.

\bibitem[Fotouhi and Golalizadeh(2012)]{Fotouhi2012}
H.~Fotouhi and M.~Golalizadeh.
\newblock Exploring the variability of dna molecules via principal geodesic
  analysis on the shape space.
\newblock \emph{Journal of Applied Statistics}, 39\penalty0 (10):\penalty0
  2199--2207, 2012.

\bibitem[Fotouhi and Golalizadeh(2015)]{Fotouhi2015}
H.~Fotouhi and M.~Golalizadeh.
\newblock Highly resistant gradient descent algorithm for computing intrinsic
  mean shape on similarity shape spaces.
\newblock \emph{Statistical Papers}, 56:\penalty0 391--410, 2015.

\bibitem[Garc{\'\i}a-Portugu{\'e}s and Verdebout(2021)]{Garcia2021}
E.~Garc{\'\i}a-Portugu{\'e}s and T.~Verdebout.
\newblock sphunif: Uniformity tests on the circle, sphere, and hypersphere.
\newblock \emph{R package version}, 1\penalty0 (1):\penalty0 8, 2021.

\bibitem[Garc{\'\i}a-Portugu{\'e}s et~al.(2020)Garc{\'\i}a-Portugu{\'e}s,
  Paindaveine, and Verdebout]{Garcia2020}
E.~Garc{\'\i}a-Portugu{\'e}s, D.~Paindaveine, and T.~Verdebout.
\newblock On optimal tests for rotational symmetry against new classes of
  hyperspherical distributions.
\newblock \emph{Journal of the American Statistical Association}, 115\penalty0
  (532):\penalty0 1873--1887, 2020.

\bibitem[Guttman(1954)]{Guttman1954}
L.~Guttman.
\newblock Some necessary conditions for common factor analysis.
\newblock \emph{Psychometrika}, 19:\penalty0 149--162, 1954.

\bibitem[Hastie and Stuetzle(1989)]{HastieStuetzle1989}
T.~Hastie and W.~Stuetzle.
\newblock Principal curves.
\newblock \emph{Journal of American Statistics Association}, 84:\penalty0
  502--516, 1989.

\bibitem[Hayashi et~al.(2007)Hayashi, Bentler, and Yuan]{Hayashi2007}
K.~Hayashi, P.M. Bentler, and K.H. Yuan.
\newblock On the likelihood ratio test for the number of factors in exploratory
  factor analysis.
\newblock \emph{Structural Equation Modeling}, 14\penalty0 (3):\penalty0
  505--526, 2007.

\bibitem[Hotelling(1933)]{Hotelling1933}
H.~Hotelling.
\newblock Analysis of a complex of statistical variables into principal
  components.
\newblock \emph{Journal of Educational Psychology}, 24:\penalty0 417--441,
  1933.

\bibitem[Hubert and Arabie(1985)]{HubertArabie1985}
L.~Hubert and P.~Arabie.
\newblock Comparing partitions.
\newblock \emph{Journal of classification}, 2:\penalty0 193--218, 1985.

\bibitem[Huckemann and Ziezold(2006)]{HuckemannZiezold2006}
S.~Huckemann and H.~Ziezold.
\newblock Principal component analysis for {R}iemannian manifolds, with
  application to triangular shape spaces.
\newblock \emph{Advances in Applied Probability}, 38:\penalty0 299--319, 2006.

\bibitem[Jackson(1991)]{Jackson1991}
J.~Jackson.
\newblock \emph{A User's Guide to Principal Component}.
\newblock John Wiley, New York, 1991.

\bibitem[Jolliffe(2002)]{Jolliffe2002}
I.~Jolliffe.
\newblock \emph{Principal Component Analysis}.
\newblock Springer, New York, 2002.

\bibitem[J\"oreskog(1967)]{Joreskog1967}
K.G. J\"oreskog.
\newblock Some contributions to maximum likelihood factor analysis.
\newblock \emph{Psychometrika}, 32:\penalty0 443--482, 1967.

\bibitem[Jung et~al.(2010)Jung, Liu, Marron, and Pizer]{Jung2010}
S.~Jung, X.~Liu, J.S. Marron, and S.~Pizer.
\newblock Generalized {PCA} via the backward stepwise approach in image
  analysis.
\newblock \emph{Brain, Body and Machine}, 83:\penalty0 111--123, 2010.

\bibitem[Jung et~al.(2012)Jung, Dryden, and Marron]{Jung2012}
S.~Jung, I.~L. Dryden, and J.S. Marron.
\newblock Analysis of principal nested spheres.
\newblock \emph{Biometrika}, 99:\penalty0 551--568, 2012.

\bibitem[Kaiser(1960)]{Kaiser1960}
H.F. Kaiser.
\newblock The application of electronic computers to factor analysis.
\newblock \emph{Educational and Psychological Measurement}, 20:\penalty0
  141--151, 1960.

\bibitem[Karcher(1977)]{Karcher1977}
H.~Karcher.
\newblock {R}iemannian center of mass and mollifier smoothing.
\newblock \emph{Communications on Pure and Applied Math}, 30\penalty0
  (5):\penalty0 509--541, 1977.

\bibitem[Kaufman and Rousseeuw(2009)]{kaufman2009}
L.~Kaufman and P.~J. Rousseeuw.
\newblock \emph{Finding groups in data: an introduction to cluster analysis}.
\newblock John Wiley \& Sons, 2009.

\bibitem[Kim et~al.(2018)Kim, Kang, Huo, Park, and Tseng]{kimETAL:2018}
S.~H. Kim, D.~Kang, Z.~Huo, Y.~Park, and G.~C. Tseng.
\newblock Meta-analytic principal component analysis in integrative omics
  application.
\newblock \emph{Bioinformatics}, 34\penalty0 (8):\penalty0 1321--1328, 2018.

\bibitem[Koutroumbas and Theodoridis(2008)]{Koutroumbas2008}
K.~Koutroumbas and S.~Theodoridis.
\newblock \emph{Pattern recognition}.
\newblock Academic Press, 2008.

\bibitem[Krzanowski(1983)]{Krzanowski1983}
W.~J. Krzanowski.
\newblock Cross-validatory choice in principal component analysis: Some
  sampling results.
\newblock \emph{Journal of Statistical Computation and Simulation},
  18:\penalty0 299--314, 1983.

\bibitem[Krzanowski(1987)]{Krzanowski1987}
W.~J. Krzanowski.
\newblock Cross-validation in principal component analysis.
\newblock \emph{Biometrics}, 43:\penalty0 575--584, 1987.

\bibitem[Lawley and Maxwell(1971)]{LawleyMaxwel1971}
D.N. Lawley and A.E. Maxwell.
\newblock \emph{Factor Analysis as a Statistical Method}.
\newblock Elsevier, New York, 1971.

\bibitem[Maadooliat et~al.(2016)Maadooliat, Zhou, Najibi, Gao, and
  Huang]{Maadooliat2016}
M.~Maadooliat, L.~Zhou, S.~M. Najibi, X.~Gao, and J.~Z. Huang.
\newblock Collective estimation of multiple bivariate density functions with
  application to angular-sampling-based protein loop modeling.
\newblock \emph{Journal of the American Statistical Association}, 111\penalty0
  (513):\penalty0 43--56, 2016.

\bibitem[Mardia et~al.(2022)Mardia, Wiechers, Eltzner, and
  Huckemann]{Mardia2021}
K.~V. Mardia, H.~Wiechers, B.~Eltzner, and S.~Huckemann.
\newblock Principal component analysis and clustering on manifolds.
\newblock \emph{Journal of Multivariate Analysis}, 188:\penalty0 104862, 2022.

\bibitem[Mardia(1972)]{Mardia1972}
K.V. Mardia.
\newblock \emph{Statistics of Directional data}.
\newblock Academic Press, London, 1972.

\bibitem[Mardia et~al.(1979)Mardia, Kent, and Bibby]{Mardia1979}
K.V. Mardia, J.T. Kent, and J.M. Bibby.
\newblock \emph{Multivariate Analysis}.
\newblock Academic Press, London, 1979.

\bibitem[Moghimbeygi and Nodehi(2022)]{moghimbeygi2022}
M.~Moghimbeygi and A.~Nodehi.
\newblock Multinomial principal component logistic regression on shape data.
\newblock \emph{Journal of Classification}, 39\penalty0 (3):\penalty0 578--599,
  2022.

\bibitem[Mu et~al.(2005)Mu, Nguyen, and Stock]{Mu2005}
Y.~Mu, P.H. Nguyen, and G.~Stock.
\newblock Energy landscape of a small peptide revealed by dihedral angles
  principal component analysis.
\newblock \emph{Proteins}, 58:\penalty0 45, 2005.

\bibitem[Nodehi et~al.(2015)Nodehi, Golalizadeh, and Heydari]{Nodehi2015}
A.~Nodehi, M.~Golalizadeh, and A.~Heydari.
\newblock Dihedral angles principal geodesic analysis using nonlinear
  statistics.
\newblock \emph{Journal of Applied Statistics}, 42:\penalty0 1962--1972, 2015.

\bibitem[Nodehi et~al.(2021)Nodehi, Golalizadeh, Maadooliat, and
  Agostinelli]{Nodehi2018}
A.~Nodehi, M.~Golalizadeh, M.~Maadooliat, and C.~Agostinelli.
\newblock Estimation of parameters in multivariate wrapped models for data on a
  p-torus.
\newblock \emph{Computational Statistics}, 36:\penalty0 193--215, 2021.

\bibitem[Novembre and Stephens(2008)]{NovembreStephens:2008}
J.~Novembre and M.~Stephens.
\newblock Interpreting principal component analyses of spatial population
  genetic variation.
\newblock \emph{Nature genetics}, 40\penalty0 (5):\penalty0 646--649, 2008.

\bibitem[Panaretos et~al.(2014)Panaretos, Pham, and Yao]{Panaretos2014}
V.M. Panaretos, T.~Pham, and Z.~Yao.
\newblock Principal flows.
\newblock \emph{Journal of the American Statistical Association}, 109:\penalty0
  424--436, 2014.

\bibitem[Pearson(1901)]{Pearson1901}
K.~Pearson.
\newblock On lines and planes of closest fit to systems of points in space.
\newblock \emph{The London, Edinburgh and Dublin Philosphical Magazine and
  Journal of Science}, 2:\penalty0 559--572, 1901.

\bibitem[Pennec(2006)]{Pennec2006}
X.~Pennec.
\newblock Intrinsic statistics on {R}iemannian manifolds: Basic tools for
  geometric measurements.
\newblock \emph{Journal of Mathematics and Imaging Vision}, 25:\penalty0
  127--154, 2006.

\bibitem[Price et~al.(2006)Price, Patterson, Plenge, Weinblatt, Shadick, and
  Reich]{PriceETAL:2006}
A.~L. Price, N.~J. Patterson, R.~M. Plenge, M.~E. Weinblatt, N.~A. Shadick, and
  D.~Reich.
\newblock Principal components analysis corrects for stratification in
  genome-wide association studies.
\newblock \emph{Nature genetics}, 38\penalty0 (8):\penalty0 904--909, 2006.

\bibitem[Priv{\'e} et~al.(2020)Priv{\'e}, Luu, Blum, McGrath, and
  Vilhj{\'a}lmsson]{PriveETAL:2020}
F.~Priv{\'e}, K.~Luu, M.~G.B. Blum, J.~J. McGrath, and B.~J. Vilhj{\'a}lmsson.
\newblock Efficient toolkit implementing best practices for principal component
  analysis of population genetic data.
\newblock \emph{Bioinformatics}, 36\penalty0 (16):\penalty0 4449--4457, 2020.

\bibitem[Riccardi et~al.(2009)Riccardi, Nguyen, and Stock]{Riccardi2009}
L.~Riccardi, P.~H. Nguyen, and G.~Stock.
\newblock Free-energy landscape of {RNA} hairpins constructed via dihedral
  angle principal component analysis.
\newblock \emph{The Journal of Physical Chemistry B}, 113\penalty0
  (52):\penalty0 16660--16668, 2009.

\bibitem[Roweis(1998)]{Roweis1998}
S.~Roweis.
\newblock {EM} algorithms for {PCA} and {SPCA}.
\newblock \emph{Advances in Neural Information Processing Systems},
  10:\penalty0 626--632, 1998.

\bibitem[Sargsyan et~al.(2012)Sargsyan, Wright, and Lim]{Sargsyan2012}
K.~Sargsyan, J.~Wright, and C.~Lim.
\newblock Geopca: a new tool for multivariate analysis of dihedral angles based
  on principal component geodesics.
\newblock \emph{Nucleic Acids Research}, 40:\penalty0 25, 2012.

\bibitem[Schwarz(1978)]{Schwarz1978}
G.~Schwarz.
\newblock Estimating the dimension of a model.
\newblock \emph{The Annals of Statistics}, 6:\penalty0 461--464, 1978.

\bibitem[Sittel et~al.(2017)Sittel, Filk, and Stock]{Sittel2017}
F.~Sittel, T.~Filk, and G.~Stock.
\newblock Principal component analysis on a torus: Theory and application to
  protein dynamics.
\newblock \emph{Journal of Chemical Physics}, 147:\penalty0
  244101.1--244101.12, 2017.

\bibitem[Tibshirani et~al.(2001)Tibshirani, Walther, and
  Hastie]{Tibshirani2001}
R.~Tibshirani, G.~Walther, and T.~Hastie.
\newblock Estimating the number of clusters in a data set via the gap
  statistic.
\newblock \emph{Journal of the Royal Statistical Society: Series B (Statistical
  Methodology)}, 63\penalty0 (2):\penalty0 411--423, 2001.

\bibitem[Tipping and Bishop(1997)]{TippingBishop1997}
M.E. Tipping and C.M. Bishop.
\newblock \emph{Probabilistic Principal Component Analysis}.
\newblock Neural Computing Research, Aston University, Birmingham, 1997.
\newblock Technical Report.

\bibitem[Tipping and Bishop(1999)]{TippingBishop1999}
M.E. Tipping and C.M. Bishop.
\newblock Probabilistic principal component analysis.
\newblock \emph{Journal of the Royal Statistical Society}, 61:\penalty0
  611--622, 1999.

\bibitem[Wiseman et~al.(2021)Wiseman, Samra, Lara, Penrice, and
  Goddard]{Wiseman2021}
D.~N. Wiseman, N.~Samra, M.~M.~R. Lara, S.~C. Penrice, and A.~D. Goddard.
\newblock The novel application of geometric morphometrics with principal
  component analysis to existing g protein-coupled receptor (gpcr) structures.
\newblock \emph{Pharmaceuticals}, 14:\penalty0 953, 2021.

\bibitem[Wold(1978)]{Wold1978}
S.~Wold.
\newblock Cross-validatory estimation of the number of components in factor and
  principal component models.
\newblock \emph{Technometrics}, 20:\penalty0 397--405, 1978.

\bibitem[Zhang and Fletcher(2013)]{ZhangFletcher2013}
M.~Zhang and P.T. Fletcher.
\newblock \emph{Probabilistic Principal Geodesic Analysis}.
\newblock In Neural Information Processing Systems (NIPS), 2013.
\newblock Nevada, United States.

\bibitem[Zoubouloglou et~al.(2022)Zoubouloglou, Garc{\'\i}a-Portugu{\'e}s, and
  Marron]{zoubouloglou2021}
P.~Zoubouloglou, E.~Garc{\'\i}a-Portugu{\'e}s, and J.S. Marron.
\newblock Scaled torus principal component analysis.
\newblock \emph{Journal of Computational and Graphical Statistics},
  32:\penalty0 1024--1035, 2022.

\end{thebibliography}

\end{document}